\documentclass[prc,amsfonts,reprint, superscriptaddress,showpacs, 
notitlepage,
longbibliography
]{revtex4-1} 

\usepackage{hyperref}
\hypersetup{colorlinks=true, linkcolor=blue, citecolor=red, urlcolor=blue}
\usepackage{tikz}
\usetikzlibrary{arrows}
\usetikzlibrary{decorations.pathmorphing}
\usepackage{hyperref}
\usepackage{amssymb}
\usepackage{amsmath}
\usepackage{color}
\usepackage{xcolor}
\usepackage{graphicx}
\usepackage{soul}
\usepackage{mathtools}
\usepackage{braket}

\newcommand{\comment}[1]{}

\newcommand{\mc}[1]{\mathcal{#1}}

\newcommand{\ii}{{\rm i}}
\newcommand{\tr}{{\rm Tr}}
\newcommand{\Tr}{{\rm Tr}}
\newcommand{\Real}{{\rm Re}}

\newcommand{\be}{\begin{equation}\begin{aligned}\label}{}
\newcommand{\bes}{\begin{equation}\small\begin{aligned}}{}

\newcommand{\ee}{\end{aligned}\end{equation}}
\usepackage{pgfplots}
\pgfplotsset{width=10cm,compat=1.9}
\usetikzlibrary{patterns}
\usetikzlibrary{decorations.markings}

\begin{document}

\title{A unified diagrammatic approach to quantum transport in few-level junctions for bosonic and fermionic reservoirs: Application to the quantum Rabi model}

\author{L. Magazz\`u}
\affiliation{Institute for Theoretical Physics, University of Regensburg, 93040 Regensburg, Germany}

\author{E. Paladino}
\affiliation{Dipartimento di Fisica e Astronomia Ettore Majorana, Universit\`a di Catania, Via S. Sofia 64, I-95123, Catania, Italy;\\
INFN, Sez. Catania, I-95123, Catania, Italy; and CNR-IMM, Via S. Sofia 64, I-95123, Catania, Italy}

\author{M. Grifoni}
\affiliation{Institute for Theoretical Physics, University of Regensburg, 93040 Regensburg, Germany}

\date{\today}

\begin{abstract}

We apply the Nakajima-Zwanzig approach to open quantum systems to study steady-state transport across generic multi-level junctions coupled to bosonic or fermionic reservoirs. The method allows for a unified diagrammatic formulation in Liouville space, with diagrams being classified according to an expansion in the coupling strength between the reservoirs and the junction. Analytical, approximate expressions are provided up to fourth order for the steady-state boson transport that generalize to multi-level systems the known results for the low-temperature thermal conductance in the spin-boson model. The formalism is applied to the problem of heat transport in a qubit-resonator junction modeled by the quantum Rabi model. Nontrivial transport features emerge as a result of the interplay between the qubit-oscillator detuning and coupling strength. For quasi-degenerate spectra, nonvanishing steady-state coherences  cause a suppression of the thermal conductance.

\end{abstract}

\maketitle

\section{Introduction}

Current advances in quantum technologies allow to explore quantum effects in the transport of charge and energy through multi-level junctions in a fully controllable fashion and in the presence of distinct transport regimes~\cite{vanderWiel2000,Nazarov2009,Andergassen2010,Heikkila2013,Segal2016,Pekola2021}. The latter are determined by the relative magnitude of few important parameters, namely the temperature, interactions or nonlinearities, and coupling between the parts of the setup that exchange particles or energy.
In a typical transport setup, a number of large, noninteracting bosonic or fermionic environments, each with their own temperature and chemical potential, are connected to a common, small central system, which displays level quantization. Examples are quantum dots, low-dimensional nanostructures where a small number of electrons are confined in many-body bound states forming artificial atoms~\cite{Hanson2007, Laird2015}, or superconducting qubits, quantum two-level systems implemented by superconducting circuits based on Josephson junctions~\cite{Vool2017}. In these two examples of junctions, the Coulomb interaction and nonlinearity of the spectrum, respectively, can give rise to nontrivial transport features and render the transport problem generally hard to treat.   
By now, several approaches have been devised, that are best suited for specific parameter regimes. For example, when the Coulomb interaction is absent or admits a mean field-like treatment, the Green's function approach provides a convenient tool for studying transport with no limitation on the coupling to the environments~\cite{Bruus2004,Haug2008,Cuevas2010}. In the opposite regime of strongly interacting/nonlinear junctions, the kinetic equations for the reduced density operator are the tool of choice in that the junction is treated exactly and the coupling to the environments is addressed in some approximate, non-necessarily perturbative, fashion~\cite{Leggett1987,Koenig1996,Schoeller2009,Donarini2024}.\\
\indent In the kinetic equations approach, the junction is considered as an open quantum system and the relevant observables are calculated via the system reduced density matrix (RDM)~\cite{Breuer2002,Li2005,Leijnse2008,Timm2008,Timm2011, Blum2012,Weiss2012,Donarini2024}. The method exploits the noninteracting nature of the environments to trace them out exactly. This can result in a formulation in terms of Feynman-Vernon influence functional~\cite{Feynman1963} which is at the basis of the numerically-exact approach of hierarchical equations of motion~\cite{Tanimura2006, Jin2008,Tanimura2020,Xu2022}. The dynamics of the RDM can then be cast in the form of a generalized master equation (GME) where environmental influences are encapsulated in a memory kernel~\cite{Weiss2012, Magazzu2022}. A different route, based on the projection operator formalism yields the dynamical equation for the RDM in the form of the Nakajima-Zwanzig (NZ) equation~\cite{Nakajima1958,Zwanzig1960,Hashitsumae1977}. In this approach, the formal expression for the memory kernel is suitable for expansion in the system-environment coupling. A leading-order expansion of the NZ equation combined with a Markovian approximation yields the weak coupling master equation (ME) of the Redfield type~\cite{Blum2012}. This is in turn suitable for further approximations, returning the celebrated master equation in the Gorini-Kossakowski-Lindblad-Sudarshan form~\cite{Gorini1976,Lindblad1976}, and improvements thereof~\cite{Nathan2020,Mozgunov2020,McCauley2020,Trushechkin2021, Nakamura2024}. An alternative route is the time-convolutionless (TCL) master equation approach, where, under appropriate conditions, a formally exact time-local expression for the reduced dynamics is obtained using the projector operator technique. The resulting time-local generator is well suited for a perturbation expansion in the system-bath coupling~\cite{VanKampen1974I, VanKampen1974II,Hashitsumae1977}. 
The NZ and TCL formalisms are also the starting points for nonperturbative approaches based on numerical evaluation of the memory kernel or the TCL generator of the dynamics~\cite{Shi2003,Zhang2006,Kidon2015}.\\ 
\indent In steady-state transport, the topic of this work, the NZ is a convenient approach, since  the finite memory time of the kernels renders the Markovian approximations not necessary. In this case, the Redfield equation, with a partial or full secular treatment, as required by consistency with perturbation theory and according to the structure of the spectrum of the junction, provides a reliable tool with the only requirement of weak coupling, in the appropriate temperature regime~\cite{Kato2015,Hartmann2020,Trushechkin2021,Xu2021}.  \\
\indent When considering transport across junctions, higher-order processes might become important, even at weak coupling. Higher-order phenomena involving the so-called \emph{cotunneling} processes, are well-studied in the context of charge transport~\cite{Sukhorukov2001, Leijnse2008, Koller2010,Eckern2020,Tesser2022, Donarini2024}: The virtual processes involved yield a nonvanishing conductance at zero voltage bias whereas the leading order (\emph{sequential tunneling}) would predict a suppressed conductance due to Coulomb interaction in the dot (Coulomb blockade). 
In the context of bosonic heat transport, a similar behavior is found at low temperatures, where the sequential tunneling current is exponentially suppressed  and the dominant contribution to the thermal conductance of a weakly-coupled two-level system junction is the cotunneling~\cite{Ruokola2011,Yamamoto2018}. However, results beyond the sequential tunneling in the context of bosonic heat transport are mostly limited to
two-level systems and are found with the generalized golden rule or the Green's function approach~\cite{Velizhanin2010,Ruokola2011,Yang2014, Yamamoto2018, Bhandari2021}. A diagrammatic formulation based on the Dyson series was applied to a harmonic oscillator junction in~\cite{Thingna2014}.
In~\cite{Ferguson2021}, a formalism based on the diagrammatic unravelling of the TCL master equation is carried out and compared with the results from Fermi's golden rule. A different approach to thermal transport, based on the reaction coordinate mapping~\cite{Iles-Smith2016, Strasberg2018, Anto-Sztrikacs2021, Anto-Sztrikacs2022, Anto-Sztrikacs2023}, can account for strong coupling to the baths.
The strong system-bath coupling is also captured by Redfield-type MEs following from on a polaron transformation~\cite{Wang2015, Xu2016, Wang2017}.
A review of several analytical and numerical approaches to transport in different coupling regimes is provided, e.g.  in~\cite{Landi2022}. An exact expression for the heat current through a qubit coupled to two bosonic baths for $\alpha_1+\alpha_2=1/2$, where $\alpha_i$ are the slopes of the linear low-frequency Ohmic behavior of the baths' spectral densities, has been found in~\cite{Bhandari2021}.\\ 
\indent In this work, we describe a generalized master equation formulation in Liouville space which allows one to treat particle and/or heat transport in multi-level junctions coupled to fermionic or bosonic setups on the same footing. Specifically, we extend to the bosonic case a diagrammatic formulation established for  interacting fermionic junctions~\cite{Donarini2024}. The approach provides a method for calculating fermionic and bosonic currents within an identical formalism in the spirit of the theory of open quantum systems in Liouville space. A similar point of view is adopted in~\cite{Schoeller2009} to derive nonperturbative renormalization group equations for the dynamics of the density operator. The starting points are, in our case, the NZ equation for the RDM and the related expression for the particle/energy current. The diagrammatic formalism in Liouville space allows for a systematic perturbation expansion in the system-environment coupling using simple diagrammatic rules.  Explicit expressions, up to fourth order, are given for bosonic heat transport in a generic junction.  Our considerations are then applied to a concrete heat transport problem, in which the junction is composed by a coupled qubit-oscillator system, a realization of the so-called quantum Rabi model~\footnote{Part of the key results on heat transport in the Rabi model are in the companion letter.}: The junction between bosonic heat baths is in this case the fundamental object of quantum electrodynamics (QED), being the archetypal system to study light-matter interaction. We consider a superconducting circuit realization of it~\cite{Blais2021} whereby a flux qubit is coupled to an LC oscillator. Superconducting circuit platforms offer the possibility to operate in a vast range of coupling strengths $g$, from the weak to the ultrastrong coupling (USC) regime~\cite{Yoshihara2017,Forn-Diaz2018review,Kockum2019,Falci2019,Giannelli2024}. In the latter case, the frequency associated to the coupling strength is of the same order of magnitude of the ones of the isolate constituents of the Rabi model and perturbative approaches in $g$, e.g. the rotating wave approximation (RWA), which are appropriate in quantum optical systems, break down. 
We show that the heat transport properties of the setup are determined by both the qubit-oscillator coupling  and detuning. The coupling induces a qubit-oscillator entanglement that dictates the low-energy spectrum of the Rabi model~\cite{Hausinger2010PRA, Ashhab2010}. Application of a bias on the qubit can tune in- and out- of resonance the two elements. When the coupling is not too strong, this feature manifests itself in the so-called heat valve behavior, namely an enhancement (suppression) of the heat current when the resonance condition is (not) attained~\cite{Ronzani2018, Pekola2021}. Quasi-degeneracies in the spectrum yield steady-state coherences that suppress the thermal conductance even at high temperatures. At low temperature, when the Rabi system is essentially in its ground state, transport occurs via virtual (cotunneling) processes. In this regime, a universal power-law dependence of the conductance on the temperature is found. In addition, quasi-degeneracies enhance multi-level interference effects that, in turn, suppress the conductance.\\
\indent This work is structured as follows: In Sec.~\ref{NZ_approach}, we introduce the transport setting described by the theory and the exact, formal results from the NZ approach to quantum transport. This is the starting point for developing, in Sections~\ref{Liouville_approach} and~\ref{projection_energy}, the diagrammatic unravelling of the RDM and current kernels, the objects that encapsulate the effects of the coupling to the environment on the system's steady state and the current. In Sec.~\ref{bosons}, the results from the diagrammatic formalism  are specialized to the steady-state bosonic heat transport, up to the fourth order and for a generic junction. Finally, in Sec.~\ref{HT_QRM}, the theory is applied to the concrete example of heat transport in the Rabi model connected to bosonic heat baths. Conclusions are left to  Sec.~\ref{conclusions}.

\section{Nakajima-Zwanzig approach to quantum transport}\label{NZ_approach}

\subsection{Quantum transport setup}
\begin{figure}[ht]
\centering
\includegraphics[width=6.5cm]{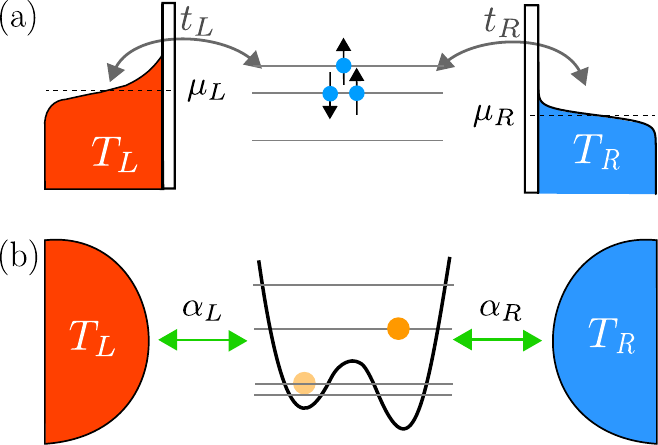}
\caption{Examples of fermionic and bosonic transport setups. An interacting quantum dot connected to fermionic leads kept at different temperatures and chemical potentials (a), and a nonlinear system coupled to two heat baths at different temperatures (b).}
\label{scheme}
\end{figure}
Our starting point is  the  partition of the time-independent total
Hamiltonian of an open quantum system \emph{linearly} coupled to an environment of  fermionic/bosonic baths (leads/heat baths) possibly kept at different chemical potentials and/or temperatures. The coupling term is denoted with $V$ and can be for example mediated by displacement operators or by fermionic  creation/annihilation operators (tunneling Hamiltonian), according to the considered setup.
The total Hamiltonian reads thus
\be{Htot}
\hat H=\hat H_S+\hat H_B+\hat H_V\;,
\ee
where the first term describes the open system ($S$), the portion of the total system we are interested in describing via the reduced density operator $\hat\varrho(t)=\Tr_B\{\hat\varrho_{\rm tot}(t)\}$. The second term accounts for one or multiple  (non-interacting) fermionic or bosonic baths ($B$). The reservoirs are indexed by $l$ in the following so that \hbox{$\hat H_B=\sum_l \hat H_l$}. For a simple transport setup often one chooses $l=L,R$ for left and right, respectively. The third term is the coupling between $S$ and $B$.
We assume it to be of the form $\hat H_V=\sum_l \hat H_{V,l}=\sum_{l,i} \hat D_{li}\otimes \hat C_{li}$, with $\hat D_i$ and $\hat C_i$ defined in the system and environment Hilbert space, respectively.\\ 
\indent For definiteness, let us consider the two paradigmatic examples of open quantum systems, of interest in the present transport context, which are depicted in Fig.~\ref{scheme}. The first is an interacting fermionic system, with creation/annihilation operators $\hat d^\dag_s, \hat d_s$, tunnel-coupled to noninteracting fermionic leads with operators $\hat c^\dag_{l\mathbf{k}\sigma}, \hat c_{l\mathbf{k}\sigma}$ where $\{\hat d^\dag_s, \hat c_{l\mathbf{k}\sigma}\}=\{\hat d_s, \hat c^\dag_{l\mathbf{k}\sigma}\}=0$. Here, $s$ is the state of $S$ where the fermion is created/annihilated and $\mathbf{k},\sigma$ denote the momenta and spin of the electrons in the leads~\footnote{In the case where $S$ is an interacting quantum dot, $\hat H_S=\sum_{s,s'}\varepsilon_{ss'} \hat d_s^\dag \hat d_{s'} + v(\{\hat d_s,\hat d^\dag_s\})$,
where $v$ is a quartic function of the fermionic operators $\hat d_s,\hat d^\dag_s$ accounting for the Coulomb interaction.}. The tunneling amplitude are thus $t_{l \mathbf{k}\sigma s}$. In the second setup, an open system with energy levels $E_n$ is coupled to bosonic heat baths. Creation/annihilation of an excitation in the bosonic mode $j$ of bath $l$ is given by $\hat b^\dag_{lj}, \hat b_{lj}$, with $[\hat b_{lj},\hat b_{lj'}]=0$, $[\hat b_{lj}^\dag,\hat b_{lj'}]=\delta_{jj'}$. We assume the coupling between system and bath $l$ to be mediated by the bath displacement operator $\hat b^\dag_{lj} +\hat  b_{lj}$ and the system operator $\hat{Q}_l$. The coupling strength with the $j$th bosonic mode of bath $l$ is quantified by the frequency $\lambda_{l j}$, see the Caldeira-Leggett model~\cite{Caldeira1983,Weiss2012}. The system Hamiltonian is left, at this stage, unspecified except that it reads in the system energy eigenbasis $\hat H_S =
\sum_{n} E_n \ket{n}\bra{n}$. 
The environment and interaction terms in the general total Hamiltonian~\eqref{Htot} are then summarized by
\bes\label{Hfermions}
\hat H_B + \hat H_V  = \sum_{l \mathbf{k}\sigma} \epsilon_{l \mathbf{k} \sigma} \hat c^\dag_{l \mathbf{k}\sigma}\hat c^\dag_{l \mathbf{k}\sigma}+\sum_{l \mathbf{k}\sigma s} [t_{l \mathbf{k}\sigma s} \hat c^\dag_{l \mathbf{k}\sigma}\hat d_s+t^*_{l \mathbf{k}\sigma s} \hat d^\dag_s \hat c_{l \mathbf{k}\sigma}]
\ee
for fermions, see  Fig.~\ref{scheme}(a), and
\bes\label{Hbosons}
\hat H_B + \hat H_V  =\sum_{l j}\hbar\omega_{l j}\hat b^\dag_{l j}\hat b_{l j}+\hbar\sum_{l j}\hat{Q}_{l}\lambda_{l j}(\hat b_{l j}+\hat b_{l j}^\dag)
\ee
for bosons, see  Fig.~\ref{scheme}(b). In the bosonic case the system couples to the collective bath operator \hbox{$\hat B_l=\sum_j\lambda_{l j}(\hat b_{l j}+\hat b_{l j}^\dag)$}. However, other types of coupling are possible as, e.g., with the momentum operators of the bath oscillators, namely \hbox{$\hat B_l=\sum_j\lambda_{l j}\ii(\hat b_{l j}^\dag-\hat b_{l j})$}. As we shall see, in the perturbative expansion with respect to the system-bath coupling, the bath enters solely  via the free correlation function $\langle \hat B_l^\dag(t)\hat B_l(0)\rangle$, where the 
thermal average and evolution refer to the free bath Hamiltonian. The approach applies to any linear bosonic collective bath interaction.\\
\indent Throughout the present work, we use the convention $\begin{bmatrix}
{\rm fermions}\\
{\rm bosons}
\end{bmatrix}$.
The environment and interaction terms can then be compactly written, for both setups, as
\bes\label{HEV}
\hat H_B&=\sum_{l}\hat H_l=\sum_{lj}
\hbar\omega_{lj}\hat N_{lj}\;,\\
\hat H_V&=\sum_{l j}\left(\hat C^+_{l j}\hat D^-_{l j} + \hat D^+_{l j}\hat C^-_{l j}\right) \equiv\sum_{p l j}\begin{bmatrix}
p\\
1
\end{bmatrix}
\hat C^{p}_{l j}\hat D^{-p}_{l j}\;,
\ee
where $p$ is the so-called Fock index $p=\pm$, with the meaning $\hat{O}^+\equiv\hat{O}^\dag$ and $\hat{O}^-\equiv\hat{O}$.
In Eq.~\eqref{HEV}, we defined the operators $\hat C,\hat D$ and $\hat N$ as follows.
For fermions, setting  $j \equiv\mathbf{k}\sigma$, we have defined
\bes\label{D_fermions}
\hat D^p_{l j}=\sum_s t^p_{l \mathbf{k}\sigma s} \hat d^p_s\;, \quad  \hat C^p_{l j}=\hat c^p_{l \mathbf{k} \sigma}\;, \quad \hat N_{lj}=\hat C^\dag_{lj}\hat C_{lj}\;,
\ee
For bosons, system and bath operators are defined as
\bes\label{D_bosons} 
&\hat D^p_{l j}=\hbar \lambda_{l j} \hat{Q}^p_{l}\;, \quad  \hat C^p_{l j}=\hat b^p_{l j}\;, \quad \hat N_{lj}=\hat C^\dag_{lj}\hat C_{lj}\;. 
\ee
Since the Fock and lead indices will appear together, both in the system and the bath operators, it is convenient to introduce the compact notation $\pm q:=(\pm p,l,j)$, namely 
\bes\label{q}
\hat{O}^q:=\hat{O}^p_{lj}\;,\qquad \hat{O}^{-q}:=\hat{O}^{-p}_{lj}\;.
\ee

\subsection{Nakajima-Zwanzig  equation and current}

In what follows superoperators are denoted by calligraphic letters.
The time evolution of the total density matrix $\hat\varrho_{\rm tot}(t)$ is provided by the Liouville-Von Neumann equation 
\be{LvN}
\dot{\hat{\varrho}}_{\rm tot}(t)=-\frac{\ii}{\hbar}[\hat H,\hat\varrho_{\rm tot}(t)]\equiv \mathcal{L}_{\rm tot}\hat\varrho_{\rm tot}(t)\;,
\ee
where $\mathcal{L}_{\rm tot}=\mathcal{L}_S+\mathcal{L}_B+\mathcal{L}_V$. The action of the Liouvillian superoperators is given by $\mathcal{L}_i\bullet=-(\ii/\hbar)[\hat H_i,\bullet]$ with $i=S,B$, or $V$. Assume that the total density matrix at time $t=0$ is in the factorized initial condition $\hat\varrho_{\rm tot}(0)=\hat\varrho(0)\otimes\hat\varrho_B$, where $\hat\varrho_B=\bigotimes_l\hat\varrho_l^{\rm th}$. Here,
$\hat\varrho_l^{\rm th}:=
\exp[-\beta_l (\hat H_l-\mu_l)]/Z_l$, with $Z_l=\Tr_B\{\exp[-\beta_l (\hat H_l-\mu_l)]\}$, is the thermal state of the reservoir $l$ at inverse temperature $\beta_l$
and chemical potential $\mu_l$ (for bosons $\mu_l=0$). Starting from Eq.~\eqref{LvN} and using the projection operator technique, the time evolution of the system reduced density operator $\hat\varrho(t)$ can be cast as the following formally exact GME
\be{GME}
\dot{\hat{\varrho}}(t)=\mathcal{L}_S\hat\varrho(t) + \int_0^t dt' \mathcal{K}(t-t')\hat\varrho(t')
\ee
of the Nakajima-Zwanzig form~\cite{Nakajima1958,Zwanzig1960}, see Appendix~\ref{appendix_GME} for details. The exact kernel superoperator acts on $\hat\varrho$ as
\be{Kexact}
\mathcal{K}(t-t')\hat\varrho(t')&=\Tr_B\{\mathcal{P}\mathcal{L}_V \mathcal{G}_{\mathcal{Q}}(t-t')\mathcal{L}_V\hat\varrho(t')\otimes\hat\varrho_B\}\;,
\\
\mathcal{G}_{\mathcal{Q}}(t)&=e^{(\mathcal{L}_S+\mathcal{L}_B+\mathcal{Q}\mathcal{L}_V\mathcal{Q})t}
\;,
\ee
where, the projection superoperators $\mathcal{P}$ and $\mathcal{Q}$ are defined by 
\be{}
\mathcal{P}\bullet &=\Tr_B\{\bullet\}\otimes\hat\varrho_B\;,\quad 
\mathcal{Q}=\mathbf{1}-\mathcal{P}\;.
\ee
The above formal results do not rest on $\hat\varrho_B$ being the tensor product of the reservoirs thermal states. It suffices for $\hat\varrho_B$ to be a fixed reference state of the environment with the property $[\hat H_B,\hat\varrho_B]=0$. An extension of Eq.~\eqref{GME} to the case of time-dependent Hamiltonians is straightforward, and yields a GME where the kernel depends on the two times $t$ and $t'$, and not just on their difference~\cite{Grifoni1998}. 

\subsubsection{Current}

The general definition of current \emph{to} the reservoir $r$ is given by the trace 
$$I_r(t)=\Tr\{\hat{I}_r\hat\varrho_{\rm tot}(t)\}\;,$$
where $\hat I_r$ is the relevant current operator.
Specifically, the particle and energy current operators are defined as 
\begin{equation}
\begin{aligned}
\hat I^{\rm p}&=d \hat N_r/dt=\ii [\hat H,\hat N_r]/\hbar\;,\\
\hat I^{\rm E}_r&=d \hat H_r/dt=\ii [\hat H,\hat H_r]/\hbar\;,    
\end{aligned}    
\end{equation}
respectively. For example, the fermionic particle current operator is $\hat{I}^{\rm p}_r=(\ii/\hbar)\sum_{j,s}(t_{rjs}\hat c^\dag_{rj}\hat d_s - t^*_{rjs}\hat d_s^\dag \hat c_{rj})$ and the corresponding charge current to reservoir $r$ reads $-e I^{\rm p}_r$. The associated heat current  is then defined as $I_r^{\rm h}=I_r^{\rm E}-\mu_r I_r^{\rm p}$. This coincides, at the steady state, with $I_r^{\rm E}$, for $\mu_L=\mu_R$,
and with $I_r^{\rm p}$, for $T_L=T_R$~\cite{Muralidharan2012,Pekola2021,Tesser2022}.
In our setting, for bosonic reservoirs, the heat current coincides with the energy current, namely $\hat{I}^{\rm h}_r=\hat{I}^{\rm E}_r$, and is given by the operator $\hat{I}^{\rm h}_r={\rm i} \hat{Q}_r\sum_{j}\lambda_{rj}\hbar\omega_{rj}[\hat b_{rj} - \hat b_{rj}^\dag]$. Let $\zeta_{rj}$ be the quantity to be transferred. Then a general expression for the particle/charge/heat current operator, in the fermionic or bosonic case, is 
\be{current}
\hat{I}_r
=\sum_{j} 
\zeta_{rj}
\frac{\ii}{\hbar}[\hat H,\hat N_{rj}]
\equiv  -\frac{\ii}{\hbar}\sum_{p j}\zeta_{rj}
 \begin{bmatrix}
1\\
p
\end{bmatrix}
\hat C^{p}_{r j}\hat D^{-p}_{r j}\;,
\ee
with $\zeta_{rj}=-e$ for charge current in electron transport (fermion reservoirs, $\zeta_{rj}=1$ for the particle current) and $\zeta_{rj}=\hbar\omega_{rj}-\mu_r$ in the  case of heat current (fermion or boson reservoirs, with $\mu_r=0$ in the latter case).
A quantity of interest in the context of heat transport is the linear thermal conductance, which gives the heat conduction properties for a small temperature bias $\Delta T$. 
Setting $T_r=T$ and the temperature of the other bath $T+\Delta T$, the thermal conductance is defined as the derivative of the \emph{forward} current with respect to the temperature bias, in the limit of vanishing bias, namely 
\be{kappa}
\kappa=\lim_{\Delta T\to 0}\frac{\partial I^{\rm h}_r}{\partial\Delta T}\Bigg|_{I^{\rm p}=0}\;,
\ee
where the constraint of zero particle current is specified for the fermionic case~\cite{Schulenborg2018}.\\
\indent Using the projection operator technique, with a similar procedure as for the GME, the current can be cast in the following form 
\be{Ir}
I_r(t)=\Tr_S\left\{\int_0^t dt'\mathcal{K}_{{\rm I}r}(t-t')\hat\varrho(t')\right\}
\ee
(details are given in Appendix~\ref{appendix_GME}). The current kernel has the exact expression 
\be{Kcurrentexact}
\mathcal{K}_{{\rm I}r}(t-t')\hat\varrho(t')&=\Tr_B\{\mathcal{I}_r \mathcal{G}_{\mathcal{Q}}(t-t')\mathcal{L}_V\hat\varrho(t')\otimes\hat\varrho_B\}\;,
\ee
which is the same as the one  for the kernel of the GME, Eq.~\eqref{Kexact}, except that the last interaction Liouvillian (to the left) is here substituted by the  current superoperator. The latter is simply the current operator acting from the left,
$\mathcal{I}_r\bullet=\hat{I}_r\bullet$.\\
\indent The steady-state RDM and current are given by the final value theorem as $\varrho^\infty=\lim_{\lambda\to 0^+} \lambda\tilde\varrho(\lambda)$ and \hbox{$I_l^\infty=\lim_{\lambda\to 0^+} \lambda \tilde I_l(\lambda)$}, where the tilde denotes the Laplace transform. Applied to the GME~\eqref{GME} and Eq.~\eqref{Ir}, this yields
\be{GME_current_ss}
0=\mathcal{L}_S\hat\varrho^\infty+\tilde{\mathcal{K}}(0)\hat\varrho^\infty\quad {\rm and}\quad I_r^\infty=\Tr\{\tilde{\mathcal{K}}_{{\rm I} r}(0)\hat\varrho^\infty\}\;.
\ee
Note that the steady state does not result from a Markovian approximation. It nevertheless coincides with the steady-state limit of a Markov-approximated GME, if the Markovian approximation is performed in the Schr\"odinger picture.

\section{Liouville space approach}\label{Liouville_approach}

Our aim is to systematically expand the GME in the system-environment coupling by associating diagrams to each power. 
This is most conveniently achieved by operating in Liouville space. To establish the connection between the system or environment operators $\hat O$ and the corresponding Liouvillian superoperators $\mathcal{O}$, it is convenient to introduce the so-called Liouville index $\nu$.  The latter indicates the position of the operators with respect to the object the Liouvillian acts upon. Specifically,
 the action of the  Liouvillian superoperators can be expressed by 
 \[ \mathcal{L}_i\bullet=-\frac{\ii}{\hbar}[\hat H_i,\bullet]\equiv -\frac{\ii}{\hbar}\sum_\nu \nu\mathcal{H}_i^\nu\bullet\;,
 \] 
 where the Liouville index $\nu=\pm$ has the effect \hbox{$\mc{H}_i^+\bullet=\hat H_i\bullet$} and $\mc{H}_i^-\bullet=\bullet \hat H_i$. 
 Applied to the specific systems considered here, the action of the interaction Liouville superoperator is then ($q=(p,l,j)$, see Eq.~\eqref{q})
\be{}
\mathcal{L}_V\bullet=-\frac{\ii}{\hbar}\sum_{\nu q}\begin{bmatrix}
p\\
\nu
\end{bmatrix}
\mc C^{q\nu}\mc D^{-q\nu}\bullet\;.
\ee
Here, $\mathcal{C}$ and $\mathcal{D}$ are the superoperators corresponding to $\hat C$ and $\hat D$, respectively.
Note that the Liouville index $\nu$ in a superoperator $\mathcal{A}^{p\nu}$ does not multiply the Fock index $p$, i.e. $\mathcal{A}^{p+}\bullet=\hat A^p\bullet$ and  $\mathcal{A}^{p-}\bullet=\bullet \hat A^p$. 
The fermionic/bosonic (anti-) commutation relations between the operators $\hat A$ and $\hat B$ translate, in Liouville space, into 
\be{CommLiouville}
\mc A^\nu \mc B^{\nu'}=
\begin{bmatrix}
-\nu \nu'\\
1
\end{bmatrix}
\mc B^{\nu'} \mc A^\nu\;.
\ee
Finally, the
current superoperator corresponding to the current to the reservoir $r$, Eqs.~\eqref{current} and~\eqref{Kcurrentexact}, in terms of Liouville indices reads
\be{current_Liouville}
\mathcal{I}_r
=-\frac{\ii}{\hbar}\sum_{q}\zeta_{lj}\delta_{l,r}
 \begin{bmatrix}
1\\
p
\end{bmatrix}
\mathcal{C}^{q+}\mathcal{D}^{-q+}\;.
\ee

\vspace{-0.25cm}

\subsection{Trace over the environments}

For a thermal reference state of the baths $\hat\varrho_B=\bigotimes_l\hat\varrho_l^{\rm th}$, the trace over the environment degrees of freedom yields, using the cyclic property of the trace, the result
%
\be{CF}
\tr_B\{\mc C^{q\nu}\mc C^{q'\nu'}\hat\varrho_B\}=&\;\delta_{p',-p}\delta_{l j, l' j'} 
n_{l}^{p\nu'}(\omega_{j}),
\\
n_{l}^+(\omega)=&\;[e^{\beta_l(\hbar\omega-\mu_l)}\pm 1]^{-1}\;,\\
n_{l}^-(\omega)=&\; 1 \mp n_{l}^+(\omega)\;.
\ee
Here $n_{l}^+(\omega)$ is either the Fermi-Dirac $f_l(\omega)$  or the Bose-Einstein distribution $n_l(\omega)$ of bath $l$, according to the environmental statistics, with $\mu_l=0$ in the latter case. Contrary to the case of a superoperator, the index $p\nu$ appearing in a function is simply the product of the two indices, e.g. $n^{+-}=n^{-}$.\\

\vspace{-0.7cm}

\subsection{Second order}

Expansion of the exact kernel superoperators, Eqs.~\eqref{Kexact} and~\eqref{Kcurrentexact}, in powers of $V$  yields the perturbation expansion of the GME and the current, see Appendix~\ref{expansion}. The interaction Hamiltonian $\hat H_V$ does not conserve the reservoirs' particle number, thus \hbox{$\mathcal{P}\mathcal{L}_V^{2n+1}\mathcal{P}=0$}. As a result, all odd-order contributions vanish and the leading order is the second. It is obtained with $\mathcal{G}_\mathcal{Q}(t)\simeq \mathcal{G}_0(t):= e^{(\mathcal{L}_S+\mathcal{L}_B)t}$. The Heisenberg evolution of the  baths operators, for non-interacting baths, gives $\hat a_j(t)=U^\dag_B(t) \hat a_j U_B(t)=\hat a_j \exp(-\ii \omega_j t)$. Then the environment propagator results in $\mathcal{G}_B(t)\hat a_j =e^{-\frac{\ii}{\hbar}[\hat H_B,\bullet] t}\hat a_j=U_B(t) \hat a_j U^\dag_B(t)=\hat a_j(-t)=\hat a_j e^{\ii \omega_j t}$, which yields for the RDM kernel
\be{}
&\mathcal{K}^{(2)}(t-t')\hat\varrho(t')=\tr_B\{\mathcal{L}_V \mathcal{G}_0(t-t')\mathcal{L}_V \hat\varrho(t')\otimes\hat\varrho_B\}\\
&=-\frac{1}{\hbar^2}\sum_{\nu_2 \nu_1}\sum_{q_2q_1}\begin{bmatrix}
p_2p_1\\
\nu_2\nu_1
\end{bmatrix}\\
&\times
\tr_B\{\mc C^{q_2\nu_2}\mc D^{-q_2\nu_2}\mathcal{G}_0(t-t')
\mc C^{q_1\nu_1}\mc D^{-q_1\nu_1}\hat\varrho(t')\otimes\hat\varrho_B\}\\
&=-\frac{1}{\hbar^2}\sum_{\nu_2 \nu_1}\sum_{q_2 q_1}
\begin{bmatrix}
-p_2p_1\nu_2\nu_1\\
\nu_2\nu_1
\end{bmatrix}\tr_B\{\mc C^{q_2\nu_2}\mc C^{q_1\nu_1}\hat\varrho_B\}\\
&\times
\mc D^{-q_2\nu_2}e^{(\mathcal{L}_S+\ii p_1\omega_{l_1j_1})(t-t')}
\mc D^{-q_1\nu_1}\hat\varrho(t')\;.
\ee
Using the result~\eqref{CF} for the trace over the environments, we obtain 
\be{K2time}
\mathcal{K}^{(2)}&(t-t')\hat\varrho(t')
=-\frac{1}{\hbar^2}\sum_{\nu_2\nu_1 q}\nu_2\nu_1\\
&\quad\times\mc D^{-q\nu_2} n^{p\nu_1}_{l}(\omega_{ j})e^{(\mathcal{L}_S+\ii p\omega_{lj})(t-t')}
\mc D^{q\nu_1} \hat\varrho(t')\;.
\ee
Note that Eq.~\eqref{GME} with this second-order kernel (GME with Born approximation) is still nonlocal. The Markov approximation $\hat\varrho(t')\rightarrow \hat\varrho(t)$ gives the Redfield I master equation, the time-local version of the GME~\eqref{GME} with time-dependent rates. Further, extending the upper integration limit to $t\rightarrow \infty$ one obtains the Redfield II master equation with constant rates~\cite{Landi2022}.

Proceeding similarly for the 2nd-order  current kernel, we obtain for the RDM and current kernel to leading order, in Laplace space, the following expressions
\bes\label{K2Laplace}
\tilde{\mathcal{K}}^{(2)}(\lambda)\tilde\varrho(\lambda)
&=-\frac{1}{\hbar^2}\sum_{\nu_2\nu_1 q}\nu_2\nu_1
\mc D^{-q\nu_2}
\frac{n_{l}^{p\nu_1}(\omega_{lj})}{\lambda -\mathcal{L}_S -\ii p \omega_{l j}}\mc D^{q\nu_1}\tilde\varrho(\lambda)\;,\\
\tilde{\mathcal{K}}^{(2)}_{{\rm I} r}(\lambda)\tilde\varrho(\lambda)
&=-\frac{1}{\hbar^2}\sum_{\nu_1 q}\delta_{l,r}\zeta_{lj}
p\nu_1
\mc D^{-q+}
\frac{n_{l}^{p\nu_1}(\omega_{l j})}{\lambda -\mathcal{L}_S -\ii p \omega_{l j}}\mc D^{q\nu_1}\tilde\varrho(\lambda)\;.
\ee
Diagrammatically, the action of the above kernel superoperators
for the steady-state GME and current, Eq.~\eqref{GME_current_ss}, can be compactly defined as
\be{KD2}
\tilde{\mathcal{K}}^{(2)}(0)\hat\varrho^\infty=&-\frac{1}{\hbar^2}\sum_q {\boldsymbol{D}}_2\hat\varrho^\infty\;,\\
\tilde{\mathcal{K}}^{(2)}_{{\rm I} r}(0)\hat\varrho^\infty=&-\frac{1}{\hbar^2}\sum_q p\zeta_{lj}\delta_{l,r}{\boldsymbol{D}}_2^+\hat\varrho^\infty\;,
\ee
where the ${\boldsymbol{D}}_2$
superoperators are rendered by the single diagram
\be{}
\boldsymbol{D}_2&=\sum_{\nu_2\nu_1}\nu_2\nu_1
\begin{gathered}
\resizebox{2.25cm}{!}{
\begin{tikzpicture}[] 
\draw[thick] (-0.,0) -- (2.,0); 
\draw[black,thick] (0,0)  node [below]{\Large{$\nu_2$}} arc (180:0:1.cm  and 1.cm)  node [below]{\Large{$\nu_1$}};
\filldraw[](2,0) circle (2pt); 
\draw[thick] (1,1) node [above]{\large{$q$}};
\end{tikzpicture}
}
\end{gathered} \;,
\ee
and ${\boldsymbol{D}}_2^+$ includes the constraint $\nu_2=+$.

\vspace{-0.2cm}

\subsection{Fourth order}

Using again $\mathcal{P}\mathcal{L}_V^{2n+1}\mathcal{P}=0$, the fourth-order kernel superoperator in Laplace space reads
\bes\label{K4th}
\tilde{\mathcal{K}}^{(4)}&(\lambda)\tilde\varrho(\lambda)
\\
=&\;
\tr_B\{\mathcal{L}_V \tilde{\mathcal{G}}_0(\lambda)\mathcal{L}_V\tilde{\mathcal{G}}_0(\lambda) \mathcal{L}_V\tilde{\mathcal{G}}_0(\lambda)\mathcal{L}_V \tilde\varrho(\lambda)\otimes\hat\varrho_B\}\\
&-
\tr_B\{\mathcal{L}_V \tilde{\mathcal{G}}_0(\lambda)\mathcal{L}_V\tilde{\mathcal{G}}_0(\lambda)\mathcal{P} \mathcal{L}_V\tilde{\mathcal{G}}_0(\lambda)\mathcal{L}_V \tilde\varrho(\lambda)\otimes\hat\varrho_B\}\;,
\ee
where the last term, due to the action of $\mathcal{P}$, gives a so-called reducible contribution.
A similar result is obtained for the current kernel by substituting the last tunneling Liouvillian $\mc L_V$ with the current, Eq.~\eqref{current_Liouville}. 
Using Eq.~\eqref{CommLiouville}, the Wick's theorem yields for the (non-interacting) baths' operators~\cite{Schoeller2009,Ferguson2021}

\bes\label{4th_order_correlator}
\langle \mathcal{C}_4 \mathcal{C}_3\mathcal{C}_2\mathcal{C}_1\rangle =& \;\langle \mathcal{C}_4 \mathcal{C}_3\rangle\langle \mathcal{C}_2\mathcal{C}_1\rangle
+\begin{bmatrix}
-\nu_3\nu_2\\
1
\end{bmatrix}
\langle \mathcal{C}_4 \mathcal{C}_2\rangle\langle \mathcal{C}_3\mathcal{C}_1\rangle\\
&+\begin{bmatrix}
\nu_3\nu_2\\
1
\end{bmatrix}
\langle \mathcal{C}_4 \mathcal{C}_1\rangle\langle \mathcal{C}_3\mathcal{C}_2\rangle\;,
\ee
where $\mc C_i\equiv\mc C^{q_i\nu_i}$.
Equation~\eqref{K4th} implies subtracting the reducible part form the correlator in Eq.~\eqref{4th_order_correlator}, namely $\langle \mathcal{C}_4\mathcal{C}_3\mathcal{C}_2\mathcal{C}_1\rangle - \langle \mathcal{C}_4\mathcal{C}_3\rangle \langle \mathcal{C}_2\mathcal{C}_1\rangle$, yielding
\bes\label{K4_formal}
\tilde{\mathcal{K}}^{(4)}(\lambda)&\tilde\varrho(\lambda)
=\left(-\frac{1}{\hbar^2}\right)^2\sum_{\{\nu_i\} \{q_i\}}\begin{bmatrix}
 p_4 p_3 p_2 p_1 \nu_4\nu_1\\
 \nu_4\nu_3\nu_2\nu_1
 \end{bmatrix}\\
&\times\mathcal{D}_4\tilde{\mathcal{G}}_{0,3}(\lambda)  \mathcal{D}_3\tilde{\mathcal{G}}_{0,2}(\lambda) \mathcal{D}_2\tilde{\mathcal{G}}_{0,1}(\lambda) \mathcal{D}_1\tilde\varrho(\lambda)\\
&\times\left[\langle \mathcal{C}_4\mathcal{C}_1\rangle \langle \mathcal{C}_3\mathcal{C}_2\rangle \mp \langle \mathcal{C}_4\mathcal{C}_2\rangle \langle \mathcal{C}_3\mathcal{C}_1\rangle\right]\;,
\ee
where $i=1,\dots,4$, $\mathcal{D}_i=\mc D^{-q_i\nu_i}$,  and
$$\tilde{\mathcal{G}}_{0,i}(\lambda)=\left(\lambda-\mathcal{L}_S - \ii \sum_{k=1}^{M_i} p_k\omega_{l_k j_k}\right)^{-1}\;,$$
with the sum extending to the number $M_i$ of overlapping lines between the vertices $i$ and $i+1$. 
Diagrammatically, the action of the kernels induced by the correlators are rendered by the two irreducible diagrams
\be{}
&\begin{gathered}
\resizebox{3.5cm}{!}{
\begin{tikzpicture}[] 
\draw[thick] (-0.,0) -- (3.,0); 
\draw[black,thick] (0,0)  node [below]{\small{$ \nu_4$}} arc (180:0:1.5cm  and 1.cm) node [below]{\small{$ \nu_1$}};
\draw[black,thick] (1,0)  node [below]{\small{$ \nu_3$}} arc (180:0:0.5cm  and 0.6cm) node [below]{\small{$ \nu_2$}} ;
\filldraw[](2,-0) circle (1.5pt); 
\filldraw[](3,-0) circle (1.5pt); 
\draw[thick] (1.5,1) node [above]{\small{$q$}};
\draw[thick] (1.5,0.5) node [above]{\small{$q'$}};
\end{tikzpicture}
}
\end{gathered} 
\quad{\rm and}\quad
\begin{gathered}
\resizebox{3.5cm}{!}{
\begin{tikzpicture}[] 
\draw[thick] (-0.,0) -- (3.,0); 
\draw[black,thick] (0,0)  node [below]{\small{$\nu_4$}} arc (180:0:1cm  and 0.85cm) node [below]{\small{$\nu_2$}};
\draw[black,thick] (1,0)  node [below]{\small{$\nu_3$}} arc (180:0:1cm  and 0.85cm) node [below]{\small{$\nu_1$}} ;
\filldraw[](2,-0) circle (1.5pt); 
\filldraw[](3,-0) circle (1.5pt); 
\draw[thick] (1,0.85) node [above]{\small{$q$}};
\draw[thick] (2,0.85) node [above]{\small{$q'$}};
\end{tikzpicture}
}
\end{gathered}\;.
\ee
Such diagrams display a system's propagation line with four vertices, pairwise joined by an environmental arc.
Using Eq.~\eqref{CF} for the correlators, one finds for the 4th-order contribution to the RDM the current kernel (the latter is obtained along similar lines) in Laplace space 

\begin{widetext}
\bes\label{K4}
\tilde{\mathcal{K}}^{(4)}(\lambda)\tilde\varrho(\lambda)
=\left(-\frac{1}{\hbar^2}\right)^2\sum_{\nu_{1,\dots,4}}\sum_{qq'}
\begin{bmatrix}
1\\
\nu_3\nu_2
\end{bmatrix}\nu_4 \nu_1
\Bigg\{
&\mc D^{-q \nu_4}\frac{1}{\lambda-\mathcal{L}_S - \ii p \omega_{l j}}\mc D^{-q' \nu_3}
\frac{n_{l'}^{p'\nu_2}(\omega_{j'})}{\lambda-\mathcal{L}_S - \ii (p \omega_{l j}+p' \omega_{l' j'})}\mc D^{q' \nu_2}
\frac{n_{l}^{p\nu_1}(\omega_{j})}{\lambda-\mathcal{L}_S - \ii p \omega_{l j}}\mc D^{q \nu_1}\\
\mp \mc D^{-q \nu_4}&\frac{1}{\lambda-\mathcal{L}_S - \ii p \omega_{l j}}\mc D^{-q' \nu_3}
\frac{n_{l}^{p\nu_2}(\omega_{j})}{\lambda-\mathcal{L}_S - \ii (p' \omega_{l' j'}+p \omega_{l j})}
\mc D^{q \nu_2}
\frac{n_{l'}^{p' \nu_1}( \omega_{j'})}{\lambda-\mathcal{L}_S - \ii p' \omega_{l' j'}}\mc D^{q' \nu_1}
\Bigg\}\tilde\varrho(\lambda)
\ee
and
\bes\label{KI4}
\tilde{\mathcal{K}}_{{\rm I} r}^{(4)}(\lambda)\tilde\varrho(\lambda)
=\left(-\frac{1}{\hbar^2}\right)^2\sum_{\nu_{1,\dots,3}}\sum_{qq'}\delta_{l,r}\zeta_{l j}
\begin{bmatrix}
 1 \; \\
 \nu_3\nu_2  
\end{bmatrix}&\nu_1 p\Bigg\{
\mc D^{-q +}\frac{1}{\lambda-\mathcal{L}_S - \ii p \omega_{l j}}\mc D^{-q' \nu_3}
\frac{n_{l'}^{p'\nu_2}(\omega_{j'})}{\lambda-\mathcal{L}_S - \ii (p \omega_{l j}+p' \omega_{l' j'})}\mc D^{q' \nu_2}
\frac{n_{l}^{p\nu_1}(\omega_{j})}{\lambda-\mathcal{L}_S - \ii p \omega_{l j}}\mc D^{q \nu_1}\\
\mp \mc D^{-q +}&\frac{1}{\lambda-\mathcal{L}_S - \ii p \omega_{l j}}\mc D^{-q' \nu_3}
\frac{n_{l}^{p\nu_2}(\omega_{j})}{\lambda-\mathcal{L}_S - \ii (p' \omega_{l' j'}+p \omega_{l j})}
\mc D^{q \nu_2}
\frac{n_{l'}^{p' \nu_1}( \omega_{j'})}{\lambda-\mathcal{L}_S - \ii p' \omega_{l' j'}}\mc D^{q' \nu_1}
\Bigg\}\tilde\varrho(\lambda)\:.
\ee
\end{widetext}

Upper/lower signs refer to fermions/bosons, respectively.
In analogy to Eq.~\eqref{KD2} for the second order, the actions of the 4th-order the kernels at the steady state are compactly written as 
\bes\label{K4laplace}
\tilde{\mathcal{K}}^{(4)}(0)\hat\varrho^\infty=&\left(-\frac{1}{\hbar^2}\right)^2\sum_{qq'}\left[{\boldsymbol{D}}_4+{\boldsymbol{X}}_4\right]\hat\varrho^\infty\;,\\
\tilde{\mathcal{K}}_{{\rm I} r}^{(4)}(0)\hat\varrho^\infty=&\left(-\frac{1}{\hbar^2}\right)^2\sum_{qq'}p\zeta_{lj}\delta_{l,r}\left[{\boldsymbol{D}}_4^+ +{\boldsymbol{X}}_4^+\right]\hat\varrho^\infty\:,
\ee
where, as for the second-order diagrams, the superscript $+$ accounts for the current constraint $\nu_4=+$.

\subsection{Diagrammatic rules for a diagram of order $2n$}

An advantage of the diagrammatic approach is that it provides a systematic way to construct diagrams of any given order according to simple diagrammatic rules~\cite{Schoeller2009,Donarini2024}. In the following, we present diagrammatic rules, valid for bosonic or fermionic systems, to construct the RDM and current kernels at any given order $2n$. They can be easily verified on the second- and fourth-order diagrams discussed in the previous subsection. 

\begin{itemize}

\item Draw a system's propagation line with $2n$ vertices connected pairwise by $n$ fermion or boson arcs. From  right to left, associate to the vertices the Liouville indices  $\nu_1,\dots,\nu_{2n}$ and to each arc a composite index index $q_i:=\{p_i,l_i,j_i\}$, with $i=1,\dots,N$.  

\item For each arc, associate to the starting (right) vertex $i$ the vertex superoperator  $\mathcal{D}^{q\nu_i}$ and the Fermi/Bose function $n^{p\nu}_{l}(\omega_{lj})$ and to the free (left) vertex $f$ the superoperator $\mathcal{D}^{-q \nu_f}$, with $-q=\{-p,l,j\}$. The starting vertex, called anchor vertex, is identified by a dot.

\item Multiply by $(-1/\hbar^2)^n\prod_{i=1}^{2n}\nu_i$ and, \emph{for fermions}, by the product $-\nu_i\nu_k$ for each permutation of vertices $i,k$ needed to bring the diagram in the fully reducible form $\resizebox{3cm}{!}{
\begin{tikzpicture}[] 
\draw[thick] (-0.,0) -- (3.5,0); 
\draw[black,thick] (0,0) arc (180:0:0.5cm  and 0.75cm);
\draw[black,thick] (2,0) arc (180:0:0.5cm  and 0.75cm) ;
\draw[thick] (3.5,0) -- node[fill=white,rotate=0,inner sep=-2.5pt,outer sep=0]{\Large{//}} (4,0);
\draw[thick] (4.0,0) -- (5.5,0); 
\draw[black,thick] (4.5,0) arc (180:0:0.5cm  and 0.75cm) ;
 \end{tikzpicture}
 }$.

\item To each segment connecting two consecutive vertices associate the propagator $[\lambda-\mathcal{L}_S - \ii \sum_{a=1}^M p_a \omega_{l_a j_a}]^{-1}$, where the sum extends to the $M$ fermion/boson arcs overlapping within the segment.

\item For the current to reservoir $r$: Multiply by $\zeta_{lj}$ and by the Fock index $p_n$ of the last fermion/boson arc and include the constraints $\delta_{\nu_n,+}\delta_{l_n,r}$, where $\nu_n$ is the \hbox{Liouville} index of the last vertex and $l_n$ is the reservoir index of the last fermion line.

\item Sum over the $2n$ Liouville indices $\{\nu_i\}$ and the $n$ composite Fock/reservoir indices $\{q_i\}$.   
\end{itemize}

In order to give higher-order diagrams in compact form, we generalize the treatment in~\cite{Magazzu2022} and introduce the boxes that account for different diagrams with swapped fermion/boson lines 

\begin{equation}\begin{aligned}\label{boxes}
\begin{gathered}
\resizebox{0.8cm}{!}{
\begin{tikzpicture}[] 
\draw[] (0.4,0.8) node[below] {\Large{$2$}};
\draw[thick] (0,0) -- (0.8,0) ;
\draw[thick] (0.8,0) -- (0.8,1) ; 
\draw[thick] (0.8,1) -- (0,1)  ;
\draw[thick] (0,1) -- (0,0) ;
 \end{tikzpicture}
}
\end{gathered}
=&
\begin{gathered}
\resizebox{0.8cm}{!}{
\begin{tikzpicture}[] 
\draw[thick] (0,0.2) -- (0.8,0.2) ;
\draw[thick] (0.8,0.8) -- (0,0.8)  ;
 \end{tikzpicture}
}
\end{gathered}
\mp
\begin{gathered}
\resizebox{0.8cm}{!}{
\begin{tikzpicture}[] 
\draw[thick] (0,0.2) -- (0.8,0.8) ;
\draw[thick] (0.8,0.2) -- (0,0.8)  ;
 \end{tikzpicture}
}
\end{gathered}\\
\begin{gathered}
\resizebox{0.8cm}{!}{
\begin{tikzpicture}[] 
\draw[] (0.4,0.8) node[below] {\Large{$3$}};
\draw[thick] (0,0) -- (0.8,0) ;
\draw[thick] (0.8,0) -- (0.8,1) ; 
\draw[thick] (0.8,1) -- (0,1)  ;
\draw[thick] (0,1) -- (0,0) ;
 \end{tikzpicture}
}
\end{gathered}
=&
\begin{gathered}
\resizebox{0.8cm}{!}{
\begin{tikzpicture}[] 
\draw[thick] (0,0.2) -- (0.8,0.2) ;
\draw[thick] (0,0.8) -- (0.8,0.8)  ;
\draw[thick] (0,1.4) -- (0.8,1.4)  ;
 \end{tikzpicture}
}
\end{gathered}
\mp
\begin{gathered}
\resizebox{0.8cm}{!}{
\begin{tikzpicture}[] 
\draw[thick] (0,0.2) -- (0.8,0.8) ;
\draw[thick] (0.8,0.2) -- (0,0.8)  ;
\draw[thick] (0,1.4) -- (0.8,1.4)  ;
 \end{tikzpicture}
}
\end{gathered}
\mp
\begin{gathered}
\resizebox{0.8cm}{!}{
\begin{tikzpicture}[] 
\draw[thick] (0,0.2) -- (0.8,1.4) ;
\draw[thick] (0,0.8) -- (0.8,0.8)  ;
\draw[thick] (0,1.4) -- (0.8,0.2)  ;
 \end{tikzpicture}
}
\end{gathered}\\
&\vdots\\
\begin{gathered}
\resizebox{0.8cm}{!}{
\begin{tikzpicture}[] 
\draw[] (0.4,0.8) node[below] {\Large{$n$}};
\draw[thick] (0,0) -- (0.8,0) ;
\draw[thick] (0.8,0) -- (0.8,1) ; 
\draw[thick] (0.8,1) -- (0,1)  ;
\draw[thick] (0,1) -- (0,0) ;
 \end{tikzpicture}
}
\end{gathered}
=&
\begin{gathered}
\resizebox{0.8cm}{!}{
\begin{tikzpicture}[] 
\draw[thick] (0,0.2) -- (0.8,0.2) ;
\draw[thick] (0,0.6) -- (0.8,0.6)  ;
\draw[loosely dotted,thick] (0,1.0) -- (0.8,1.0)  ;
\draw[thick] (0,1.4) -- (0.8,1.4)  ;
 \end{tikzpicture}
}
\end{gathered}
\mp
\begin{gathered}
\resizebox{0.8cm}{!}{
\begin{tikzpicture}[] 
\draw[thick] (0,0.2) -- (0.8,0.6) ;
\draw[thick] (0,0.6) -- (0.8,0.2)  ;
\draw[loosely dotted,thick] (0,1.0) -- (0.8,1.0)  ;
\draw[thick] (0,1.4) -- (0.8,1.4)  ;
 \end{tikzpicture}
}
\end{gathered}
\mp
\dots
\mp
\begin{gathered}
\resizebox{0.8cm}{!}{
\begin{tikzpicture}[] 
\draw[thick] (0,0.2) -- (0.8,1.4) ;
\draw[thick] (0,0.6) -- (0.8,0.6)  ;
\draw[loosely dotted,thick] (0,1.0) -- (0.8,1.0)  ;
\draw[thick] (0,1.4) -- (0.8,0.2)  ;
 \end{tikzpicture}
}
\end{gathered}\;,
\end{aligned}\end{equation}
where the upper (lower) sign refers to fermion (boson) reservoirs.
With this notation, the \emph{irreducible} diagrams of order $2n$ are compactly generated by enclosing with a fermion/boson line each diagram of order $2n-2$ (reducible and irreducible) and swapping the lines via the boxes as follows:


\begin{equation}\begin{aligned}\nonumber
2^{\rm nd}&\quad
\resizebox{2.25cm}{!}{
\begin{tikzpicture}[] 
\draw[thick] (-0.,0) -- (3.,0); 
\draw[black,thick] (0,0) arc (180:0:1.5cm  and 1cm);
 \end{tikzpicture}
 }\;,\\
 4^{\rm th}&\quad
 \begin{bmatrix}
 \nu_3\nu_2 \\
 1
 \end{bmatrix}
\resizebox{2.25cm}{!}{
\begin{tikzpicture}[] 
\draw[thick] (-0.,0) -- (3.,0); 
\draw[black,thick] (0,0) arc (180:0:1.5cm  and 1cm);
\draw[black,thick] (1,0) arc (180:0:0.5cm  and 0.7cm) ;
\fill[white] (1.25,0.4) rectangle (1.75,1.15);
\draw [] (1.25,0.4) rectangle (1.75,1.15);
\draw[](1.5,0.75) node[] {\Large{$2$}};
 \end{tikzpicture}
 }\;,\\
6^{\rm th}&\quad
 \begin{bmatrix}
 \nu_5\nu_4\nu_3\nu_2 \\
 1
 \end{bmatrix}
 \resizebox{3.25cm}{!}{
\begin{tikzpicture}[] 
\draw[thick] (-0.,0) -- (5.,0); 
\draw[black,thick] (0,0) arc (180:0:2.5cm  and 1.cm);
\draw[black,thick] (1,0) arc (180:0:0.5cm  and 0.7cm) ;
\fill[white] (1.25,0.4) rectangle (1.75,1.15);
\draw [] (1.25,0.4) rectangle (1.75,1.15);
\draw[](1.5,0.75) node[] {\Large{$2$}};
\draw[black,thick] (3,0) arc (180:0:0.5cm  and 0.75cm) ;
\fill[white] (3.75,0.4) rectangle (3.25,1.15);
\draw [] (3.75,0.4) rectangle (3.25,1.15);
\draw[](3.5,0.75) node[] {\Large{$2$}};
 \end{tikzpicture}
 }\\
 &\quad
 \begin{bmatrix}
 \nu_5\nu_2 \\
 1
 \end{bmatrix}
 \resizebox{3.25cm}{!}{
\begin{tikzpicture}[] 
\draw[thick] (-0.,0) -- (5.,0); 
\draw[black,thick] (0,0) arc (180:0:2.5cm  and 1.3cm);
\draw[black,thick] (1,0) arc (180:0:1.5cm  and .95cm) ;
\fill[white] (3.25,0.5) rectangle (3.75,1.35);
\draw [] (3.25,0.4) rectangle (3.75,1.35);
\draw[](3.5,0.85) node[] {\Large{$2$}};
\draw[black,thick] (2,0) arc (180:0:0.5cm  and 0.75cm) ;
\fill[white] (2.25,0.4) rectangle (2.75,1.5);
\draw [] (2.25,0.4) rectangle (2.75,1.5);
\draw[](2.5,0.9) node[] {\Large{$3$}};
\end{tikzpicture}
}\;,\\
8^{\rm th}&\quad
\begin{bmatrix}
 \nu_7\nu_6\nu_5\nu_4\nu_3\nu_2 \\
 1
 \end{bmatrix}
\resizebox{4cm}{!}{
\begin{tikzpicture}[] 
\draw[thick] (-1.,0) -- (6.,0);
\draw[black,thick] (-1,0) arc (180:0:3.5cm  and 1.3cm);
\draw[black,thick] (0,0) arc (180:0:0.5cm  and 0.75cm);
\draw[black,thick] (2,0) arc (180:0:0.5cm  and 0.75cm) ;
\draw[black,thick] (4,0) arc (180:0:0.5cm  and 0.75cm) ;
\fill[white] (0.25,0.4) rectangle (0.75,1.5);
\draw [] (0.25,0.4) rectangle (0.75,1.5);
\draw[](0.5,0.9) node[] {\Large{$2$}};
\fill[white] (2.25,0.4) rectangle (2.75,1.5);
\draw [] (2.25,0.4) rectangle (2.75,1.5);
\draw[](2.5,0.9) node[] {\Large{$2$}};
\fill[white] (4.25,0.4) rectangle (4.75,1.5);
\draw [] (4.25,0.4) rectangle (4.75,1.5);
\draw[](4.5,0.9) node[] {\Large{$2$}};
 \end{tikzpicture}
 }\\
 &\quad
\begin{bmatrix}
 \nu_7\nu_4\nu_3\nu_2 \\
 1
 \end{bmatrix}
 \resizebox{4cm}{!}{
\begin{tikzpicture}[] 
\draw[thick] (-1.,0) -- (6.,0); 
\draw[black,thick] (-1,0) arc (180:0:3.5cm  and 1.5cm);
\draw[black,thick] (0,0) arc (180:0:1.5cm  and 1cm);
\draw[black,thick] (1,0) arc (180:0:0.5cm  and 0.7cm) ;
\draw[black,thick] (4,0) arc (180:0:0.5cm  and 0.75cm) ;
\fill[white] (2.,0.6) rectangle (2.5,1.7);
\draw [] (2.,0.6) rectangle (2.5,1.7);
\draw[](2.25,1.2) node[] {\Large{$2$}};
\fill[white] (1.25,0.4) rectangle (1.75,1.75);
\draw [] (1.25,0.4) rectangle (1.75,1.75);
\draw[](1.5,1.2) node[] {\Large{$3$}};
\fill[white] (4.25,0.4) rectangle (4.75,1.5);
\draw [] (4.25,0.4) rectangle (4.75,1.5);
\draw[](4.5,0.9) node[] {\Large{$2$}};
 \end{tikzpicture}
 }\\
 &\quad
 \begin{bmatrix}
 \nu_7\nu_6\nu_5\nu_2 \\
 1
 \end{bmatrix}
 \resizebox{4cm}{!}{
\begin{tikzpicture}[] 
\draw[thick] (-3.,0) -- (4.,0); 
\draw[black,thick] (-3,0) arc (180:0:3.5cm  and 1.5cm);
\draw[black,thick] (0,0) arc (180:0:1.5cm  and 1cm);
\draw[black,thick] (1,0) arc (180:0:0.5cm  and 0.7cm) ;
\draw[black,thick] (-2,0) arc (180:0:0.5cm  and 0.75cm) ;
\fill[white] (-1.25,0.4) rectangle (-1.75,1.5);
\draw [] (-1.25,0.4) rectangle (-1.75,1.5);
\draw[](-1.5,0.9) node[] {\Large{$2$}};
\fill[white] (2.,0.6) rectangle (2.5,1.6);
\draw [] (2.,0.6) rectangle (2.5,1.6);
\draw[](2.25,1.) node[] {\Large{$2$}};
\fill[white] (1.25,0.4) rectangle (1.75,1.75);
\draw [] (1.25,0.4) rectangle (1.75,1.75);
\draw[](1.5,1.1) node[] {\Large{$3$}};
\end{tikzpicture}
 }\\
 &\quad
  \begin{bmatrix}
 \nu_7\nu_2 \\
 1
 \end{bmatrix}
  \resizebox{4cm}{!}{
\begin{tikzpicture}[] 
\draw[thick] (-1.,0) -- (6.,0); 
\draw[black,thick] (-1,0) arc (180:0:3.5cm  and 1.5cm);
\draw[black,thick] (0,0) arc (180:0:2.5cm  and 1.cm);
\draw[black,thick] (1,0) arc (180:0:0.5cm  and 0.7cm) ;
\draw[black,thick] (3,0) arc (180:0:0.5cm  and 0.75cm) ;
\fill[white] (1.25,0.4) rectangle (1.75,1.75);
\draw [] (1.25,0.4) rectangle (1.75,1.75);
\draw[](1.5,1.2) node[] {\Large{$3$}};
\fill[white] (3.75,0.4) rectangle (3.25,1.75);
\draw [] (3.75,0.4) rectangle (3.25,1.75);
\draw[](3.5,1.2) node[] {\Large{$3$}};
\fill[white] (4.25,0.4) rectangle (4.75,1.5);
\draw [] (4.25,0.4) rectangle (4.75,1.5);
\draw[](4.5,1.) node[] {\Large{$2$}};
 \end{tikzpicture}
 }\\
 &\quad
  \begin{bmatrix}
 \nu_7\nu_5\nu_4\nu_2 \\
 1
 \end{bmatrix}
  \resizebox{4cm}{!}{
\begin{tikzpicture}[] 
\draw[thick] (-1.,0) -- (6.,0); 
\draw[black,thick] (-1,0) arc (180:0:3.5cm  and 1.8cm);
\draw[black,thick] (0,0) arc (180:0:2.5cm  and 1.3cm);
\draw[black,thick] (1,0) arc (180:0:1.5cm  and .95cm) ;
\draw[black,thick] (2,0) arc (180:0:0.5cm  and 0.75cm) ;
\fill[white] (2.25,0.3) rectangle (2.75,2);
\draw [] (2.25,0.3) rectangle (2.75,2);
\draw[](2.5,1.2) node[] {\Large{$4$}};
\fill[white] (3.25,0.4) rectangle (3.75,1.9);
\draw [] (3.25,0.4) rectangle (3.75,1.9);
\draw[](3.5,1.2) node[] {\Large{$3$}};
\fill[white] (4.75,0.5) rectangle (4.25,1.75);
\draw [] (4.75,0.5) rectangle (4.25,1.75);
\draw[](4.5,1.2) node[] {\Large{$2$}};
\end{tikzpicture}
}\\
&\qquad\qquad\qquad\qquad \dots
\end{aligned}\end{equation}
Here, we also accounted for the products of $\nu$'s required, in the fermionic case, to bring the non-crossing versions of these diagrams into the fully reducible form. Additional signs due to the crossings are contained in the boxes $\boxed{n}$, see Eq.~\eqref{boxes}. To order $2$, $4$, $6$, and $8$, there correspond $1$, $2$, $10$, and $74$ irreducible diagrams, respectively. For example, to 6th order there are $2\times 2 + 2\times 3=10$ irreducible diagrams. 

\section{Projection in the system energy eigenbasis}
\label{projection_energy}

For practical calculations, it is useful to project the equations for the RDM and the current in the system energy eigenbasis $\{\ket{n}\}$, where \hbox{$\hat H_S\ket{n}=\hbar\omega_n\ket{n}$}. Resolving Eq.~\eqref{GME_current_ss} in this basis via $\bra{n}\tilde{\mathcal{K}}(0)\hat\varrho^\infty\ket{m}=\sum_{n'm'}\tilde{\mathcal{K}}_{nmn'm'}(0)\varrho_{n'm'}^\infty$, where $\tilde{\mathcal{K}}_{nmn'm'}(0)=\bra{n}(\tilde{\mathcal{K}}(0)\ket{n'}\bra{m'})\ket{m}$, we have for the NZ equation and the current in Laplace space, the formal, exact expressions
\be{GME_energy}
0=&-\ii\omega_{nm}\varrho_{nm}^\infty+\sum_{n'm'}\tilde{\mathcal{K}}_{n m n'm'}(0)\varrho_{n'm'}^\infty\;,
\ee
where $\omega_{nm}:=\omega_n-\omega_m$ are the Bohr frequencies of the system, and
\be{current_energy}
I_r^\infty&=\sum_{nn'm'}\tilde{\mathcal{K}}_{{\rm I}r,n n n'm'}(0)\varrho_{n'm'}^\infty\;,
\ee
respectively. 
The kernel $\tilde{\mathcal{K}}_{n m n'm'}$ fulfills the sum rule $\sum_n \tilde{\mathcal{K}}_{n n n'm'}=0$, $\forall n',m'$, dictated by probability conservation,  and the property $\tilde{\mathcal{K}}_{n m n'm'}=[\tilde{\mathcal{K}}_{m n m'n'}]^*$, given by the hermiticity of the RDM.

\subsection{Perturbation theory}

In the energy eigenbasis, the system operators are expressed as $\hat D^q=\sum_{nm}D^p_{nm}\ket{n}\bra{m}$ with $D^{-q}_{mn}=(D^q_{nm})^*$. From Eq.~\eqref{K2Laplace}, this yields for the elements of the  GME kernel tensor, to lowest order,   
\bes\label{K2general}
\tilde{\mathcal{K}}_{n m n'm'}^{(2)}&(\lambda)=
-\frac{1}{\hbar^2}\sum_{q}\Bigg\{ \sum_{k}\Big[D^{-q}_{nk} D^q_{kn'}\tilde G^{q,+}_{km'}(\lambda)\delta_{m'm}\\
&\qquad\qquad\qquad\quad+[D^{-q}_{mk} D^q_{km'}\tilde G^{q,+}_{kn'}(\lambda)]^*\delta_{n'n}\Big]\\
&-D^{-q}_{m'm} D^q_{nn'}  \tilde G^{q,+}_{nm'}(\lambda) -[ D^{-q}_{n'n}D^q_{mm'} \tilde G^{q,+}_{mn'}(\lambda)]^*\Bigg\}\;.
\ee
Here, we have introduced, cf. Eq.~\eqref{K2Laplace}, the function $\tilde G^{q,\nu}_{nm}(\lambda):=\tilde G^{q}_{nm}(\lambda) n_{l}^{p\nu}(\omega_{l j})$, where $\tilde  G^{q}_{nm}(\lambda)$ is the free propagator resolved in the system energy basis
\be{G_qnu}
\tilde G^{q}_{nm}(\lambda):=\frac{1}{\lambda +\ii\omega_{nm} -\ii p \omega_{l j}}\;,
\ee
and used $\tilde G^{-q,-\nu}_{mn}(\lambda)=[\tilde  G^{q,\nu}_{nm}(\lambda)]^*$. 
Note that $n_{l}^{p\nu}(\omega_{l j})$ is either the Fermi or the Bose function, according to the statistics of the bath $l$, see Eq.~\eqref{CF}. In particular the (real) rates connecting the populations $\rho_{nn}$ explicitly read
\be{rates2nd}
\tilde {\mathcal{K}}_{n n mm}^{(2)}(\lambda)&=\frac{1}{\hbar^2}2 \Real \sum_{q} D^q_{nm} D^{-q}_{mn} \tilde G^{q,+}_{nm}(\lambda)\quad m\neq n\;,\\
\tilde{\mathcal{K}}_{n n nn}^{(2)}(\lambda)&=-\sum_{m(\neq n)} \tilde {\mathcal{K}}_{n n mm}^{(2)}(\lambda)\;. 
\ee
One similarly obtains the second-order current kernel tensor elements 
\bes\label{K2nd_general}
\tilde{\mathcal{K}}_{{\rm I}r,n m n'm'}^{(2)}(\lambda)=&-\frac{1}{\hbar^2}\sum_{q}\delta_{lr}\zeta_{lj}\Bigg[ \sum_{k}pD^{-q}_{nk} D^q_{kn'}\tilde G^{q,+}_{km'}(\lambda)\delta_{m'm}\\
&\qquad\qquad\qquad+p[ D^{-q}_{n'n}D^q_{mm'} \tilde G^{q,+}_{mn'}(\lambda)]^* \Bigg]\;.
\ee
With this, the steady-state current to the reservoir $r$ reads, to lowest order, 
\bes\label{I2nd_general}
I^{(2)}_r
&=-\frac{1}{\hbar^2}2\Real\sum_{q}
\delta_{lr}\zeta_{lj}
\sum_{nn'm'} pD^{-q}_{m'n} D^q_{nn'}\tilde G^{q,+}_{nm'}(0^+)
\varrho_{n'm'}^\infty\\
&=2\Real\sum_{nm'n'(\neq n)}\tilde{ \mathcal{K}}_{{\rm I}r,n n m'n'}^{(2)}(0)\varrho_{n'm'}^\infty\;.
\ee
The projection in the system energy eigenbasis of the fourth-order RDM and current kernels is obtained starting form the fourth-order kernel superoperators given in Eq.~\eqref{K4laplace}. The explicit expression for the current kernel is provided in Appendix~\ref{4th_order}.

\subsection{Partial and full secular 2nd-order master equation}

The starting point is the 2nd-order steady-state Redfield ME, see Eqs.~\eqref{GME_current_ss},~\eqref{GME_energy}, and~\eqref{K2general} 
\be{RME}
0=-\ii\omega_{nm}\varrho_{nm}^\infty+\sum_{n'm'}\tilde{\mathcal{K}}_{n m n'm'}^{(2)}(0)\varrho_{n'm'}^\infty\;,
\ee
where the only approximation is leading order perturbation theory in the system-environment coupling: No Markovian or secular approximation have been invoked in its derivation.
Assume that we can separate the transition frequencies in two classes. The first collects the Bohr frequencies $\omega_{nm}$ much larger than the scale $\gamma$ set by the relevant Redfield tensor elements. To the second class belong the frequencies that are of the same order of magnitude of $\gamma$ or the ones with identical indices, i.e. $\omega_{nn}=0$. 
Consistency with perturbation theory to leading order entails
that Eq.~\eqref{RME} splits into the set of equations 
\be{BR_partial_secular}
&\omega_{nm}\rho_{nm}^\infty=\;0\rightarrow \rho_{nm}^\infty =0\;,\quad n\neq m \quad{\rm and}\quad \omega_{nm}\gg \gamma\;,\\
&0=\;-{\rm i}\omega_{nm}\rho_{nm}^\infty + \sum_{n',m'}\tilde{\mathcal{K}}^{(2)}_{nmn'm'}(0)\rho_{n'm'}^\infty\;,\quad  \omega_{nm}\lesssim\gamma\;.
\ee
Within this partial secular approximation~\cite{Cattaneo2019,Trushechkin2021,Potts2021,Ivander2022}, the surviving coherences at the steady state are those associated to the sub-spaces defined by the indices $n,m$ of the small transition frequencies. The remaining coherences vanish to leading order. \\
\indent For a non-degenerate spectrum with well separated energy levels, a full secular master equation is appropriate, where populations and coherences are fully decoupled (the latter vanish at the steady state) according to
\be{BR_full_secular}
\rho_{nm}^{(0)}=&\;0\qquad n\neq m\;,\\
{\rm and}\qquad\;\sum_m\Gamma_{nm} \rho_{mm}^\infty=&0 \;.
\ee
The second of these equations gives the steady-state populations $\rho_{nn}^\infty$ to leading order.
 Here, we have used the definition $\Gamma_{nm}=\sum_l\Gamma^l_{nm}:=\tilde{\mathcal{K}}^{(2)}_{nnmm}(0)$. 
Note that these rates are real and satisfy \hbox{$\Gamma^l_{nn}=-\sum_{m(\neq n)}\Gamma^l_{mn}$}, which guarantees probability conservation.

\section{Steady-state bosonic heat current}\label{bosons}

\begin{figure}[ht!]
\includegraphics[width=8.cm]{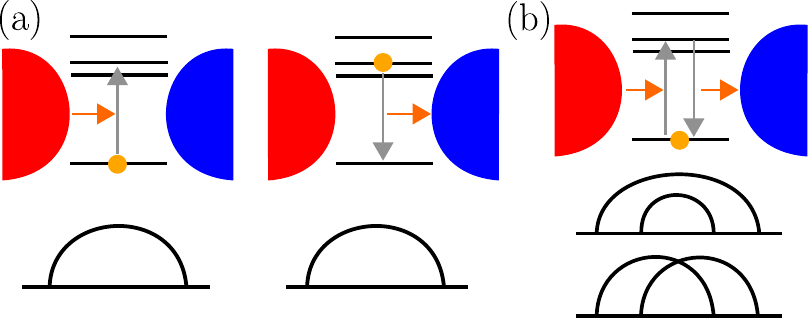}
\caption{Relevant heat transfer mechanisms for a weakly-coupled multi-level junction and the corresponding diagrammatic contributions to the current kernel. (a) - Sequential (second order) processes are dominant at intermediate to high temperature, compared to the relevant energy scale of the system.  (b) - Virtual (fourth order) processes constitute the main transport mechanism at low temperature.}
\label{fig_processes}
\end{figure}

We now specialize the general formalism developed in previous sections to the problem of heat transport through a central system connected to bosonic reservoirs. A simple application to fermionic currents is given for completeness in Appendix~\ref{fermions}. The method will be applied, in Sec.~\ref{HT_QRM}, to the specific model of a qubit-resonator system (Rabi model) coupled to Ohmic heat baths.\\
\indent In Fig.~\ref{fig_processes}, a scheme is provided that summarizes the dominant transport mechanisms at weak coupling with the heat baths. At intermediate to high temperatures, heat transfer occurs mainly via second-order processes whereby the system is sequentially excited and then decays absorbing and re-emitting with a net  of energy from the hot to the cold bath. At low temperature though, the probability for the system to be in an excited state is suppressed and the dominant mechanisms are of fourth order: Energy transfer occurs via virtual processes of the \emph{cotunneling} type. Second- and fourth-order contributions to the heat current are explicitly calculated in the following two subsections.

\subsection{Second-order GME and heat current}

In the boson transport setting, the operators $\hat D$ are independent of the Fock index $p$ (the system coupling operators $\hat{Q}_l$ are Hermitian). We assume $Q_{lnm}\in\mathbb{R}$. Then the GME kernel, to second order, specializes to
\be{K2_bosons}
&\tilde{\mathcal{K}}_{n m n'm'}^{(2)}(0)=-\frac{1}{\hbar^2}\sum_{l }\Bigg\{\sum_{k}\Big[Q_{l  nk}
Q_{l k n'}W_{l k m'}\delta_{m'm}\\
\quad&+
Q_{l m'k}Q_{l km}W_{l k n'}^*\delta_{n'n}\Big]-Q_{l  nn'}
Q_{l m'm}[W_{l nm'}+W_{l mn'}^*]
\Bigg\}\;,
\ee
where, introducing the bath spectral density function $J_l(\omega)=\sum_j\lambda_{l j}^2 \delta(\omega-\omega_j)$, the sum over the reservoir  states $j$ becomes the integral
\be{}
W_{l nm}=&\lim_{\lambda \to 0^+}\sum_p \hbar^2\int_0^\infty d\omega \frac{n_{l}^{p+}(\omega)J_l(\omega)}{\lambda+\ii\omega_{nm} -\ii p\omega}\\
=&\int_{0}^{\infty}dt\langle \hat{B}_l(t)\hat{B}_l(0)\rangle 
e^{-\ii\omega_{nm} t}
\;.
\ee
Here, $\langle \hat{B}_l(t)\hat{B}_l(0)\rangle$ is the correlation function of the baths operators $\hat{B}_l:=\sum_{ j}\hbar\lambda_{l j}(\hat b_{l j}+\hat b_{l j}^\dag)$, the time evolution being with respect to the free bath Hamiltonian. The rates $W_{nm}$ are explicitly calculated in Appendix~\ref{PVintegrals}.\\ 
\indent Similarly, according to Eq.~\eqref{I2nd_general}, we have for the bosonic heat current to bath $r$ the expression
\be{I2}
I_r^{{\rm h}(2)}
&=-\frac{1}{\hbar^2}2{\rm Re}\sum_{nmm'}Q_{r m n}Q_{r n m'}\bar{W}_{r n m}\varrho^\infty_{m'm}\;,
\ee
where
\be{}
\bar{W}_{lnm}=&\lim_{\lambda \to 0^+}\sum_p\hbar^2\int_0^\infty d\omega \frac{\hbar\omega p\; n_{l}^{p+}(\omega)J_l(\omega)}{\lambda+\ii\omega_{nm} -\ii p\omega}\\
=&\;\hbar\omega_{nm}W_{l,nm}+\ii\hbar\langle \hat{B}_l(0)\hat{B}_l(0)\rangle
\;,
\ee
see also Ref.~\cite{Thingna2012}. From Eq.~\eqref{I2}, the zero-time  correlation function enters the current  via  $(2/\hbar)\sum_{mm'}Q_{l m n}Q_{l n m'}\langle \hat{B}_l(0)\hat{B}_l(0)\rangle\varrho^{''\infty}_{m'm}$, where $\varrho^{''\infty}_{m'm}$ is the imaginary part of $\varrho^{\infty}_{m'm}$. This term vanishes due to the hermiticity of $\varrho^{\infty}$, which implies $\varrho^{''\infty}_{m'm}=-\varrho^{''\infty}_{mm'}$ and consequently $\langle \hat{B}_l(0)\hat{B}_l(0)\rangle$ does not contribute to the current.\\ 
\indent In the full secular ME, coherences vanish to lowest order and the heat current to bath $r$ acquires the simple form
\be{current_secular}
I_r^{{\rm h}(2)}=\sum_{n,m}\hbar\omega_{mn}\Gamma^r_{nm}\rho_{mm}^\infty\;.
\ee
Note that, due to $\omega_{nn}=0$, the nonvanishing contributions are given by the rates $\Gamma_{nm}$ with $n\neq m$. According to Eqs.~\eqref{rates2nd} and~\eqref{Wab},
$$
\Gamma^r_{nm(\neq n)}=\frac{1}{\hbar^2}Q_{r  nm}^2 2\Real W_{r nm}=2\pi J_r(\omega_{nm})Q_{r  nm}^2  n_r(\omega_{nm})\;,
$$
which is the transition rate $m\rightarrow n$ induced by the bath $r$ as given by Fermi's golden rule~\cite{Blum2012}.

\subsection{Fourth-order heat current: Low temperature regime}

At low temperature, the dominant contribution to the bosonic heat current is given by the 4th order, as the second-order current is exponentially suppressed.
As shown in Appendix~\ref{4th_order_heat_current}, the 4th-order current kernel has the symmetry
\be{}
\tilde{\mathcal{K}}_{{\rm I} r,nnnn}^{(4)}(\lambda)
=&\sum_{m\neq n}[\tilde{\mathcal{K}}_{{\rm I} r,mmnn}^{(4)}(\lambda)]^*\;,
\ee
which gives (assuming the coherences to vanish at the steady state),
\be{I4heat_current}
I^{{\rm h}(4)}_r &=2\Real\sum_{n,m(\neq n)}\tilde{\mathcal{K}}_{{\rm I} r,mmnn}^{(4)}(0)\rho_{nn}^\infty\;.
\ee
Note that this same result holds for the second-order current, see Eq.~\eqref{I2nd_general}, for vanishing coherences.
The low-temperature expression for the real part of the 4th-order current kernel, in Eq.~\eqref{I4heat_current} is found in Appendix~\ref{4th_order_heat_current} to be ($m\neq n$)
\be{CKernel4th}
2\Real\tilde{\mathcal{K}}_{{\rm I} r,mmnn}^{(4)}(0)
\simeq &
8\pi\hbar
\int_0^\infty d\omega\omega[n_{\bar{r}}(\omega)-n_{r}(\omega)] J_r(\omega)J_{\bar{r}}(\omega)\\
&\times\sum_{k\neq n} \frac{1}{\omega_{mn}\omega_{kn}}Q_{r,mn}Q_{\bar{r},nm}Q_{\bar{r},nk}Q_{r,kn}
\;,
\ee
where $\bar{r}=L$ ($R$) if $r=R$ ($L$).\\
\indent Provided that the ground state energy is well separated from the excited levels, which is the case in the systems considered below, we can set  $\rho_{nn}^\infty\simeq \delta_{n0}$, namely only the ground state of the system is populated. This yields for the current \hbox{$I^{(4)}_r \simeq2\Real\sum_{n\neq 0}\tilde{\mathcal{K}}_{{\rm I} r,nn00}^{(4)}(0)$}.
Using 
\be{dndDt}
\frac{\partial n_L(\omega)}{\partial\Delta T}\Big|_{\Delta T=0}=\frac{\hbar\omega}{4k_B T^2}\frac{1}{\sinh^2(\beta\hbar\omega/2)}
\ee
and $\int_0^\infty dx x^4/\sinh^2(x/2)=16\pi/15$, we readily find for the fourth-order contribution to the linear conductance
\be{kappa4th}
\kappa^{(4)}
\simeq &
\frac{32\alpha^2\pi^5 k_B^4 T^3}{15\hbar^3} \sum_{k,m\neq 0}
\frac{Q_{R,m0}Q_{L,0m}Q_{L,0k}Q_{R,k0}}{\omega_{m0}\omega_{k0}}\;,
\ee
where, due to the low-$\omega$ cutoff imposed by $\sinh^2(\beta\hbar\omega/2)$ at low temperature, we used $J_l(\omega)\simeq \alpha\omega$ (for Ohmic heat baths).
Note that the matrix elements of the coupling operators $\hat{Q}_l$ play an important role in Eq.~\eqref{kappa4th} in that they determine the overall sign for each contribution to the sum. This yields interference effects -- and a resulting coherent suppression of the conductance -- in the presence of quasi-degenerate transitions, as shown below for specific examples of multi-level junctions. Equation~\eqref{kappa4th} constitutes a major result of the application of the Liouville-space formalism to bosonic heat transport with Ohmic heat baths.

\vspace{-0.25cm}

\subsection{Recovering special cases: Two-level system and two coupled oscillators}\label{HT_TLS2HO}
In view of applying to the quantum Rabi model the general theory formulated in the previous sections, it is helpful to summarize the known results for the simpler cases. We consider a weakly-coupled spin-boson model (two-level system coupled to harmonic heat baths) and also apply the formalism to a multi-level system that admits and exact solution for the heat current, namely a junction composed of two coupled quantum oscillators.
\subsubsection{Spin-boson model}\label{SBM}

\indent Consider a two-level system (TLS) junction, with energy eigenstates $\ket{0},\ket{1}$ and of frequency $\omega_{10}$, weakly coupled to bosonic heat baths.
The populations of the TLS are obtained by solving the steady-state full secular master equation, Eq.~\eqref{BR_full_secular}, specialized to the TLS with rates $$\Gamma_{01}^l=\gamma^l[n_l(\omega_{10})+1]\;, \quad\Gamma_{10}^l=\Gamma_{01}^l e^{-\beta_l\hbar\omega_{10}}\;,$$
where $\gamma^l:=2\pi J_l(\omega_{10})  Q_{l01}^2$. Using Eq.~\eqref{current_secular}, we obtain for the current to the reservoir $r$~\cite{Segal2005, Segal2005PRL}
\be{I2R_TLS}
I^{{\rm h}(2)}_{r,{\rm TLS}}=\frac{\hbar\omega_{10} \gamma^R\gamma^L[n_{\bar{r}}(\omega_{10})-n_r(\omega_{10})]}{\gamma^R[1+2n_R(\omega_{10})]+\gamma^L[1+2n_L(\omega_{10})]}\;.
\ee
To second order and for identical Ohmic baths with high-frequency cutoff, $J_l(\omega_{10})\simeq\alpha\omega_{10}$, giving $ \gamma^l \simeq 2\pi\alpha \omega_{10} Q_{l01}^2$. The conductance of the TLS is obtained applying Eq.~\eqref{dndDt} and reads
\be{kappa2TLS}
\kappa^{(2)}_{\rm TLS}
&=\alpha \eta k_B^2 \omega_{10}\frac{(\hbar\omega_{10}/k_BT)^2}{2\hbar\sinh(\hbar\omega_{10}/k_BT)}\;,
\ee
where $\eta :=2\pi Q_{R01}^2 Q_{L01}^2/(Q_{R01}^2+Q_{L01}^2)$ is a dimensionless asymmetry factor. It has been shown that, in the spin-boson model, expansion of the conductance up to fourth order covers the whole temperature regime provided that the coupling to the baths is weak~\cite{Yamamoto2018}.  
Equation~\eqref{kappa4th} recovers, for the 4th-order conductance of the TLS, the known result~\cite{Yamamoto2018}
\begin{figure}[ht!]
\includegraphics[width=8.cm]{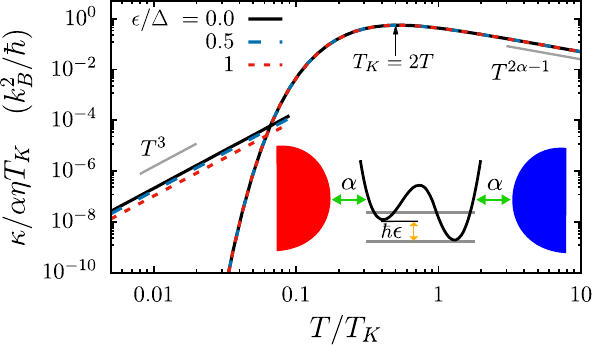}
\caption{Thermal conductance of the spin-boson model \emph{vs}. the temperature, rescaled with the Kondo-like temperature $k_B T_K=\hbar\omega_{10}$, for three values of the applied bias $\epsilon$. The second-order conductance $\kappa^{(2)}_{\rm TLS}$ shows a scaling behavior and is exponentially suppressed at low temperature. In the latter regime, $\kappa$ is dominated by $\kappa^{(4)}_{\rm TLS}$, which displays the $T^3$ universal behavior, with a bias-dependent prefactor.}
\label{fig_kappa_vs_T_qubit}
\end{figure}
\be{kappa4TLS}
\kappa^{(4)}_{\rm TLS}
\simeq &
\frac{32\alpha^2\pi^5 k_B^4 T^3}{15\hbar^3}
\frac{Q_{R,10}^2Q_{L,10}^2}{\omega_{10}^2}\;.
\ee
As a concrete example, consider a TLS with Hamiltonian 
$$\hat H_{\rm TLS}=-\frac{\hbar}{2}(\epsilon\hat\sigma_z+\Delta\hat\sigma_x)\;,$$
coupled to Ohmic heat baths according to $\hat H=\hat H_{\rm TLS}+\hat H_B+\hat H_V$, see Fig.~\ref{fig_kappa_vs_T_qubit}. Here, $\hat H_B$ and $\hat H_V$ are given by Eq.~\eqref{Hbosons} and the qubit coupling operators are $\hat{Q}_l=\hat \sigma_z$.
The TLS is defined by the lowest energy doublet of a double-well potential. This description is appropriate for a flux qubit~\cite{Rasmussen2021}, see Fig.~\ref{Rabi_model} below, where the energy minima correspond to the flux states associated to clockwise/anti-clockwise circulating current (circular arrows in Fig.~\ref{Rabi_model}(a)). These states are coupled by a tunneling amplitude $\Delta$, which is the qubit frequency gap for an unbiased potential.
An energy bias $\hbar\epsilon$ can be induced  via an  applied external magnetic flux $\phi_{\rm ext}$.
We choose identical Ohmic-Drude spectral densities $J_l(\omega)=\alpha \omega/(1+\omega^2/\omega_c^2)$ for the heat baths with coupling strength $\alpha=10^{-3}$ and cutoff frequency $\omega_c=5~\omega_r$.\\ 
\indent In Fig.~\ref{fig_kappa_vs_T_qubit} the conductance as a function of the temperature is shown for three values of the applied bias $\epsilon$.
It is insightful to introduce the Kondo temperature scale $T_K$, defined in terms of the TLS frequency scale $\omega_{10}=\sqrt{\Delta^2+\epsilon^2}$ via $k_B T_K=\hbar\omega_{10}$. In general, this frequency scale is renormalized by the coupling $\alpha$ to the baths, whereby \hbox{$\Delta \to \Delta_r=[\Gamma(1-2\alpha)\cos(\pi\alpha)]^{(1/2(1-\alpha))}\Delta(\Delta/\omega_c)^{\alpha/(1-\alpha)}$}, where $\Gamma(x)$ is the Gamma function~
\cite{Saito2013}. In our case this effect is negligible, as we assume the coupling to be weak. 
\begin{figure}[ht!]
\includegraphics[width=8.cm]{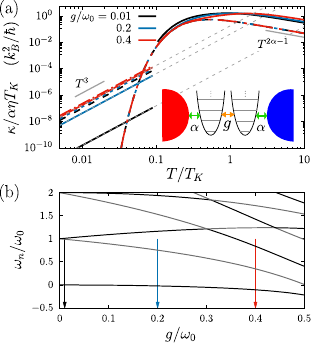}
\caption{Thermal conductance of a two-oscillator junction (a) - Thermal conductance of two coupled oscillators of frequency $\omega_0$ for three values of the \emph{internal} coupling strength $g$. Dashed lines: $\kappa^{(2)}_{\rm TLS}$ and $\kappa^{(4)}_{\rm TLS}$ for a TLS truncation of the two-oscillator system. Dashed-gray lines: Analytical expression from~\cite{Freitas2014} in the limit of low temperature and weak coupling, Eq.~\eqref{kappa2HO}. (b) - Spectrum of the two-oscillator system as a function of their coupling strength. The arrows point at the values of $g$ chosen in panel (a). At small $g$ the first and second excited states are quasi-degenerate. This results in the coherent suppression of the low-temperature conductance, as given by the interference effect captured by Eq.~\eqref{kappa4th}.}
\label{fig_kappa_vs_T_osc}
\end{figure}
The conductance displays two distinct regimes: At intermediate to high temperature, from $T\lesssim 0.1~T_K$ upwards, it is dominated by $\kappa^{(2)}_{\rm TLS}$, Eq.~\eqref{kappa2TLS}. The latter gives a maximum at $T/T_K=1/2$. With the present scaling, the curves corresponding to the different values of bias collapse. At high temperature the conductance scales as $T^{-1}$ (a very good approximation of the nonperturbative result $T^{2\alpha-1}$). At low temperature, $T\ll T_K$, the second-order conductance is exponentially suppressed, and $\kappa$ is dominated by the fourth-order contribution $\kappa^{(4)}_{\rm TLS}$, Eq.~\eqref{kappa4TLS}, which decays algebraically as $T^3$.  
Note that the latter does not scale with the asymmetry factor $\eta$ as $\kappa^{(2)}_{\rm TLS}$ due to the different combination of the matrix elements of the coupling operators.
As a result, the universal behavior $T^3$, as shown in Fig.~\ref{fig_kappa_vs_T_qubit}, is modulated by a bias-dependent prefactor: This is seen in the decreasing conductance for increased $\epsilon$.\\

\subsubsection{Two coupled oscillators}\label{TCO}

As a second application, to benchmark our multi-level formula~\eqref{kappa4th}, we consider the case where the junction is formed of two coupled oscillators placed in series between the heat baths, see Fig.~\ref{fig_kappa_vs_T_osc}(a). Given this configuration, we identify each oscillator using the bath label $l=L,R$ of the bath it directly couples to. We assume the same Ohmic baths and system-bath couplings as for the TLS considered above. 
The linearity of the system allows for an exact solution for the heat current, as demonstrated in~\cite{Freitas2014},  where general expressions are given for the heat current through harmonic networks. 
The Hamiltonian of the two-oscillator junction reads
\be{}
H_{\rm osc}=\sum_{l=L,R}\hbar\omega_l a_l^\dag a_l+\hbar g \hat Q_L\hat Q_R\;,
\ee
where $\hat Q_l=a_l^\dag + a_l$ and $g$ is the coupling strength. The spectrum is characterized by non-equally-spaced levels and, for $\omega_L=\omega_R$, by quasi-degeneracies at small $g$ (the excited levels form doublets with frequency splitting $2g$), see Fig.~\ref{fig_kappa_vs_T_osc}(b). The baths' Hamiltonian $\hat H_B$ and the system-bath coupling term $\hat H_V$ are given by Eq.~\eqref{Hbosons} with $\hat Q_l$ the system operator that couples to bath $l$. Form Ref.~\cite{Freitas2014}, in the weak coupling limit, the exact expression for the current to bath $r$ through an harmonic network specializes to 
\be{}
I^{h}_r\simeq 8\pi\hbar\alpha^2\frac{4g^2}{(\omega_L\omega_R-4g^2)^2}\int_0^\infty d \omega \omega^3[n_{\bar r }(\omega)-n_r(\omega)]\;.
\ee
The corresponding conductance, in the low-temperature limit, reads
\be{kappa2HO}
\kappa\simeq\frac{32 \pi^5 k_B^4 T^3\alpha^2}{15 \hbar^3}
\frac{4g^2}{(\omega_L\omega_R-4g^2)^2}\;,
\ee
which is to be compared to the cotunneling formula~\eqref{kappa4th}.\\
\indent In Fig.~\ref{fig_kappa_vs_T_osc}(a), the above expression is rendered, for \hbox{$\omega_L=\omega_R=\omega_0$}, by the dashed-grey lines with $T^3$ behavior and displays agreement with our general multi-level expression~\eqref{kappa4th} (solid lines with $T^3$ behavior). 
As shown in Fig.~\ref{fig_kappa_vs_T_osc}(b), for weak internal coupling the first and second excited state are quasi-degenerate. In this regime, the cotunneling formula for the conductance, Eq.~\eqref{kappa4th}, describes an interference effect that results in a large coherent suppression of the conductance. This prediction is confirmed by the weak-coupling, low-temperature limit, Eq.~\eqref{kappa2HO}, of the exact formula for coupled oscillators~\cite{Freitas2014}.
At intermediate to high temperatures, the conductance, solid curves in the center and right part of Fig.~\ref{fig_kappa_vs_T_qubit}(b), is given by $\kappa\simeq \kappa^{(2)}$, obtained numerically from Eq.~\eqref{kappa} with $I^{\rm h}_r\simeq I^{\rm h(2)}_r$, the current being calculated within the full secular approximation, Eqs.~\eqref{BR_full_secular} and~\eqref{current_secular}. The dashed lines are obtained considering a TLS truncation formed of the ground and first excited state of the two-oscillator system and using again Eqs.~\eqref{kappa2TLS} and~\eqref{kappa4TLS}. Note that at strong coupling the TLS truncation reproduces the conductance of the system up to intermediate temperatures.

\section{Application: heat transport in the quantum Rabi model}\label{HT_QRM}

\begin{figure}[ht!]
\includegraphics[width=7cm]{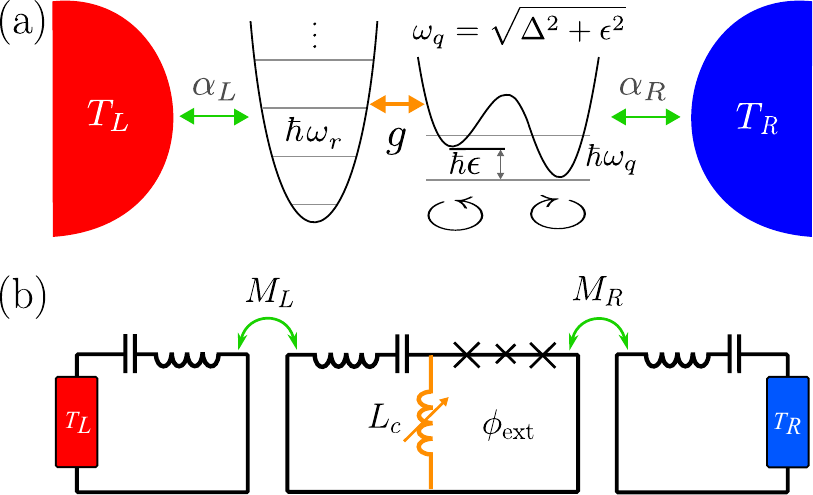}
\caption{Heat transport setup where a qubit-oscillator system (Rabi model), constitutes  the junction between two heat baths. (a) - Scheme of the setup. (b) Simplified circuit model of the corresponding superconducting circuit realization. The qubit bias $\epsilon$ is induced via the applied flux $\phi_{\rm ext}$.}
\label{Rabi_model}
\end{figure}
In this section, we apply the theory to the case of heat transport in a qubit-oscillator system described by the quantum Rabi model, see also~\cite{companion}. In particular, we wish to outline which features are unique to the Rabi model and which are universal, in line with the findings in the two archetypal examples of the previous section. This system is of high relevance for current  heat transport experiments involving coupled qubit-resonator systems~\cite{Pekola2021}.

\subsection{Model}

In panel (a) of Fig.~\ref{Rabi_model}, a schematic depiction of the heat transport setup is given, where the junction between the bosonic heat baths is formed of a qubit and an oscillator (resonator), of frequencies $\omega_q$ and $\omega_r$, respectively, coupled with coupling strength $g$. This system is described by the quantum Rabi model~\cite{Rabi1936,Rabi1937}. 
We consider a realization of the setup based on a superconducting circuit~\cite{Blais2021}.
Figure~\ref{Rabi_model}(b) shows a simplified circuit model. The qubit resonator system is realized by a flux qubit (the symbol $\times$ indicates a Josephson junction) sharing a common tunable inductance $L_c$~\cite{Forn-Diaz2010,Yoshihara2017,Magazzu2021} with an LC circuit, the oscillator. The resulting Rabi model is inductively coupled to RLC circuits that realize the heat baths, the mutual inductance $M_l$ quantifying the coupling strength~\cite{Ojanen2008,Karimi2016,Karimi2017,Iorio2021}. The parameters of the RLC circuits realizing the heat baths should be chosen such that their spectral density functions are  approximately of the Ohmic type at low frequency. 
The assumption of weak system-baths coupling is realistic experimentally and allows for a perturbative treatment of the system-baths interaction.
The resulting Hamiltonian of the setup is given by $\hat H=\hat H_{\rm Rabi}+\hat H_B+\hat H_V$, where the baths and interaction Hamiltonians $\hat H_B$ and $\hat H_V$ are defined in Eq.~\eqref{Hbosons} while
\be{HRabi}
\hat H_{\rm Rabi}=-\frac{\hbar}{2}(\epsilon\hat\sigma_z+\Delta\hat\sigma_x)+\hbar\omega_r \hat{a}^\dag \hat{a} +\hbar g \hat\sigma_z(\hat{a}^\dag + \hat{a})\;.
\ee
The first and second terms of the junction's Hamiltonian describe the flux qubit (in the basis of persistent current states, see also Sec.~\ref{HT_TLS2HO}) and the resonator, respectively, with $\sigma_i$ the Pauli spin operators and $\hat{a}^\dag$ and $\hat{a}$ bosonic creation and annihilation operators. The third term gives the qubit-oscillator coupling. The interaction with the heat baths $L$ and $R$ is mediated by the oscillator and qubit coupling operators $\hat{Q}_L=\hat{a}^\dag + \hat{a}$ and $\hat{Q}_R=\hat\sigma_z$, respectively. In the literature on ultrastrong coupling, the coupling regime is established by comparing $g$ to both the loss rates of the system and its characteristic frequencies, see e.g.~\cite{Zueco2019}. Here, as in~\cite{companion}, the coupling $\alpha$ to the heat baths is assumed to be very small so that, for simplicity, we identify the coupling regimes based on the value of the ratio $g/\omega_r$.
At weak coupling, the counter-rotating terms in the coupling Hamiltonian 
expressed in the qubit energy eigenbasis, namely $\hat\sigma_+\hat a^\dag$ and $\hat\sigma_- \hat a$, can be safely neglected. The resulting RWA yields an easily solvable block-diagonal Hamiltonian~\cite{Jaynes1963,Shore1993}. In the opposite regime of USC the counter-rotating terms are relevant and lead to peculiar nonperturbative effects~\cite{Forn-Diaz2018review, Kockum2019}. This regime is captured, for example, by perturbation theory in the qubit splitting $\Delta$, renormalized by the coupling to the oscillator~\cite{Irish2007,Ashhab2010, Hausinger2010PRA,Zhang2013}.  

\subsection{Thermal conductance of the quantum Rabi model}
\label{results}

In the following, we analyze the  steady-state heat transport properties of the setup via the thermal conductance $\kappa$, defined in Eq.~\eqref{kappa}. Note that the formalism developed in the previous sections has no restrictions on the temperature/chemical potential bias of the quantum transport setup and is able to describe the nonlinear transport regime at no additional cost. As in Sec.~\ref{HT_TLS2HO}, we assume identical Ohmic-Drude spectral densities $J_l(\omega)=\alpha \omega/(1+\omega^2/\omega_c^2)$  with $\alpha=10^{-3}$ and $\omega_c=5~\omega_r$. The conductance is calculated up to fourth order. The leading-order contribution $\kappa^{(2)}$ is given by  Eq.~\eqref{kappa} with $I^{\rm h}_r\simeq I^{\rm h(2)}_r$, the current being calculated using the partial secular, Eqs.~\eqref{BR_partial_secular} and~\eqref{I2}, or the full secular ME, Eqs.~\eqref{BR_full_secular} and~\eqref{current_secular}, depending on the spectral properties of the system. For the fourth-order contribution $\kappa^{(4)}$ we use the analytical limiting expression~\eqref{kappa4th}.\\
\begin{figure}[ht!]
\includegraphics[width=8.cm]{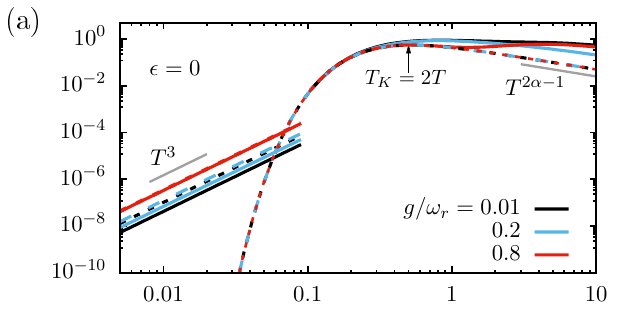}\\
\vspace{-0.25cm}
\includegraphics[width=8.cm]{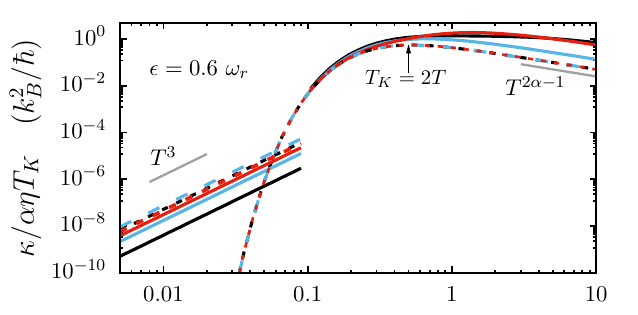}\\
\vspace{-0.0cm}
\hspace{0.195cm}
\includegraphics[width=6.97cm]{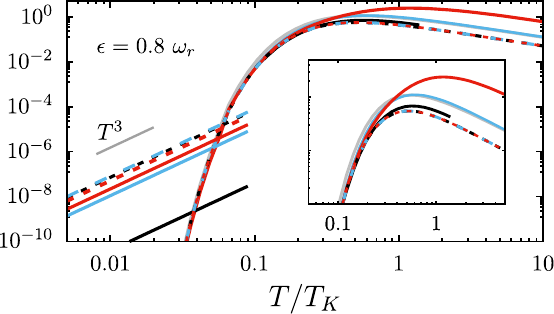}\\
\includegraphics[width=8.25cm]{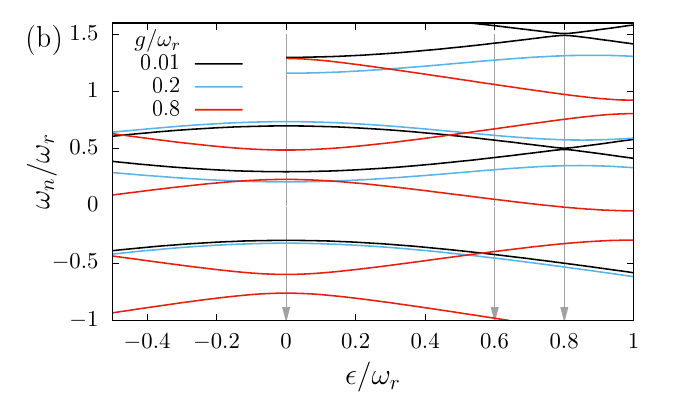}
\caption{Universality and interference effects in the Rabi model for $\Delta=0.6~\omega_r$. (a)  Thermal conductance \emph{vs}. the temperature rescaled with the Kondo-like temperature $ T_K(g)=\hbar\omega_{10}/k_B$ for three values of the qubit-resonator coupling. The three panels show the conductance for different values of the qubit bias $\epsilon$, see the arrows in panel (b). The conductance of the full Rabi model is (solid lines) is compared to the one of its TLS truncation (dashed lines). The gray solid line at $\epsilon=0.8~\omega_r$ shows the result of the the full secular approximation for $g=0.01~\omega_r$. The inset highlights the difference with the partial secular ME (black solid line). (b) Spectra of the Rabi model as a function of the qubit bias. Quasi degeneracies for low $g$ under qubit-resonator resonance conditions
(here for $\epsilon/\omega_r=0.8$) correspond to suppression of the low-temperature conductance.}
\label{fig_kappa_vs_T}
\end{figure}

\subsubsection{Conductance as a function of the temperature: Universality, interference and the USC regime}

\indent We start by considering the temperature behavior of the conductance of the Rabi model up to fourth order. The results, scaled by the appropriate Kondo temperature $T_K(g)$, are shown in Fig.~\ref{fig_kappa_vs_T}(a) (solid lines). A comparison is made with the conductance $\kappa_{\rm TLS}$ of the TLS truncation of the Rabi model, Eqs.~\eqref{kappa2TLS} and~\eqref{kappa4TLS}, namely the TLS formed by the ground and first excited state of the full Rabi model (dashed lines). Similarly to the cases of TLS and coupled-oscillators junctions, Figs.~\ref{fig_kappa_vs_T_qubit} and~\ref{fig_kappa_vs_T_osc}, we introduced also here a Kondo-like temperature scale $T_K(g)=\hbar\omega_{10}/k_B$, defined in terms of the frequency scale $\omega_{10}=\omega_1-\omega_0$ associated to the lowest energy doublet of the quantum Rabi model. As shown below and in~\cite{companion}, for $T\lesssim T_K$ the Rabi model is well described by its TLS truncation and a scaling behavior emerges in $\kappa^{(2)}$ as a function of $T$ when both quantities are scaled with $T_K$. Indeed $\kappa^{(2)}_{\rm TLS}/T_K$ is a function of $T/T_K$, see Eq.~\eqref{kappa2TLS}.
For $T\ll T_K$ the system is essentially in the ground state. In this case, $\kappa^{(2)}_{\rm TLS}$ displays an exponential suppression at low temperature. Indeed, in this temperature regime, virtual processes, accounted for by the 4th-order expression for the current kernel, Eq.~\eqref{CKernel4th},  allow for energy transfer with the system staying in its ground state~\cite{Saito2013,Yamamoto2018,Bhandari2021, companion}. This results in the algebraic low-$T$ behavior $\kappa \propto T^3$. The three panels in Fig.~\ref{fig_kappa_vs_T}(a) show different values of the applied qubit bias for $\Delta=0.6~\omega_r$ and for the same three chosen values of coupling strength $g$, ranging from the weak to the USC regime.  In particular, in moving from the first to the third panel, the qubit and oscillator frequencies approach resonance $\omega_q = \omega_r$, starting from far detuned. This is reflected in Fig.~\ref{fig_kappa_vs_T}(b) where the spectrum at weak coupling displays, at resonance, quasi-degenerate doublets of excited levels. The latter in turn result in a large \emph{coherent} suppression effect of the low-$T$ conductance, with respect to the TLS counterpart.  The same effect is obtained at zero bias for $\Delta=\omega_r$, see~\cite{companion}. This suppression is due to inherently multi-level interference effects induced by the sum over the states in Eq.~\eqref{kappa4th}, which is absent in the TLS truncation, cf. Eq.~\eqref{kappa2TLS}. The analytical RWA treatment of the coupling matrix elements, see the comment below Eq.~\eqref{Q_RWA}, provides a simple explanation for this.
At resonance, USC restores the separation of the energy levels and yields an increased $\omega_{10}$ with $\kappa^{(4)}$ approaching $\kappa^{(4)}_{\rm TLS}$.\\
\indent As anticipated above, in the intermediate temperature regime, where the conductance is dominated by $\kappa^{(2)}$ but the temperature is low enough so that $\kappa^{(2)}\simeq \kappa^{(2)}_{\rm TLS}$, the curves at different $g$ collapse for all values of detuning. The second-order conductance $\kappa^{(2)}_{\rm TLS}$ of the TLS truncation of the Rabi model has a maximum when the condition $T_K=2T$ is met, as derived from Eq.~\eqref{kappa2TLS}. 
Note that, in the USC regime, the scaling behavior $\kappa^{(2)}\simeq \kappa^{(2)}_{\rm TLS}$ extends to higher temperatures.
Increasing further the temperature, higher energy levels are involved and the TLS truncation breaks down: The curves depend on the details of the spectrum (and the matrix elements of the coupling operators) which are peculiar to the different values of the coupling. Deviations from the behavior $\kappa\sim T^{2\alpha-1}$, predicted for a TLS and well approximated by the $T^{-1}$ limiting behavior of $\kappa^{(2)}_{\rm TLS}$, are found. 
\begin{figure}[ht]
\includegraphics[width=8.5cm]{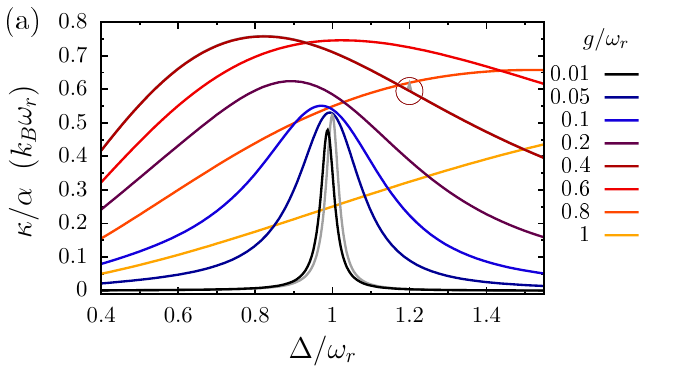}
\includegraphics[width=8.5cm]{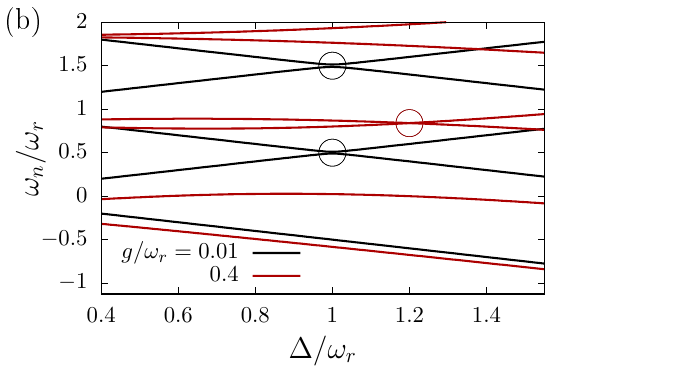}
\caption{Thermal conductance as a function of the qubit-resonator detuning from weak to USC coupling at zero bias, $\epsilon=0$.  (a) Conductance \emph{vs.} the splitting $\Delta$ at $k_B T=0.2~\hbar\omega_r$.  In this temperature regime the conductance is dominated by $\kappa^{(2)}$. 
Weak qubit-oscillator coupling, at resonance, gives $\omega_{21}\ll\omega_r$. This, in turn, yields a lower peak-conductance and a shift of the peak (Lamb shift) in the result from the partial secular master equation with respect to the full secular approximated version (gray lines). (b) Spectra of the Rabi model as a function of $\Delta$ for two values of $g$. Circles highlight the (quasi-) degeneracies.}
\label{fig_kappa_vs_D}
\end{figure}
It is worth noting that also the second-order conductance is suppressed, at resonance and for weak $g$, due to steady-state coherences, see also~\cite{Ivander2022}. 
Indeed, the conductance predicted by the partial secular master equation, solid, black curve for $\epsilon=0.8~\omega_r$ in Fig.~\ref{fig_kappa_vs_T}(a), is smaller than the one calculated in the full secular approximation (solid gray line in the same panel, see the inset). 
Due to a truncation to the first five levels used in the partial secular ME, its high-temperature behavior is not reliable and thus omitted. An illustration of the emergent steady-state coherences in the presence of quasi-degenerate levels at second order is provided in Appendix~\ref{analytical_solution} for a three-level truncation of the full Rabi model. Likewise, the efficiency of quantum thermal machines is impacted by coherences, as shown in~\cite{Pekola2019,Gramajo2023}. \\  

\subsubsection{Conductance \emph{vs.} qubit-oscillator detuning: Transition from resonant  to shifted, broadened peaks}

\begin{figure}[ht!]
\includegraphics[width=7.8cm]{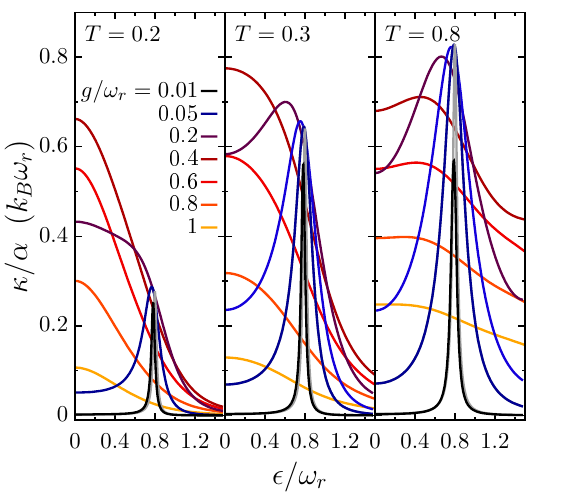}
\caption{Transition from resonant to USC regime. Thermal conductance \emph{vs.} the qubit bias $\epsilon$ for $\Delta/\omega_r=0.6$ and three values of the temperature (in units of $\hbar\omega_r/k_B$).  In this temperature regime the conductance is dominated by $\kappa^{(2)}$. For weak qubit-oscillator coupling, at resonance, $\omega_{21}\ll\omega_r$, namely the first and second excited states of the Rabi Hamiltonian are quasi-degenerate, see Fig.~\ref{fig_kappa_vs_T}(b). This in turn gives a lower peak-conductance and a Lamb-shift of the resonant peak in the partial secular master equation (black lines) with respect to the full secular approximated version (gray lines).}
\label{fig_kappa_vs_eps}
\end{figure}
In Figs.~\ref{fig_kappa_vs_D} and~\ref{fig_kappa_vs_eps}, the conductance is shown as a function of the qubit-oscillator detuning in a temperature regime where $\kappa\simeq\kappa^{(2)}$. 
The transport features shown in these figures result from the interplay between the qubit-resonator coupling strength $g$ and the detuning and also include coherent effects at the level of second order.\\
\indent In Fig.~\ref{fig_kappa_vs_D}(a), $\kappa$ is plotted as a function of the qubit splitting $\Delta$, at zero qubit bias, for several values of the qubit-oscillator coupling strength, well into the USC regime. 
At weak coupling, the conductance is sharply peaked at resonance, and suppressed elsewhere. At resonance, the weak-coupling spectrum displays quasi-degeneracies, indicated by the black circles in  Fig.~\ref{fig_kappa_vs_D}(b). As already seen in Fig.~\ref{fig_kappa_vs_T}, these entail a deviation of the conductance from the full secular approximation. The latter is depicted as a grey line. The partial secular ME yields a suppressed peak and a Lamb shift (black line).
Increasing the coupling, the peaks broaden and move towards values of $\Delta$ smaller than $\omega_r$. However, in the nonperturbative coupling regime, the maxima are found at $\Delta>\omega_r$. 
Intuition on these three regimes across the range of coupling $g$ considered in Fig.~\ref{fig_kappa_vs_D}, namely resonant peak, broadened peaks for $\Delta<\omega_r$, and maxima for $\Delta>\omega_r$, is given by the approximate treatments of the quantum Rabi model summarized in Appendix~\ref{Rabi_analytical}. At weak coupling, the RWA  shows that the matrix elements of the coupling operators $Q_{l,01}$ are suppressed everywhere except at the resonance condition, see Eq.~\eqref{QLR_RWA}. At larger $g$, Van Vleck perturbation theory in $g$, a second-order treatment which includes in a perturbative way the counter-rotating terms in the qubit-oscillator coupling Hamiltonian~\cite{Hausinger2008}, predicts a shift of the resonance condition to negative detuning $\Delta<\omega_r$, as pointed out below 
Eq.~\eqref{uv_VVPT}. Finally, at large coupling, in the nonperturbative regime, the generalized rotating-wave approximation applies~\cite{Irish2007}. It requires a large bare splitting, $\Delta>\omega_r$, in order to achieve the resonance condition due to an effective dressing of the bare qubit splitting, see Eq.~\eqref{QLR_GRWA}. 
For values of $g$ larger than $0.01~\omega_r$, the results from the full secular ME essentially reproduce those of the partial secular ME, with the exception of the degeneracy at $\Delta=1.2~\omega_r$ for $g=0.4~\omega_r$, see the dark-red circle in Fig.~\ref{fig_kappa_vs_D}(b). In this case, the partial secular approximation corrects an unphysical jump given by the full secular approximation, as highlighted by the dark-red circle in Fig.~\ref{fig_kappa_vs_D}(a).\\
\indent The behavior of the conductance as a function of the qubit bias $\epsilon$ with fixed $\Delta<\omega_r$ displays, at low $g$, resonant peaks when the qubit frequency $\omega_q=\sqrt{\Delta^2+\epsilon^2}$ matches $\omega_r$. Upon increasing $g$, $\kappa$
undergoes a transition from  resonant peaks to a broadened zero-bias maxima~\cite{companion}. 
This is shown in Fig.~\ref{fig_kappa_vs_eps}, where the case $\Delta=0.6~\omega_r$ is considered. 
Similarly to Figs.~\ref{fig_kappa_vs_T} and~\ref{fig_kappa_vs_D}, at resonance and weak coupling, the spectrum presents quasi-degeneracies, see Fig.~\ref{fig_kappa_vs_T}(b), that entail nonzero steady-state coherences. The latter result in a suppression of the conduction peak, with respect to the full secular approximation (gray curves) and a Lamb-shift to negative detuning, see also~\cite{companion}. 
This transition from resonant peaks to broad zero-bias maxima is reproduced at different temperatures: The suppression of the weak-coupling resonance peak is proportionally larger at higher temperature, consistently with  increased steady-state coherences, see Fig.~\ref{fig_rho12} in Appendix~\ref{analytical_solution},  and the onset of the zero-bias maxima occurs at larger $g$ the larger the temperature.\\
\indent These results show how the conduction properties of the quantum Rabi model, in a realistic implementation based on superconducting circuits, can be tuned by acting on the qubit parameters and on the qubit-oscillator coupling. The application showcases the nontrivial features of quantum heat transport of a system weakly coupled to heat baths which can be easily accessed experimentally.

\section{Conclusions}
\label{conclusions}

In this work, we have presented a unified treatment of quantum transport in nanojunctions coupled to fermionic or bosonic baths. It is based on a generalized master equation approach in Liouville space to open quantum systems and is suitable for a perturbation expansion in the system-baths coupling. Nevertheless, nonperurbative results are also possible with suitable diagram selections to all order~\cite{Donarini2024}. 
While the fermionic case has already being extensively investigated in the context of charge transport in  quantum dot systems, bosonic heat transport beyond the leading order  in multi-level systems has not yet received the same attention. 
The diagrammatic method described in this work allowed us to obtain analytical, approximate expressions for the steady-state bosonic heat transport in multi-level systems up to the so-called cotunneling level, namely to fourth order in the system-baths coupling.  At the level of second order, projection in the system eigenbasis reproduces the Redfield equation.\\
\indent Though the method can be directly applied to the nonlinear transport regime, we focused on linear transport: Known results for the spin-boson model and for a two-oscillator junction are reproduced, and novel results for generic multi-level systems are obtained.
The low-temperature thermal conductance is dominated by fourth-order processes where heat transfer between the baths occurs via virtual transitions in the system. Generically, a power-law behavior of the linear conductance $\kappa\sim T^3$, is found. For quasi-degenerate excited states, interference effects yield a large suppression of the conductance. This is an exquisite multi-level effect which is not displayed by two-level-system junctions. Besides this, system-specific properties are found, which depend on the details of the multi-level spectrum.\\
\indent This behavior is clearly observed for the case in which the junction is formed of a qubit-oscillator system described by the quantum Rabi model. The features of thermal transport in this system are dictated by the qubit-oscillator detuning and coupling strength. At resonance and weak coupling, the multi-level structure of the system presents quasi-degeneracies in the excited levels that induce steady-state coherences at second order and multi-level interference effects at the level of cotunneling. This is shown by studying the conductance as a function of the temperature at different detunings and by changing the qubit splitting and bias in a temperature regime where the sequential tunneling is the dominant heat transfer mechanism. Heat transport regimes are crucially influenced by the qubit-resonator coupling strength. Indeed, coherent effects are removed at large coupling due to the pronounced   repulsion of the excited levels and a behavior that converge to the one of the two-level system truncation of the Rabi model is found in the USC regime. The conductance as a function of the temperature displays in this case a scaling behavior which is highlighted by introducing the Kondo-like temperature scale corresponding to the separation internal to the lowest energy doublet.\\
\indent The thermal conductance as a function of the detuning shows instead a transition from a resonant peak behavior at weak coupling to a broadened, shifted peaks in the USC regime. In particular, when increasing the coupling at zero applied bias on the qubit, the resonant peak broadens and moves to negative detuning, i.e. the maxima occur at values of the qubit splitting smaller than the oscillator frequency. Upon a further increase of the coupling, in the non perturbative regime, the maxima move to positive detunings as the Rabi system behaves as an effective two-level system undergoing a strong down-renormalization of its energy. When adjusting the applied bias, the weak-coupling resonant peaks make a transition to zero-bias maxima upon increasing the coupling.\\
\indent The results obtained and the specific application to the quantum Rabi model, which describes the core element of circuit QED, are of relevance given the level of experimental control and extreme parameter regime achieved in superconducting setups. These are among the leading platforms for quantum information and simulation and, more broadly, provide appealing technological applications~\cite{Vepsalainen2016,Blais2020,Blais2021, Pekola2021}.  Moreover, in our model, the qubit-oscillator system is placed in series between the heat baths. This constitutes an inherently asymmetric, nonlinear junction and thus satisfies the criteria for displaying heat rectification~\cite{Segal2005, Senior2020}.

\section{Acknowledgements}

The authors thank A. Donarini, G. Falci, and J. Pekola for fruitful discussions.
LM and MG acknowledge financial support from BMBF (German Ministry for Education and Research), Project No. 13N15208, QuantERA SiUCs. The research is part of the Munich Quantum Valley, which is supported by the Bavarian state government with funds from the Hightech Agenda Bavaria. EP acknoweldges financial support from PNRR MUR project
PE0000023-NQSTI and from COST ACTION SUPQERQUMAP, CA21144.

\appendix

\section{Nakajima-Zwanzig formalism for quantum transport}
\label{appendix_GME}

The solution of the Liouville-von Neumann equation for the total density operator is $\hat\varrho_{\rm tot}(t)=e^{\mathcal{L} t} \hat\varrho_{\rm tot}(0)$, where \hbox{$\mathcal{L}=\mathcal{L}_0 +\mathcal{L}_V $}, with $\mathcal{L}_0=\mathcal{L}_S+\mathcal{L}_B$. 
In Laplace space
\be{rho_lambda}
\tilde\varrho_{\rm tot}(\lambda) &= \frac{1}{\lambda - \mathcal{L}}\hat\varrho_{\rm tot}(0)\equiv  \tilde{\mathcal{G}}(\lambda) \hat\varrho_{\rm tot}(0)\;.
\ee
Let us introduce the projection superoperators
\be{projectors}
\mathcal{P}\bullet =\Tr_B\{\bullet\}\otimes\varrho_B\qquad{\rm and}\qquad
\mathcal{Q} =\mathbf{1}-\mathcal{P}\;.
\ee
These definitions imply $\mathcal{P}\mathcal{Q}=\mathcal{Q}\mathcal{P}=0$ and $\mathcal{P}^2=\mathcal{P}$, $\mathcal{Q}^2=\mathcal{Q}(1-\mathcal{P})=\mathcal{Q}$.
Using $(\lambda-\mathcal{L})\tilde{\mathcal{G}}(\lambda)=\mathbf 1$ we can obtain the useful relation
\be{result}
\mathcal{Q}\mathcal{P}=&0=\mathcal{Q}(\lambda-\mathcal{L})\tilde{\mathcal{G}}(\lambda)\mathcal{P}\\
=&\mathcal{Q}(\lambda-\mathcal{L})(\mathcal{P}+\mathcal{Q})\tilde{\mathcal{G}}(\lambda)\mathcal{P}\\
=&\lambda\mathcal{Q}\mathcal{P}\tilde{\mathcal{G}}(\lambda)\mathcal{P}+\lambda\mathcal{Q}^2 \tilde{\mathcal{G}}(\lambda)\mathcal{P}\\
&-\mathcal{Q}\mathcal{L}\mathcal{P}\tilde{\mathcal{G}}(\lambda)\mathcal{P}-\mathcal{Q}\mathcal{L}\mathcal{Q}\tilde{\mathcal{G}}(\lambda)\mathcal{P}\\
\implies \mathcal{Q}\tilde{\mathcal{G}}(\lambda)\mathcal{P} =& \mathcal{Q}\frac{1}{\lambda-\mathcal{Q}\mathcal{L}\mathcal{Q}} \mathcal{Q}\mathcal{L}_V\mathcal{P}\tilde{\mathcal{G}}(\lambda)\mathcal{P} 
\ee

\subsection{Nakajima-Zwanzig equation for the reduced density matrix}

From Eq.~\eqref{rho_lambda}
 \be{}
(\lambda - \mathcal{L})\tilde\varrho_{\rm tot}(\lambda)&=\hat\varrho_{\rm tot}(0)\\
\lambda\tilde\varrho_{\rm tot}(\lambda)-\hat\varrho_{\rm tot}(0) &= \mathcal{L}\tilde\varrho_{\rm tot}(\lambda)= \mathcal{L} \tilde{\mathcal{G}}(\lambda) \hat\varrho_{\rm tot}(0)\;.
\ee
For a factorized initial condition $\hat\varrho_{\rm tot}(0)=\hat\varrho(0)\otimes\hat\varrho_B$,  the definitions~\eqref{projectors} entail $\mathcal{P} \hat\varrho_{\rm tot}(0)=\hat\varrho_{\rm tot}(0)$ so that $\mathcal{Q} \hat\varrho_{\rm tot}(0)=(1-\mathcal{P}) \hat\varrho_{\rm tot}(0)=0$.
 Thus, in Laplace space, the system RDM $\hat\varrho=\Tr_B\{\hat\varrho_{\rm tot}\}$ obeys the equation
\be{}
\lambda\tilde\varrho&(\lambda)-\hat\varrho(0)=\Tr_B\{\mathcal{L} \tilde{\mathcal{G}}(\lambda) \hat\varrho_{\rm tot}(0)\}\\
=&\Tr_B\{\mathcal{L}(\mathcal{P}+\mathcal{Q}) \tilde{\mathcal{G}}(\lambda) (\mathcal{P}+\mathcal{Q})
\hat\varrho_{\rm tot}(0)\}\\
=&\Tr_B\Bigg\{\mathcal{L}\left(1+\mathcal{Q}\frac{1}{\lambda-\mathcal{Q}\mathcal{L}\mathcal{Q}} \mathcal{Q}\mathcal{L}_V\right)\mathcal{P}\tilde{\mathcal{G}}(\lambda)\mathcal{P} \hat\varrho_{\rm tot}(0)\Bigg\}\\
=&\Tr_B\Big\{\mathcal{P}\mathcal{L}_0\mathcal{P}\tilde\varrho_{\rm tot}(\lambda)\Big\}\\
&+\Tr_B\Bigg\{\mathcal{P}\mathcal{L}_V
\mathcal{Q}\frac{1}{\lambda-\mathcal{Q}\mathcal{L}_0
-\mathcal{Q}\mathcal{L}_V\mathcal{Q}}(1-\mathcal{P})\mathcal{L}_V\mathcal{P} \tilde\varrho_{\rm tot}(\lambda)\Bigg\}\\
=&\mathcal{L}_S\tilde\varrho(\lambda)+\Tr_B\Bigg\{\mathcal{P}\mathcal{L}_V\mathcal{Q}\frac{1}{\lambda-\mathcal{Q}\mathcal{L}_0-\mathcal{Q}\mathcal{L}_V\mathcal{Q}}\mathcal{L}_V \tilde\varrho(\lambda)\otimes\hat\varrho_B\Bigg\}\\
=&\mathcal{L}_S\tilde\varrho(\lambda)+\tilde{\mathcal{K}}(\lambda)\tilde\varrho(\lambda)\\
\ee
where we used Eq.~\eqref{result} for $\mathcal{Q} \tilde{\mathcal{G}}(\lambda) \mathcal{P}$,   $\mathcal{P}\mathcal{L}_V\mathcal{P}=0$,  and $\Tr_B\{\mathcal{Q}\bullet\}=0$, along with $\mathcal{L}_B\mathcal{P}\bullet=0$ if $[\hat H_B,\hat\varrho_B]=0$. Also, since $\mathcal{P}\mathcal{L}_S=\mathcal{L}_S\mathcal{P}$, $\mathcal{Q}\mathcal{P}=0$, and $\mathcal{Q}^2=\mathcal{Q}$, we have $\mathcal{Q}\mathcal{L}\mathcal{Q}=\mathcal{Q}(\mathcal{L}_0+\mathcal{Q}\mathcal{L}_V\mathcal{Q})$. 
In the time domain
\be{}
\dot{\hat{\varrho}}(t)&=\mathcal{L}_S\hat\varrho(t) + \int_0^t dt'\mathcal{K}(t-t')\hat\varrho(t')\\
\mathcal{K}(t)\bullet
&=\Tr_B\Big\{\mathcal{P}\mathcal{L}_V\mathcal{Q} e^{(\mathcal{Q}\mathcal{L}_0+\mathcal{Q}\mathcal{L}_V\mathcal{Q})t}\mathcal{L}_V\bullet \otimes\hat\varrho_B\Big\}\\
&=\Tr_B\Big\{\mathcal{L}_V e^{(\mathcal{L}_0+\mathcal{Q}\mathcal{L}_V\mathcal{Q})t}\mathcal{L}_V\bullet \otimes\hat\varrho_B\Big\}\;,
\ee
where we used $\Tr_B\{\mathcal{P}\mathcal{L}_V(1-\mathcal{P} )\bullet\}=\Tr_B\{\mathcal{P}\mathcal{L}_V\bullet\}=\Tr_B\{\mathcal{L}_V\bullet\}$.

\subsection{Current}

The expectation value of an operator $\hat{O}$ reads
\be{}
\langle \tilde{O}(\lambda)\rangle &=\Tr\{\hat{O}\tilde\varrho_{\rm tot}(\lambda)\}=
\Tr\{\hat{O} \tilde{\mathcal{G}}(\lambda)\hat\varrho_{\rm tot}(0)\}\\
&=\Tr\{\hat{O}(\mathcal{P}+\mathcal{Q}) \tilde{\mathcal{G}}(\lambda) \mathcal{P}\hat\varrho_{\rm tot}(0)\}\;.
\ee
If $\hat{O}$ does not conserve the particle number in the reservoirs, as for the current $\hat{I}$, then $\Tr_B\{\hat{O}\mathcal{P}\bullet\}=0$. Using again Eq.~\eqref{result}, similarly as for the Nakajima-Zwanzig equation for $\hat\varrho$, we have
\be{}
\langle \tilde{I}(\lambda)\rangle 
&=\Tr_S\Tr_B\{\hat{I}\mathcal{Q} \tilde{\mathcal{G}}(\lambda) \mathcal{P}\hat\varrho_{\rm tot}(0)\}\\
&=\Tr_S\big\{ \tilde{\mathcal{K}}_{\rm I}(\lambda)\tilde\varrho(\lambda)\big\}\;.
\ee
In the time domain,
\be{}
\langle \hat{I}(t)\rangle &=\Tr_S\Bigg\{ \int_0^t dt'\mathcal{K}_{\rm I}(t-t')\hat\varrho(t')\Bigg\}\;,\\
\mathcal{K}_{\rm I}(t)\bullet &=\Tr_B\Big\{\hat{I} e^{(\mathcal{L}_0+\mathcal{Q}\mathcal{L}\mathcal{Q})t}\mathcal{L}_V\bullet \otimes\hat\varrho_B\Big\}\;.
\ee

\section{Series expansion of the propagator}
\label{expansion}

Let us define $\mathcal{G}_\mathcal{Q}(t)=\exp[(\mathcal{L}_0+\mathcal{Q}\mathcal{L}_V\mathcal{Q})t]$ and \hbox{$\mathcal{G}_0(t):=\exp(\mathcal{L}_0t)$}.
We have
\[
\partial_t \mathcal{G}_\mathcal{Q}(t)=(\mathcal{L}_0+\mathcal{Q}\mathcal{L}_V\mathcal{Q})\mathcal{G}_\mathcal{Q}(t)\;.
\]
Now consider
\be{}
\partial_t [\mathcal{G}_0(-t)\mathcal{G}_\mathcal{Q}(t)] =&-\mathcal{L}_0\mathcal{G}_0(-t)\mathcal{G}_\mathcal{Q}(t)\\
&+\mathcal{G}_0(-t)(\mathcal{L}_0+\mathcal{Q}\mathcal{L}_V\mathcal{Q})\mathcal{G}_\mathcal{Q}(t)\\
=&\mathcal{G}_0(-t)\mathcal{Q}\mathcal{L}_V\mathcal{Q} \mathcal{G}_\mathcal{Q}(t)\;.
\ee
Integrating and multiplying from the left by $\mathcal{G}_0(t)$ one obtains
\be{}
\mathcal{G}_\mathcal{Q}(t)=\mathcal{G}_0(t) + \int_0^t dt' \mathcal{G}_0(t-t')\mathcal{Q}\mathcal{L}_V\mathcal{Q} \mathcal{G}_\mathcal{Q}(t')
\ee
which qualifies $\mathcal{Q}\mathcal{L}_V\mathcal{Q}$ as the self-energy in the above Dyson equation. In Laplace space
\be{}
\tilde{\mathcal{G}}_\mathcal{Q}(\lambda) =& \tilde{\mathcal{G}}_0(\lambda) +  \tilde{\mathcal{G}}_0(\lambda)\mathcal{Q}\mathcal{L}_V\mathcal{Q} \tilde{\mathcal{G}}_\mathcal{Q}(\lambda)\\
=& \tilde{\mathcal{G}}_0(\lambda) +  \tilde{\mathcal{G}}_0(\lambda)\mathcal{Q}\mathcal{L}_V\mathcal{Q} \tilde{\mathcal{G}}_0(\lambda)\\
&+ \tilde{\mathcal{G}}_0(\lambda)\mathcal{Q}\mathcal{L}_V\mathcal{Q} \tilde{\mathcal{G}}_0(\lambda)\mathcal{Q}\mathcal{L}_V\mathcal{Q} \tilde{\mathcal{G}}_0(\lambda) + \dots\\
=& \tilde{\mathcal{G}}_0(\lambda)\sum_{n=0}^\infty\left[\mathcal{Q}\mathcal{L}_V\mathcal{Q} \tilde{\mathcal{G}}_0(\lambda)\right]^n\;,
\ee
giving the perturbative expansion in the interaction $V$.

\newpage
\begin{widetext}

\section{4th-order current kernel in the energy eigenbasis}
\label{4th_order}.

The $D$- and $X$-diagrams of the 4th-order current kernel, Eq.~\eqref{K4laplace}, applied to $\ket{n}\bra{n}$, the projector in the system state $\ket{n}$, result in

\bes\label{}
&\left[{\boldsymbol{D}}_4^++{\boldsymbol{X}}_4^+\right] \ket{n}\bra{n}\\
=&\sum_{\nu_3\nu_2\nu_1}\begin{bmatrix}
 1 \; \\
 \nu_3\nu_2  
\end{bmatrix}
\nu_1 \mathcal{D}^{-q+} \tilde{\mathcal{G}}_{0,3}(0^+) \mathcal{D}^{-q'\nu_3}  \left[ \tilde{\mathcal{G}}_{0,2}^{q',\nu_2}  (0^+)\mathcal{D}^{q'\nu_2}  \tilde{\mathcal{G}}_{0,1}^{q,\nu_1}  (0^+)\mathcal{D}^{q\nu_1}\mp\tilde{\mathcal{G}}_{0,2}^{q,\nu_2}  (0^+)\mathcal{D}^{q\nu_2}   \tilde{\mathcal{G}}_{0,1}^{q',\nu_1} (0^+) \mathcal{D}^{q'\nu_1}\right]\ket{n}\bra{n}\\
=&\quad\sum_{k,k_1,k_2,k_3} D^{-q}_{k_3k_2}\tilde G_{k_2 n}^{q}(0^+)D^{-q'}_{k_2k_1}\left[\tilde G_{k_1 n}^{q',+}(0^+-\ii p\omega)D^{q'}_{k_1k}\tilde G_{k n}^{q,+}(0^+)D^{q}_{k n}
\mp \tilde G_{k_1 n}^{q,+}(0^+-\ii p'\omega') D^{q}_{k_1k}\tilde G_{k n}^{q',+}(0^+)D^{q'}_{k n} \right]\ket{k_3}\bra{n}\\
&\pm\sum_{k,k_1,k_2,k_3} D^{-q}_{k_3k_1}\tilde G_{k_1 k_2}^{q}(0^+)D^{-q'}_{n k_2}\left[  \tilde G_{k_1 n}^{q',+}(0^+-\ii p\omega)D^{q'}_{k_1k}\tilde G_{k n}^{q,+}(0^+)D^{q}_{k n}
\mp \tilde G_{k_1 n}^{q,+}(0^+-\ii p'\omega')D^{q}_{k_1k}\tilde G_{k n}^{q',+}(0^+)D^{q'}_{k n} \right]\ket{k_3}\bra{k_2}\\
&\pm\sum_{k,k_1,k_2,k_3} D^{-q}_{k_3k_2}\tilde G_{k_2 k_1}^{q}(0^+)D^{-q'}_{k_2 k}\left[  \tilde G_{k k_1}^{q',-}(0^+-\ii p\omega)D^{q'}_{n k}\tilde G_{k n}^{q,+}(0^+)D^{q}_{k n}
\mp \tilde G_{k k_1}^{q,-}(0^+-\ii p'\omega')D^{q}_{n k}\tilde G_{k n}^{q',+}(0^+)D^{q'}_{k n} \right]\ket{k_3}\bra{k_1}\\
&+\sum_{k,k_1,k_2,k_3} D^{-q}_{k_3k}\tilde G_{k k_2}^{q}(0^+)D^{-q'}_{k_1 k_2}\left[\tilde G_{k k_1}^{q',-}(0^+-\ii p\omega)D^{q'}_{n k}\tilde G_{k n}^{q,+}(0^+)D^{q}_{k n}
\mp \tilde G_{k k_1}^{q,-}(0^+-\ii p'\omega')D^{q}_{n k}\tilde G_{k n}^{q',+}(0^+)D^{q'}_{k n} \right]\ket{k_3}\bra{k_2}\\
&-\sum_{k,k_1,k_2,k_3} D^{-q}_{k_3k_2}\tilde G_{k_2 k}^{q}(0^+)D^{-q'}_{k_2 k_1}\left[\tilde G_{k_1 k}^{q',+}(0^+-\ii p\omega)D^{q'}_{k_1 n}\tilde G_{n k}^{q,-}(0^+)D^{q}_{n k}
\mp \tilde G_{k_1 k}^{q,+}(0^+-\ii p'\omega')D^{q}_{k_1 n}\tilde G_{n k}^{q',-}(0^+)D^{q'}_{n k} \right]\ket{k_3}\bra{k}\\
&\mp\sum_{k,k_1,k_2,k_3} D^{-q}_{k_3k_1}\tilde G_{k_1 k_2}^{q}(0^+)D^{-q'}_{k k_2}\left[  \tilde G_{k_1 k}^{q',+}(0^+-\ii p\omega)D^{q'}_{k_1 n}\tilde G_{n k}^{q,-}(0^+)D^{q}_{n k}
\mp \tilde G_{k_1 k}^{q,+}(0^+-\ii p'\omega')D^{q}_{k_1 n}\tilde G_{n k}^{q',-}(0^+)D^{q'}_{n k} \right]\ket{k_3}\bra{k_2}\\
&\mp\sum_{k,k_1,k_2,k_3} D^{-q}_{k_3k_2}\tilde G_{k_2 k_1}^{q}(0^+)D^{-q'}_{k_2 n}\left[  \tilde G_{n k_1}^{q',-}(0^+-\ii p\omega)D^{q'}_{k k_1}\tilde G_{n k}^{q,-}(0^+)D^{q}_{n k}
\mp \tilde G_{n k_1}^{q,-}(0^+-\ii p'\omega')D^{q}_{k k_1}\tilde G_{n k}^{q',-}(0^+)D^{q'}_{n k} \right]\ket{k_3}\bra{k_1}\\
&-\sum_{k,k_1,k_2,k_3} D^{-q}_{k_3 n}\tilde G_{n k_2}^{q}(0^+)D^{-q'}_{k_1 k_2}\left[\tilde G_{n k_1}^{q',-}(0^+-\ii p\omega)D^{q'}_{k k_1}\tilde G_{n k}^{q,-}(0^+)D^{q}_{n k}
\mp \tilde G_{n k_1}^{q,-}(0^+-\ii p'\omega')D^{q}_{k k_1}\tilde G_{n k}^{q',-}(0^+)D^{q'}_{n k} \right]\ket{k_3}\bra{k_2}\;,
\ee
where, according to the definitions in Eqs.~\eqref{G_qnu} and~\eqref{K4_formal},
\be{}
\tilde{\mathcal{G}}_{0,i}^{q,\nu}(\lambda):=\tilde {\mathcal{G}}_{0,i}(\lambda) n_{l}^{p\nu}(\omega_{l j})\;,\quad
\tilde G^{q,\nu}_{nm}(\lambda-\ii p' \omega')=\frac{n_{l}^{p\nu}(\omega_{l j})}{\lambda +\ii\omega_{nm} -\ii(p'\omega_{l 'j'}+ p \omega_{l j})}\equiv \tilde G^{q}_{nm}(\lambda-\ii p' \omega')n_{l}^{p\nu}(\omega_{l j})\;.
\ee

In the continuum limit, the sums over $j$ and $j'$ turn into integrals over $\omega$ and $\omega'$, respectively. At low $T$, the Bose-Einstein functions are different from zero only for small arguments. Thus, the processes with $\tilde G_{k=n,n}^{q}(\lambda-\ii p'\omega')$ give the leading contributions by approximating the central fractions of Eq.~\eqref{K4} as 
$$[\omega_1-\omega_2 +p'\ii0^+]^{-1}=\lim_{\eta\to 0^+}[\omega_1-\omega_2 +p'\ii\eta]^{-1}\simeq -\ii p'\pi\delta(\omega_1-\omega_2)$$ which easily solves the integral over $\omega'$ (neglecting the principal part). The contributions to $\tilde{\mathcal{K}}^{(4)}_{{\rm I}r,mmnn}(\lambda)$ with the constraint that the system transition frequency in the central free propagators vanish, i.e. the system is in a \emph{diagonal} state between the second and third vertexes, are thus given by
\be{DXmmnn}
\bra{m}&\left\{\Big[{\boldsymbol{D}}_4^++{\boldsymbol{X}}_4^+\Big]_{\rm diag} \ket{n}\bra{n}\right\}\ket{m}\\
=&\sum_{k,k_2} D^{-q}_{n k_2}\tilde G_{k_2 n}^{q}(0^+)D^{-q'}_{k_2n}\left[  \tilde G_{n n}^{q',+}(0^+-\ii p\omega)D^{q'}_{n k}\tilde G_{k n}^{q,+}(0^+)D^{q}_{k n}
\mp \tilde G_{n n}^{q,+}(0^+-\ii p'\omega') D^{q}_{nk}\tilde G_{k n}^{q',+}(0^+)D^{q'}_{k n} \right]\delta_{m,n}\\
\pm&\sum_{k} D^{-q}_{mn}\tilde G_{nm}^{q}(0^+)D^{-q'}_{nm}\left[ \tilde G_{n n}^{q',+}(0^+-\ii p\omega)D^{q'}_{nk}\tilde G_{k n}^{q,+}(0^+)D^{q}_{k n}
\mp \tilde G_{n n}^{q,+}(0^+-\ii p'\omega') D^{q}_{nk}\tilde G_{k n}^{q',+}(0^+)D^{q'}_{k n} \right]\\
\pm &\sum_{k} D^{-q}_{mk}\tilde G_{km}^{q}(0^+)D^{-q'}_{km}\left[ \tilde G_{mm}^{q',-}(0^+-\ii p\omega)D^{q'}_{n m}\tilde G_{m n}^{q,+}(0^+)D^{q}_{m n}
\mp \tilde G_{mm}^{q,-}(0^+-\ii p'\omega')D^{q}_{n m}\tilde G_{m n}^{q',+}(0^+)D^{q'}_{m n} \right]\\
+&\sum_{k} D^{-q}_{mk}\tilde G_{km}^{q}(0^+)D^{-q'}_{km}\left[ \tilde G_{mm}^{q',-}(0^+-\ii p\omega)D^{q'}_{n m}\tilde G_{m n}^{q,+}(0^+)D^{q}_{m n}
\mp \tilde G_{mm}^{q,-}(0^+-\ii p'\omega')D^{q}_{n m}\tilde G_{m n}^{q',+}(0^+)D^{q'}_{m n} \right]\\
-&\sum_{k} D^{-q}_{mk}\tilde G_{km}^{q}(0^+)D^{-q'}_{km}\left[  \tilde G_{mm}^{q',+}(0^+-\ii p\omega)D^{q'}_{m n}\tilde G_{n m}^{q,-}(0^+)D^{q}_{n m}
\mp \tilde G_{mm}^{q,+}(0^+-\ii p'\omega')D^{q}_{m n}\tilde G_{n m}^{q',-}(0^+)D^{q'}_{n m} \right]\\
\mp&\sum_{k} D^{-q}_{mk}\tilde G_{km}^{q}(0^+)D^{-q'}_{km}\left[ \tilde G_{mm}^{q',+}(0^+-\ii p\omega)D^{q'}_{m n}\tilde G_{n m}^{q,-}(0^+)D^{q}_{n m}
\mp \tilde G_{mm}^{q,+}(0^+-\ii p'\omega')D^{q}_{m n}\tilde G_{n m}^{q',-}(0^+)D^{q'}_{n m} \right]\\
\mp&\sum_{kk_2} D^{-q}_{nk_2}\tilde G_{k_2 n}^{q}(0^+)D^{-q'}_{k_2 n}\left[\tilde G_{n n}^{q',-}(0^+-\ii p\omega)D^{q'}_{k n}\tilde G_{n k}^{q,-}(0^+)D^{q}_{n k}
\mp \tilde G_{n n}^{q,-}(0^+-\ii p'\omega')D^{q}_{k n}\tilde G_{n k}^{q',-}(0^+)D^{q'}_{n k}\right]\delta_{m,n}\\
-&\sum_{k} D^{-q}_{mn}\tilde G_{nm}^{q}(0^+)D^{-q'}_{nm}\left[ \tilde G_{n n}^{q',-}(0^+-\ii p\omega)D^{q'}_{k n}\tilde G_{n k}^{q,-}(0^+)D^{q}_{n k}
\mp \tilde G_{n n}^{q,-}(0^+-\ii p'\omega')D^{q}_{k n}\tilde G_{n k}^{q',-}(0^+)D^{q'}_{n k}\right]\:.
\ee

\section{Bosonic heat current to 4th order}
\label{4th_order_heat_current}

In the bosonic case, the 3rd and 4th lines in Eq.~\eqref{DXmmnn} cancel each other as well as the 5th and 6th lines. 
At the steady state, $\lambda\rightarrow 0^+$, we can set $\tilde G_{nn}^{q,\nu}(0^+ -\ii p'\omega')= n^{q\nu}_{l}(\omega)\pi\delta(\omega'+pp'\omega)$ and $\tilde G_{nn}^{q',\nu'}(0^+ -\ii p\omega)= 
 n^{q'\nu'}_{l}(\omega')\pi\delta(\omega'+pp'\omega)$. Moreover, since both the integrals are over the positive frequencies $\omega'$ and $\omega$, this results in the constraint $p'=-p$. Finally, for the heat current to bath $r$, we fix the bath index of the last transition to $r$. This yields
\be{KI4mmnn_b}
\bra{m}&\left\{\Big[{\boldsymbol{D}}_4^+ +{\boldsymbol{X}}_4^+\Big]_{\rm diag} \ket{n}\bra{n}\right\}\ket{m}\\
=\pi&\sum_{k,k_2\neq n} D^{-q}_{r,n k_2}\tilde G_{r,k_2 n}^{p}(0^+)D^{p}_{l,k_2n}\left[  n_l^{-p}(\omega)D^{-p}_{l,n k}\tilde G_{r,k n}^{p,+}(0^+)D^{p}_{r,k n}
+ n_r^{p}(\omega)D^{p}_{r,n k}\tilde G_{l,k n}^{-p,+}(0^+)D^{-p}_{l,k n} \right]\delta_{m,n}\\
+\pi&\sum_{k,k_2\neq n} D^{-p}_{r,nk_2}\tilde G_{r,k_2 n}^{p}(0^+)D^{p}_{l,k_2 n}\left[  n_l^{p}(\omega)D^{-p}_{l,kn }\tilde G_{r,nk}^{p,-}(0^+)D^{p}_{r,nk}
+    n_r^{-p}(\omega)D^{p}_{r,kn }\tilde G_{l,nk}^{-p,-}(0^+)D^{-p}_{l,nk} \right]\delta_{m,n}\\
-\pi &\sum_{k} D^{-p}_{r,mn}\tilde G_{r,nm}^{p}(0^+)D^{p}_{l,nm}\left[  n_l^{-p}(\omega)D^{-p}_{l,n k}\tilde G_{r,k n}^{p,+}(0^+)D^{p}_{r,k n}
+  n_r^{p}(\omega)D^{p}_{r,n k}\tilde G_{l,k n}^{-p,+}(0^+)D^{-p}_{l,k n} \right]_{m\neq n}\\
-\pi &\sum_{k} D^{-p}_{r,mn}\tilde G_{r,nm}^{p}(0^+)D^{p}_{l,nm}\left[  n_l^{p}(\omega)D^{-p}_{l,kn }\tilde G_{r,nk}^{p,-}(0^+)D^{p}_{r,nk}
+    n_r^{-p}(\omega)D^{p}_{r,kn }\tilde G_{l,nk}^{-p,-}(0^+)D^{-p}_{l,nk} \right]_{m\neq n}\;,
\ee
where the cases $k_2=n$ in the first two lines and $m=n$ in the third and fourth lines cancel each other and we conveniently excluded them.
Including also $\sum_p p$ and swapping $p\rightarrow -p$ in the last two lines of Eq.~\eqref{KI4mmnn_b},
\be{}
\sum_p p\bra{m}\left\{\Big[{\boldsymbol{D}}_4^+ +{\boldsymbol{X}}_4^+\Big]_{\rm diag} \ket{n}\bra{n}\right\}&\ket{m}\\
=\pi\sum_p p\Big\{ \sum_{k,k_2\neq n} D^{-p}_{r,n k_2}\tilde G_{r,k_2 n}^{p}(0^+)D^{p}_{l,k_2n}&\Big[  n_l^{-p}(\omega)D^{-p}_{l,n k}\tilde G_{r,k n}^{p,+}(0^+)D^{p}_{r,k n}
+ n_r^{p}(\omega)D^{p}_{r,n k}\tilde G_{l,k n}^{-p,+}(0^+)D^{-p}_{l,k n}\\
+& n_l^{p}(\omega)D^{-p}_{l,kn }\tilde G_{r,nk}^{p,-}(0^+)D^{p}_{r,nk}
+  n_r^{-p}(\omega)D^{p}_{r,kn }\tilde G_{l,nk}^{-p,-}(0^+)D^{-p}_{l,nk} \Big]\delta_{m,n}\\
+\sum_{k} D^{q}_{r,mn}\tilde G_{r,nm}^{-p}(0^+)D^{-p}_{l,nm}&\Big[  n_l^{p}(\omega)D^{p}_{l,n k}\tilde G_{r,k n}^{-p,+}(0^+)D^{-p}_{r,k n}
+  n_r^{-p}(\omega)D^{-p}_{r,n k}\tilde G_{l,k n}^{p,+}(0^+)D^{p}_{l,k n} \\
+& n_l^{-p}(\omega)D^{p}_{l,kn }\tilde G_{r,nk}^{-p,-}(0^+)D^{-p}_{r,nk}
+    n_r^{p}(\omega)D^{-p}_{r,kn }\tilde G_{l,nk}^{p,-}(0^+)D^{p}_{l,nk} \Big]_{m\neq n}\Big\}\;.
\ee
Use of the symmetry  properties $D^{-p}_{mn}=(D^{p}_{nm})^*$
 and $\tilde G_{mn}^{-p,-}(\lambda)=[\tilde G_{nm}^{p,+}(\lambda)]^*$ shows directly that
\be{}
\tilde{\mathcal{K}}_{{\rm I} r,nnnn}^{(4)}(\lambda)
=&\sum_{m\neq n}[\tilde{\mathcal{K}}_{{\rm I} r,mmnn}^{(4)}(\lambda)]^*\;.
\ee
Using this result, we have for the 4th-order current (assuming the coherences to vanish at the steady state)
\be{}
I^{(4)}_r=&\sum_{m,n}\tilde{\mathcal{K}}_{{\rm I} r,mmnn}^{(4)}(0)\rho_{nn}^\infty\\
=&\sum_{n,m\neq n}\tilde{\mathcal{K}}_{{\rm I} r,mmnn}^{(4)}(0)\rho_{nn}^\infty+\sum_{n}\tilde{\mathcal{K}}_{{\rm I} r,nnnn}^{(4)}(0)\rho_{nn}^\infty\\
=&2\Real\sum_{n,m\neq n}\tilde{\mathcal{K}}_{{\rm I} r,mmnn}^{(4)}(0)\rho_{nn}^\infty\;.
\ee

Thus, assuming $Q_{l,nm}=Q_{l,mn}\in \mathbb{R}$, we have for $m\neq n$
\be{}
\tilde{\mathcal{K}}_{{\rm I} r,mmnn}^{(4)}(0)
\simeq&-\pi\sum_{\nu p} p\sum_{l} \sum_{k} Q_{r,mn}Q_{l,nm}Q_{l,nk}Q_{r,kn}
\int_0^\infty d\omega\hbar\omega J_l(\omega)J_r(\omega)n_{l }^{p\nu}(\omega)n_{r}^{-p\nu}(\omega)\\
&\times \frac{1}{\ii0^+ +\omega_{mn} - p\omega}\Bigg[
\frac{1}{\ii0^+ -\nu\omega_{kn} - p  \omega } +
\frac{1}{\ii0^+ -\nu\omega_{kn} +  p\omega }\Bigg]\\
=&\pi\sum_{\nu}\sum_{l}\sum_{k} Q_{r,mn}Q_{l,nm}Q_{l,nk}Q_{r,kn}
\int_0^\infty d\omega\hbar\omega J_l(\omega)J_r(\omega)n_{l }^{-\nu}(\omega)n_{r}^{\nu}(\omega)\\
&\times \Bigg\{
\frac{1}{\ii0^+ +\omega_{mn} +  \omega}
\Bigg[\frac{1}{\ii0^+ -\nu\omega_{kn} +  \omega }+
\frac{1}{\ii0^+ -\nu\omega_{kn} -  \omega }\Bigg]\\
&\quad-
\frac{1}{\ii0^+ +\omega_{mn} - \omega}\Bigg[
\frac{1}{\ii0^+ +\nu\omega_{kn} - \omega }+
\frac{1}{\ii0^+ +\nu\omega_{kn} + \omega }
\Bigg]\Bigg\}
\;.
\ee
After some algebra, using the properties of the Bose-Einstein function, which gives $\sum_l\sum_{\nu}\nu n_{l}^{-\nu}n_{r}^{\nu}=\sum_l(n_{r}^+-n_{l}^+)=n_{r}-n_{\bar{r}}$, where to $r=R$ ($L$) corresponds $\bar{r}=L$ ($R$), we obtain for the real part ($\lambda\rightarrow 0^+$)
\be{}
\Real\tilde{\mathcal{K}}_{{\rm I} r,mmnn}^{(4)}(0)
\simeq &
\pi\hbar \sum_{k} Q_{r,mn}Q_{\bar{r},nm}Q_{\bar{r},nk}Q_{r,kn}
\int_0^\infty d\omega\omega J_r(\omega)J_{\bar{r}}(\omega)[n_{\bar{r}}(\omega)-n_{r}(\omega)] \\
&\times \sum_{a,b=\pm1}\frac{\omega_{mn}+a\omega}{(\omega_{mn}+a\omega)^2+\lambda^2}\frac{\omega_{kn}+b\omega}{(\omega_{kn}+b\omega)^2+\lambda^2}\qquad (m\neq n)\\
(k_B T\ll \hbar\omega_{ij})\quad\simeq &
4\pi\hbar \sum_{k\neq n} \frac{1}{\omega_{mn}\omega_{kn}}Q_{r,mn}Q_{\bar{r},nm}Q_{\bar{r},nk}Q_{r,kn}
\int_0^\infty d\omega\omega J_r(\omega)J_{\bar{r}}(\omega)[n_{\bar{r}}(\omega)-n_{r}(\omega)] 
\;.
\ee
Note that if $k=n$ or $m=n$, then the resulting expression vanishes due to the sum over $b=\pm1$, or $a=\pm1$, respectively, in the first equality.

\section{Steady-state heat current for fermionic baths}\label{fermions}

Charge transport in fermionic systems has been thoroughly investigated with diagrammatic techniques (see~\cite{Donarini2024} and references therein). Much less attention, however, has been devoted to bosonic heat transport in multi-level systems beyond the leading order. For this reason, in applying the general machinery developed so far, our main focus in the present work will be the bosonic heat transport.  
Nevertheless, we provide here, as an illustration, some basic results of the theory specialized to fermionic transport.\\
\indent From the definition of the system coupling operators $\hat D$ in the fermionic case, Eq.~\eqref{D_fermions}, the second-order rates in Eq.~\eqref{rates2nd} specialize to
\be{rates2nd_fermions}
\Gamma_{nm}=\tilde{\mathcal{K}}_{n n mm}^{(2)}(0)&=\frac{2\pi}{\hbar^2}\sum_l{\rm D}_l|t_l|^2|d_{nm}|^2 f_l(\omega_{nm})\;,
\ee
where ${\rm D}_l$ is the density of state of bath $l$, which we assume energy-, spin-, and state-independent for simplicity and $d_{nm}=\sum_s\langle n|\hat d_s|m\rangle$, with $s,n,m$ denoting the states of the system, see also~\eqref{integral_fermions}.\\
\indent Let us consider now a quantum dot described by the Hamiltonian $\hat H_{\rm dot}$ coupled to fermionic leads with $\hat H_B+\hat H_V$ given by Eq.~\eqref{Hfermions}.
For definiteness, along the lines of one example discussed in~\cite{Tesser2022}, we choose
$$\hat H_{\rm dot}=\hbar\Delta\sum_\sigma \hat n_\sigma+U\hat n_{\uparrow}\hat n_{\downarrow}\;,$$ 
with $U$ much larger than the other setup parameters, so that the energetically unfavorable doubly-occupied configuration is essentially forbidden. For this system, the transition rates between the unoccupied state $\ket{0}$ and the singly-occupied dot states $\ket{1_\sigma}$ read
$\Gamma^l_{\sigma 0}=\gamma^l f_l(\Delta)$ and $\Gamma^l_{0\sigma}=\gamma^l [1-f_l(\Delta)]$, with $\gamma^l=2\pi{\rm D}_l|t_l|^2/\hbar^2$. The states $\ket{1_\uparrow}$ and $\ket{1_\downarrow}$ cannot be connected by a rate to second order, namely $\sum_s\langle 1_\uparrow|\hat d_s|1_\downarrow\rangle=0$, as the resulting spin-flip would require higher-order processes.
The full secular treatment applied to Eq.~\eqref{I2nd_general} yields for the current to bath $r$, to leading order,
\be{I2nd_fermions}
I^{(2)}_r=2\Real\sum_{n,m(\neq n)=0,\uparrow,\downarrow}\tilde{\mathcal{K}}_{{\rm I}r,n n mm}^{(2)}(0)\varrho^\infty_{mm}\;,
\ee
where, from Eq.~\eqref{K2nd_general},  
\be{}
2\Real\tilde{\mathcal{K}}_{{\rm I}r,\sigma\sigma 00}^{(2)}(0)&=-\zeta\gamma^r f_r(\Delta)\;,\\
2\Real\tilde{\mathcal{K}}_{{\rm I}r, 00\sigma\sigma}^{(2)}(0)&=\zeta\gamma^r [1- f_r(\Delta)]\;,
\ee
with $\zeta=1$ ($\zeta=\hbar\Delta$) for particle (energy) current. Solving the full secular master equation~\eqref{BR_full_secular} with the symmetry $\rho^\infty_{\uparrow\uparrow}=\rho^\infty_{\downarrow\downarrow}$ and the conservation of the total probability, we obtain for the steady-state populations of the dot $\rho^\infty_{00}=\Gamma_{0\sigma}/(\Gamma_{0\sigma}+2\Gamma_{\sigma 0})$ and $\rho^\infty_{\sigma\sigma}=\Gamma_{\sigma 0}/(\Gamma_{0\sigma}+2\Gamma_{\sigma 0})$. Using these results, the heat current $I^{{\rm h}(2)}_r=I^{{\rm E}(2)}_r-\mu_r I^{{\rm p}(2)}_r$ to bath $r$ is found to be
\be{I2fermions}
I^{{\rm h}(2)}_r
&=2(\hbar\Delta-\mu_r)\frac{\gamma^L\gamma^R[f_{\bar r}(\Delta)-f_r(\Delta)]}{\gamma^L[1+f_L(\Delta)]+\gamma^R[1+f_R(\Delta)]}\;,
\ee
where $\bar r=L$ ($R$) for $r=R$ ($L$). 
More general results for heat transport in interacting quantum dots weakly coupled fermionic baths are provided e.g. in~\cite{Muralidharan2012, Schulenborg2018,Tesser2022}. Let us now set $\mu_L=\mu_R=0$ with $T_r=T$ and $T_{\bar r}=T+\Delta T$. The thermal conductance, Eq.~\eqref{kappa}, corresponding to the current formula~\eqref{I2fermions} is obtained using $\partial_{\Delta T} f_{\bar r}(\Delta)|_{\Delta T=0}=\hbar\Delta[4k_B T^2 \cosh^2(\hbar\Delta/2k_B T)]^{-1}$ and reads (at zero thermal bias $f_L=f_R=f$)
\be{kappa2fermions}
\kappa^{(2)}=\frac{\hbar\Delta}{2k_B T^2}  \frac{\gamma^L\gamma^R}{\gamma^L+\gamma^R}\frac{1}{[1+f(\Delta)]\cosh^2(\hbar\Delta/2k_B T)}\;.
\ee
As for the bosonic case, see  Eq.~\eqref{kappa2TLS}, this expression also shows exponential suppression at low temperature, $k_BT\ll\hbar\Delta$.

\section{Evaluation of $W_{lnm}$ and $\langle\hat{B}_l(0)\hat{B}_l(0)\rangle$}
\label{PVintegrals}

Specializing the general expression~\eqref{K2nd_general} for the 2nd-order kernel to the bosonic case, the rates $W_{lnm}$ are defined as 
$$
W_{l nm}=\lim_{\lambda \to 0^+}\sum_{pj}(\hbar\lambda_{lj})^2 \frac{n_{l}^{p+}(\omega_j)}{\lambda+\ii\omega_{nm} -\ii p\omega_j}\;.
$$
Using the definition of bath spectral density function $J_l(\omega)=\sum_j\lambda_{l j}^2 \delta(\omega-\omega_j)$, the sum over the reservoir  states $j$ becomes the integral
\be{}
W_{l nm}=&\lim_{\lambda \to 0^+}\sum_p \hbar^2\int_0^\infty d\omega \frac{n_{l}^{p+}(\omega)J_l(\omega)}{\lambda+\ii\omega_{nm} -\ii p\omega}\\
=&\sum_p\hbar^2\int_0^\infty d\omega n_{l}^{p+}(\omega)J_l(\omega) \lim_{\lambda \to 0^+}\int_{0}^{\infty}dt e^{-\lambda t} e^{\ii(p\omega-\omega_{nm})t}\\
=&\hbar^2\int_{0}^{\infty}dt\int_0^\infty d\omega J_l(\omega) \left\{n_l(\omega)e^{\ii\omega t}+[n_l(\omega)+1]e^{-\ii\omega t}\right\}e^{-\ii\omega_{nm} t}\\
=&\hbar^2\int_{0}^{\infty}dt\int_0^\infty d\omega J_l(\omega)\left[\coth\left(\frac{\beta_l\hbar\omega}{2}\right)\cos(\omega t)-{\rm i}\sin(\omega t)\right]e^{-\ii\omega_{nm} t}\\
=&\int_{0}^{\infty}dt\langle \hat{B}_l(t)\hat{B}_l(0)\rangle 
e^{-\ii\omega_{nm} t}
\;.
\ee
where,
$$\langle \hat{B}_l(t)\hat{B}_l(0)\rangle=\hbar^2\int_0^\infty d\omega J_l(\omega)\left[\coth\left(\frac{\beta_l\hbar\omega}{2}\right)\cos(\omega t)-{\rm i}\sin(\omega t)\right]$$ is the correlation function of the baths operators $\hat{B}_l:=\sum_{ j}\hbar\lambda_{l j}(b_{l j}+ b_{l j}^\dag)$, the time evolution being with respect to the free bath Hamiltonian.

The quantity $\bar{W}_{nm}$ which enters the expression for the 2nd-order current, Eq.~\eqref{I2}, is analogously  
obtained as 
\be{}
\bar{W}_{lnm}=&\lim_{\lambda \to 0^+}\sum_p\hbar^2\int_0^\infty d\omega \frac{\hbar\omega p\; n_{l}^{p+}(\omega)J_l(\omega)}{\lambda+\ii\omega_{nm} -\ii p\omega}\\
=&\sum_p\hbar^3\int_0^\infty d\omega\; J_l(\omega) \lim_{\lambda \to 0^+}\int_{0}^{\infty}dt e^{-\lambda t} e^{\ii (p\omega -\omega_{nm}) t}p\omega\; n_{l}^{p+}(\omega)\\
=&\lim_{\lambda\to 0^+}-\ii\hbar^3\int_0^\infty d\omega J_l(\omega)\int_{0}^{\infty}dt e^{-\lambda t}e^{-\ii\omega_{nm} t}\partial_t\left[\coth\left(\frac{\beta_l\hbar\omega}{2}\right)\cos(\omega t)-{\rm i}\sin(\omega t)\right]\\
=&\hbar\omega_{nm}W_{lnm}+\ii\hbar\langle \hat{B}_l(0)\hat{B}_l(0)\rangle
\;.
\ee
In practice, to calculate the rates $W_{nm}$ (we skip here the bath index $l$) we start from the expression in Laplace space 
\be{Wnm_general}
W_{nm}/\hbar^2=&\lim_{\lambda \to 0^+} \int_0^\infty d\omega \frac{\sum_p n^{p+}(\omega)J(\omega)}{\lambda+\ii\omega_{nm} -\ii p\omega}\\
\equiv &\lim_{\lambda \to 0^+} \int_0^\infty d\omega J(\omega)g_{nm}(\omega,\lambda)\;,
\ee
where
\be{}
g_{nm}(\omega,\lambda)=n(\omega)\frac{\lambda+\ii(\omega-\omega_{nm})}{\lambda^2+(\omega-\omega_{nm})^2} +  [n(\omega)+1]\frac{\lambda-\ii(\omega+\omega_{nm})}{\lambda^2+(\omega+\omega_{nm})^2}\equiv g_{1,nm}(\omega,\lambda)+g_{2,nm}(\omega,\lambda)\;.
\ee
Given that $n(-\omega)=-[n(\omega)+1]$ the following symmetries hold $g_{nm}(-\omega,\lambda)=-g_{nm}(\omega,\lambda)$.\\
\indent In the Ohmic case, $J(\omega)=\alpha\omega \Theta_c(\omega)$, with an even cutoff function, the integrand is even and we can extend the integration domain to $\mathbb{R}$. Using $\lim_{\lambda \to 0^+} \lambda/(\lambda^2 + x^2)=\pi\delta(x)$ we obtain
\be{Wab}
W_{nm}/\hbar^2= &\lim_{\lambda \to 0^+} \frac{1}{2}\int_{-\infty}^\infty d\omega J(\omega)g_{nm}(\omega)\\
=& \pi J(\omega_{nm})n(\omega_{nm})+\ii\lim_{\lambda \to 0^+} \int_{-\infty}^\infty d\omega J(\omega)g_{1,nm}''(\omega)\\
\equiv& W^{(a)}_{nm}/\hbar^2+\ii W^{(b)}_{nm}/\hbar^2\;.
\ee

\subsection{Analytical evaluation for Ohmic-Drude spectral density}
\label{WnmDrude}

Assuming a cutoff function such that its extension to the complex plane   $\Theta_c(z)$ is holomorphic except for the poles, $W_{nm}^b$ is given by
\be{}
W^{(b)}_{nm}/\hbar^2 =& {\rm P.V.}\int_{-\infty}^\infty d\omega J(\omega)
\frac{  n(\omega)}{\omega - \omega_{nm}}\\
=&2\pi{\rm i}\sum_j{\rm Res}_j\left\{\frac{J(z)n(z)}{z-\omega_{nm}}\right\}_{\includegraphics[width=0.05\textwidth,angle=0]{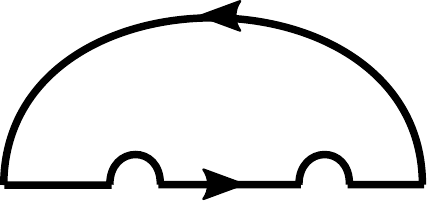}}-
\begin{gathered}
\resizebox{1.5cm}{!}{
\begin{tikzpicture}[]
\draw[black,thick,<-] (0,0) arc (180:0:0.5cm  and 0.5cm);
 \end{tikzpicture}
 }
\end{gathered} 
- 
\begin{gathered}
\resizebox{0.7cm}{!}{
\begin{tikzpicture}[]
\draw[black,thick,->] (0,0) arc (180:0:0.5cm  and 0.5cm); 
\draw[](0.5,0.) node[] {\Large{$\times$}};
 \end{tikzpicture}
 }
\end{gathered} \;,
\ee
where $\begin{gathered}
\resizebox{0.5cm}{!}{
\begin{tikzpicture}[]
\draw[black,thick,->] (0,0) arc (180:0:0.5cm  and 0.5cm); 
\draw[](0.5,0.) node[] {\Large{$\times$}};
 \end{tikzpicture}
 }
\end{gathered}=-\ii \pi J(\omega_{nm})n(\omega_{nm})$~\footnote{This can be proven by setting $z=z(\theta)=\omega_{nm}(1+sr e^{\ii\theta})$, where $s=sgn(\omega_{nm})$ and integrating over $\theta$ from $\pi$ to 0 with $dz(\theta)=\ii\omega_{nm}sr e^{\ii\theta} d\theta$, and taking the limit $r\rightarrow 0$.} is the contribution from the infinitely small detour to avoid the singularity $\times\equiv \omega_{nm}$ in the integration path. We can also chose to include the pole at $z=\omega_{nm}$, with a counterclockwise detour below the pole. In this case we sum the residuum $2\ii\pi J(\omega_{nm})n(\omega_{nm})$ and subtract the detour which contributes as $\ii\pi J(\omega_{nm})n(\omega_{nm})$. Alternatively, one can keep the $\lambda\neq 0$, perform the contour integral, and take the limit afterward, as done in Eq.~\eqref{ImW} below. The pole expansion of the Bose-Einstein distribution reads
\be{}
n(\omega)&=\frac{1}{2}
\coth\left(\frac{\beta\hbar\omega}{2}\right)-\frac{1}{2}
=\frac{1}{\beta\hbar\omega} +\frac{1}{\beta\hbar}\sum_{n=1}^\infty
\frac{2\omega}{(\omega+\ii\nu_n)(\omega-\ii\nu_n)}-\frac{1}{2}\;.
\ee
Here, $\nu_n:=2\pi n k_B T/\hbar $ are the Matsubara frequencies. For a Ohmic-Drude spectral density function $J(\omega)=\alpha\omega[1+(\omega/\omega_c)^2]^{-1}$
\be{ImW}
W^{(b)}_{nm}/\hbar^2
=&2\pi\ii{\rm Res}_{\ii\omega_c}\frac{\alpha z \omega_c^2 n(z)}{(z+\ii\omega_c)(z-\ii\omega_c)(z-\omega_{nm})}+\frac{2\pi\ii}{\hbar\beta}\sum_{n=1}^\infty{\rm Res}_{\ii\nu_n}\left\{\frac{2\alpha z^2 \omega_c^2}{(z+\ii\omega_c)(z-\ii\omega_c)(z+\ii\nu_n)(z-\ii\nu_n)(z-\omega_{nm})}
\right\}_{
\begin{gathered}
\resizebox{1.cm}{!}{
\begin{tikzpicture}[]
\draw[thick,->] (0,0) -- (0.5,0);
\draw[thick] (0.4,0) -- (1.0,0);
\draw[black,thick,<-] (0,0) arc (180:0:0.5cm  and 0.5cm);
 \end{tikzpicture}
 }
\end{gathered}}\\
&+\lim_{\lambda\to 0^+}2\pi\ii{\rm Res}_{\omega_{nm}+\ii\lambda}\frac{\alpha z \omega_c^2 n(z)(z-\omega_{nm})}{(z^2+\omega_c^2)[z-(\omega_{nm}+\ii\lambda)][z-(\omega_{nm}-\ii\lambda)]}\\
=&\pi\frac{\alpha \omega_c^2 n(\ii\omega_c)}{(\omega_c^2+\omega_{nm}^2)}(\omega_c-\ii\omega_{nm})
+\frac{2\pi\alpha\omega_c^2}{\hbar\beta}\sum_{n=1}^\infty
\frac{\nu_n(\omega_{nm}+\ii\nu_n)}{(\omega_c^2-\nu_n^2)(\omega_{nm}^2+\nu_n^2)}+
\ii \pi J(\omega_{nm})n(\omega_{nm})\;.
\ee
From Eq.~\eqref{Wab}, noting that the last term cancels with $W^{(a)}_{nm}$, we get
\be{}
W_{nm}/\hbar^2
=&\frac{\pi}{2}J(\omega_{nm})\frac{\omega_c}{\omega_{nm}}\left[\cot\left(\frac{\beta\hbar\omega_c}{2}\right)-\frac{\omega_{nm}}{\omega_c}\right]-\frac{2\pi\alpha\omega_c^2}{\hbar\beta}\sum_{n=1}^\infty
\frac{\nu_n^2}{(\omega_c^2-\nu_n^2)(\omega_{nm}^2+\nu_n^2)}\\
&-\ii\frac{\pi}{2}J(\omega_{nm})\left[\cot\left(\frac{\beta\hbar\omega_c}{2}\right)+\frac{\omega_c}{\omega_{nm}}\right]+\ii\frac{2\pi\alpha\omega_c^2}{\hbar\beta}\sum_{n=1}^\infty
\frac{\nu_n \omega_{nm}}{(\omega_c^2-\nu_n^2)(\omega_{nm}^2+\nu_n^2)}\;.
\ee
The calculation of the bath correlation function at $t=0$ goes similarly as above using the pole expansion of $\coth(x)$.
\be{}
\langle \hat{B}_l(0)\hat{B}_l(0)\rangle/\hbar^2=&\int_0^\infty d\omega J(\omega)\coth\left(\frac{\beta\hbar\omega}{2}\right)\\
=&\int_{-\infty}^\infty d\omega J(\omega)\frac{1}{2}\coth\left(\frac{\beta\hbar\omega}{2}\right)\\
=&
2\pi\ii \sum_j{\rm Res}_j \left\{J(z)\frac{1}{2}\coth\left(\frac{\beta\hbar z}{2}\right)\right\}_{
\begin{gathered}
\resizebox{1.cm}{!}{
\begin{tikzpicture}[]
\draw[thick,->] (0,0) -- (0.5,0);
\draw[thick] (0.4,0) -- (1.0,0);
\draw[black,thick,<-] (0,0) arc (180:0:0.5cm  and 0.5cm);
 \end{tikzpicture}
 }
\end{gathered}}
-
\begin{gathered}
\resizebox{1.5cm}{!}{
\begin{tikzpicture}[]
\draw[black,thick,<-] (0,0) arc (180:0:0.5cm  and 0.5cm);
 \end{tikzpicture}
 }
\end{gathered}\\
=&
2\pi\ii{\rm Res}_{\ii\omega_c}\frac{\alpha z \omega_c^2}{(z+\ii\omega_c)(z-\ii\omega_c)}\frac{1}{2}\coth\left(\frac{\beta\hbar\omega}{2}\right)+
\frac{2\pi\ii}{\hbar\beta}\sum_n{\rm Res}_{\ii\nu_n}\frac{2\alpha z^2 \omega_c^2}{(z^2+\omega_c^2)(z+\ii\nu_n)(z-\ii\nu_n)}\\
=&\frac{\pi\alpha\omega_c^2}{2}\cot\left(\frac{\beta\hbar\omega_c}{2}\right)-
\frac{2\pi\alpha\omega_c^2}{\hbar\beta}\sum_{n=1}^\infty\frac{ \nu_n }{(\omega_c^2-\nu_n^2)}\;.
\ee
Note that the the sum over the Matsubara frequencies diverges. As shown in the main text however, the terms containing the above bath correlation function in the expression for the current, Eq.~\eqref{I2}, cancel out.

\subsection*{Fermions}

For completeness, we show how the calculation of the rates, analogous to the one of $W_{nm}$, is carried out in the fermionic case.
Setting $x:=\beta(E-\mu)$, close to the poles $z_k = 2\pi \ii(k+1/2)$, the Fermi function $
f(x)$ can be written
as $-(z-z_k^+)^{-1}$.
Choosing for the electronic density of states in the leads $D(\epsilon)\simeq 1$ (wide-band limit) and considering that along the upper semi circle $f(z)\rightarrow  1$ for $\pi/2<\phi<\pi$, and 0 between $\phi=0$ and $\pi/2$, we obtain
\be{}
\int_{-\infty}^\infty d\epsilon \frac{D(\epsilon) f(\epsilon)}{\ii 0^+ - E + \epsilon}=&-2\pi{\rm i}\sum_{k=0}^{\infty}\frac{1}{2\pi{\rm i}(k+1/2)-\beta(E-\mu)}
 -
 \begin{gathered}
\resizebox{1.5cm}{!}{
\begin{tikzpicture}[]
\draw[black,thick,<-] (0,0) arc (180:0:0.5cm  and 0.5cm);
 \end{tikzpicture}
 }
\end{gathered} \\
=&-\sum_{k=0}^{\infty}\frac{1}{k+1/2+{\rm i}(E-\mu)/(2\pi k_{\rm B}T)}-{\rm i}\frac{\pi}{2}\;.
\ee
Following Ref.~\cite{Leijnse2008},   we single out the $k=0$ term in the sum over $k$ and add and subtract the Euler-Mascheroni constant $\gamma_E=\lim_{K \to \infty}\sum_{k=1}^K 1/k-\ln(K)$. At this point, using the definition of digamma function $\psi(z)=-\gamma_E-1/z-\sum_{k=1}^\infty[1/(k+z)-1/k]$, we get
\be{integral_fermions}
\lim_{W\to\infty}\int_{-W}^W d\epsilon\frac{D(\epsilon)f(\epsilon)}{\ii 0^+ - E + \epsilon}=& {\rm Re}\psi\left(\frac{1}{2}+{\rm i}\frac{E-\mu}{2\pi k_{\rm B}T}\right)
-\ln\frac{W}{2\pi k_{\rm B}T}-{\rm i}\left[\frac{\pi}{2}-{\rm Im}\psi\left(\frac{1}{2}+{\rm i}\frac{E-\mu}{2\pi k_{\rm B}T}\right)\right]\;.
\ee
Here, $W$ is the bandwidth of the fermionic reservoir. Note that the third term on the right-hand side is equal to $-\pi f(E)$. This expression diverges logarithmically with $W$. The rates stay however finite due to an exact cancellation of the second term on the right-hand side when performing the sum over the Fock/Liouville  indices~\cite{Magazzu2022,Donarini2024}.  

\section{Steady-state coherence for a three-level truncation of the Rabi model} \label{analytical_solution}

The starting point is the partial secular master equation~\eqref{BR_partial_secular}. To obtain analytical results for the steady-state RDM, we first perform a truncation of the system to a three-level system spanned by the energy eigenstates $n=0,1,2$. Further we consider the weak qubit-oscillator coupling case close to the resonance, namely $\omega_{21}\ll \omega_{10},\omega_{20}$. Proceeding along the same lines of Ref.~\cite{Ivander2022}, the elements 12, 11, and 22 of the partial secular master equation give 
\be{}
0&=-\ii\omega_{12}\rho_{12}+\sum_m\tilde{\mathcal{K}}^{(2)}_{12mm}(0)\rho_{mm}
+\tilde{\mathcal{K}}^{(2)}_{1212}(0)\rho_{12}
+\tilde{\mathcal{K}}^{(2)}_{1221}(0)\rho_{21}\\
0&=\sum_m\Gamma_{1m}\rho_{mm}
+\tilde{\mathcal{K}}^{(2)}_{1112}(0)\rho_{12}
+\tilde{\mathcal{K}}^{(2)}_{1121}(0)\rho_{21}\\
0&=\sum_m \Gamma_{2m}\rho_{mm}
+\tilde{\mathcal{K}}^{(2)}_{2212}(0)\rho_{12}
+\tilde{\mathcal{K}}^{(2)}_{2221}(0)\rho_{21}\;,
\ee
where $\Gamma_{nm}:=\tilde{\mathcal{K}}^{(2)}_{nnmm}(0)$ are the rates connecting the populations.
Using the conservation of total probability, along with the property $\tilde{\mathcal{K}}^{(2)}_{mnm'n'}(0)=[\tilde{\mathcal{K}}^{(2)}_{nm n'm'}(0)]^*$, the system can be written as
\be{}
0&=\quad\omega_+\rho_{12}''+\Omega_+\rho_{12}'+
\tilde{\mathcal{K}}^{(2)'}_{1200}(0)\rho_{00}+ \tilde{\mathcal{K}}^{(2)'}_{1211}(0)\rho_{11}+\tilde{\mathcal{K}}^{(2)'}_{1222}(0)\rho_{22}\\
0&=-\omega_-\rho_{12}'+\Omega_-\rho_{12}''+
\tilde{\mathcal{K}}^{(2)''}_{1200}(0)\rho_{00}+ \tilde{\mathcal{K}}^{(2)''}_{1211}(0)\rho_{11}+\tilde{\mathcal{K}}^{(2)''}_{1222}(0)\rho_{22}
\\
0&=\Gamma_{10} + (\Gamma_{11}-\Gamma_{10})\rho_{11}+ (\Gamma_{12}-\Gamma_{10})\rho_{22}+2\Real
\tilde{\mathcal{K}}^{(2)}_{1112}(0)\rho_{12}\\
0&=\Gamma_{20} + (\Gamma_{21}-\Gamma_{20})\rho_{11}+ (\Gamma_{22}-\Gamma_{20})\rho_{22}+2\Real
\tilde{\mathcal{K}}^{(2)}_{2212}(0)\rho_{12}\;,
\ee
where $\omega_\pm:=\omega_{12}-\tilde{\mathcal{K}}^{(2)''}_{1212}(0)\pm\tilde{\mathcal{K}}^{(2)''}_{1221}(0)$ and $\Omega_\pm:=\tilde{\mathcal{K}}^{(2)'}_{1212}(0)
\pm\tilde{\mathcal{K}}^{(2)'}_{1221}(0)$. The solution for the coherence is 
\be{}
\rho_{12}'&=a_0\rho_{00}+a_1\rho_{11}+a_2\rho_{22}\qquad a_i:=\frac{\tilde{\mathcal{K}}^{(2)''}_{12ii}(0)-\Omega_- b_i}{\omega_-}\\
\rho_{12}''&=-b_0\rho_{00}-b_1\rho_{11}-b_2\rho_{22}\qquad b_i:=\frac{\omega_-\tilde{\mathcal{K}}^{(2)'}_{12ii}(0)+\Omega_+\tilde{\mathcal{K}}^{(2)''}_{12ii}(0)}{\omega_+\omega_-+\Omega_+\Omega_-}\;,
\ee
while the populations are the solution of 
\be{Pop}
0&=\tilde\Gamma_{10} + (\tilde\Gamma_{11}-\tilde\Gamma_{10})\rho_{11}+ (\tilde\Gamma_{12}-\tilde\Gamma_{10})\rho_{22}\\
0&=\tilde\Gamma_{20} + (\tilde\Gamma_{21}-\tilde\Gamma_{20})\rho_{11}+ (\tilde\Gamma_{22}-\tilde\Gamma_{20})\rho_{22}\\
\rho_{00}&=1-\rho_{11}-\rho_{22}\;,
\ee
with $\tilde\Gamma_{ni}:=\Gamma_{ni}+2[
\tilde{\mathcal{K}}^{(2)'}_{nn12}(0)a_i+
\tilde{\mathcal{K}}^{(2)''}_{nn12}(0)b_i]$.
\begin{figure}[ht!]
    \centering
\includegraphics[width=8.5cm]{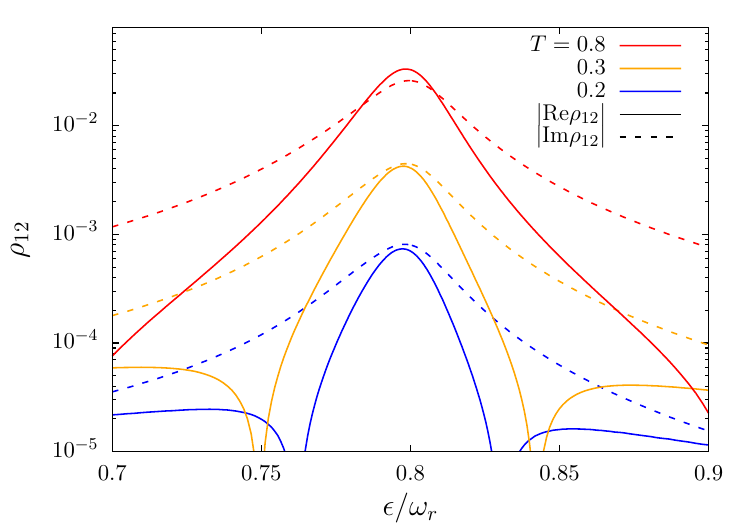}
\caption{Non-vanishing steady-state coherence at equilibrium. Coherence  $\rho_{12}$ vs. the qubit bias for the three-level truncation of the Rabi model for $\Delta=0.6~\omega_r$ and three temperatures (in units of $\hbar\omega_r/k_B$). Other parameters are $\alpha_L=\alpha_R=0.001$, $\omega_c=5~\omega_r$,  $g=0.01~\omega_r$.}
\label{fig_rho12}
\end{figure}

The condition of quasi-degenerate levels 1 and 2 allows for approximating $\tilde{\mathcal{K}}^{(2)}_{1211}(0)\simeq \tilde{\mathcal{K}}^{(2)}_{1222}(0)\equiv K\simeq \ii K''$. The solution for the coherence reads now
\be{}
\rho_{12}'\simeq &=\frac{\bar\omega K_0''}{\bar\omega^2+\Omega^2}+\frac{\bar\omega (K''-K_0'')}{\bar\omega^2+\Omega^2}(\rho_{11}+\rho_{22})\quad{\rm and}\quad
\rho_{12}''=-\frac{\Omega}{\bar\omega}\rho_{12}'\;,
\ee
with $\hat{\mathcal{K}}^{(2)}_{1200}(0)\equiv K_0\simeq \ii K_0''$, $\omega_\pm\simeq \bar\omega:=\omega_{12}-\tilde{\mathcal{K}}^{(2)''}_{1212}(0)$, and $\Omega_\pm\simeq \Omega:=\tilde{\mathcal{K}}^{(2)'}_{1212}(0)$,
and the solution for the populations is given by Eq.~\eqref{Pop} with $\tilde\Gamma_{ni}\simeq\Gamma_{ni}$. The exact solution for the steady-state coherence $\rho_{12}$ at equilibrium in the present three-level system truncation of the Rabi model is shown in Fig.~\ref{fig_rho12} as a function of the detuning between the qubit and the oscillator at weak internal coupling, $g/\omega_r\ll 1$, and at different temperatures.

\section{Analytical treatments of the Rabi model}
\label{Rabi_analytical}

\subsection{Second-order Van Vleck perturbation theory (VVPT) in $g$}
\label{sec_VVPTg}

Let us define the frequency $\bar\omega=\omega_q+\omega_r$ and the couplings $g_z= g \epsilon/\omega_q$ and $g_x= g \Delta/\omega_q$. The latter are the longitudinal and transverse coupling, respectively, of the Rabi Hamiltonian expressed in the qubit energy basis.
Within second-order VVPT in the qubit-resonator coupling $g$~\cite{Hausinger2008} the eigenenergies $E_n=\hbar\omega_n$ of the Rabi model read
 ($n\geq 1$) 
\begin{equation}
\begin{aligned}\label{eigenenergies_VVPTg}
\omega_0 &=\,-\omega_q^{(1)}/2 - g_z^2/\omega_r\;,\\
\omega_{2n-1}&=\,-\frac{\omega_q^{(n)}}{2}+\frac{\delta_n}{2} +n\omega_r - \frac{g_z^2}{\omega_r}  - \frac{g_x^2}{\bar\omega}-\frac{1}{2}\sqrt{\delta_n^2+ 4n g_x^2 }\;, \\
\omega_{2n}&=\,-\frac{\omega_q^{(n)}}{2}+\frac{\delta_n}{2} +n\omega_r - \frac{g_z^2}{\omega_r}  - \frac{g_x^2}{\bar\omega}+\frac{1}{2}\sqrt{\delta_n^2+ 4n g_x^2 }\;,
\end{aligned}
\end{equation}
where $\omega_q^{(n)}=\omega_q+2ng_x^2/\bar\omega$ and $\delta_n=\omega_q^{(n)}-\omega_r$. 
Note that this indexing provides the correct ordering of the eigenfrequencies only for small enough $g$, $\epsilon$.
The corresponding eigenvectors are 
\begin{equation}
\begin{aligned}
\label{eigenstates_VVPTg}
|0\rangle &= |\widetilde{{\rm g},0}\rangle^{(2)}\,,\\
|2n-1\rangle &= {\rm u}_{n}^-|\widetilde{{\rm e},n-1}\rangle^{(2)} + {\rm v}_{n}^-|\widetilde{{\rm g},n}\rangle^{(2)}\;,\\
|2n\rangle &= {\rm u}_{n}^+|\widetilde{{\rm e},n-1}\rangle^{(2)} + {\rm v}_{n}^+|\widetilde{{\rm g},n}\rangle^{(2)}\;,
\end{aligned}    
\end{equation}
where the coefficients are given by
\begin{equation}
\begin{aligned}
\label{}
{\rm u}_{n}^\pm &:=\frac{\delta_n \pm \sqrt{\delta_n ^2+4n g_x^2}}{\sqrt{\left(\delta_n \pm \sqrt{\delta_n ^2+4n g_x^2}\right)^2+4n g_x^2}}\;,\qquad{\rm and}\qquad
{\rm v}_{n}^\pm  :=\frac{-2\sqrt{n}g_x}{\sqrt{\left(\delta_n \pm \sqrt{\delta_n ^2+4n g_x^2}\right)^2+4n g_x^2}}\;.
\end{aligned}    
\end{equation}

The explicit form of the transformed states $ |\widetilde{e/g,n}\rangle^{(2)}$ in terms of the uncoupled energy eigenbasis $\{\Ket{{\rm g},n},\Ket{{\rm e},n}\}$ is
\be{}
\frac{|\widetilde{{\rm g},0}\rangle^{(2)}}{\mathcal{N}_{g,0}}&=\,
\Ket{{\rm g},0}+f(1)\Ket{{\rm e},0}+\frac{g_z}{\omega_r}\Ket{{\rm g},1}+\frac{g_x}{\bar\omega}\Ket{{\rm e},1}\;,\\
\frac{|\widetilde{{\rm g},n}\rangle^{(2)}}{\mathcal{N}_{g,n}}&=\,
-\sqrt{n}\frac{g_z}{\omega_r}\Ket{{\rm g},n-1}+\Ket{{\rm g},n}+f(n+1)\Ket{{\rm e},n}+\sqrt{n+1}\frac{g_z}{\omega_r}\Ket{{\rm g},n+1}+\sqrt{n+1}\frac{g_x}{\bar\omega}\Ket{{\rm e},n+1}\quad(n\geq 1)\;,\\
\frac{|\widetilde{{\rm e},0}\rangle^{(2)}}{\mathcal{N}_{e,0}}&=\,-f(1)\Ket{{\rm g},0}+\Ket{{\rm e},0}-\frac{g_z}{\omega_r}\Ket{{\rm e},1},\\
\frac{|\widetilde{{\rm e},n}\rangle^{(2)}}{\mathcal{N}_{e,n}}&=\,-\sqrt{n}\frac{g_x}{\bar\omega}\Ket{{\rm g},n-1}+\sqrt{n}\frac{g_z}{\omega_r}\Ket{{\rm e},n-1}-f(n+1)\Ket{{\rm g},n}+\Ket{{\rm e},n}-\sqrt{n+1}\frac{g_z}{\omega_r}\Ket{{\rm e},n+1}
\quad(n\geq 1)\;,\\
\ee
where $\mathcal{N}_i$ are the normalization factors and where $f(n):=(g_zg_x/\omega_r)[n/\omega_r-(n-1)/\bar\omega]$. 

\subsubsection{Zero bias, $\epsilon=0$}

At zero bias, the ground and first excited state read
\begin{equation}
\begin{aligned}
\label{eigenstates_VVPTg}
|0\rangle &\propto \,
\Ket{{\rm g},0}+\frac{g}{\Delta+\omega_r}\Ket{{\rm e},1}\;,\\
|1\rangle &\propto {\rm u}_{1}^-\Ket{{\rm e},0}+ {\rm v}_{1}^-   \left(
\Ket{{\rm g},1}+\frac{\sqrt{2}g}{\Delta+\omega_r}\Ket{{\rm e},2}\right)\;.
\end{aligned}    
\end{equation}
The matrix elements of the system's coupling operators ($\hat{Q}_L=\hat{a}+\hat{a}^\dag$ and $\hat{Q}_R=\hat\sigma_z\epsilon/\omega_q-\hat\sigma_x \Delta/\omega_q$, in the qubit energy basis) between the ground and first excited state read
\be{QLR_VVPT}
Q_{L,01}\propto{\rm v}_1^-\left(1+\frac{\sqrt{2}g}{\Delta+\omega_r}
\right)+{\rm u}_1^-
\frac{g}{\Delta+\omega_r}
\;,\qquad Q_{R,01}\propto{\rm u}_1^-
+
{\rm v}_1^-
\frac{g}{\Delta+\omega_r}\;.
\ee
The explicit expression for the coefficients is
\begin{equation}
\begin{aligned}
\label{uv_VVPT}
{\rm u}_1^- &:=\frac{\delta_1 - \sqrt{\delta_1 ^2+4 g^2}}{\sqrt{\left(\delta_1 - \sqrt{\delta_1 ^2+4 g^2}\right)^2+4 g^2}}\;,\qquad{\rm and}\qquad
{\rm v}_{1}^-  :=\frac{-2g}{\sqrt{\left(\delta_1 - \sqrt{\delta_1 ^2+4 g^2}\right)^2+4 g^2}}\;,
\end{aligned}    
\end{equation}
where $\delta_1=\Delta+2g^2/(\Delta+\omega_r)-\omega_r$. Note that, at zero bias, these coefficients peak at the value of $\Delta<\omega_r$ for which the condition $\delta_1=0$ is satisfied.

\subsection{Rotating-wave approximation (RWA)}
\label{sec_RWA}

To first order in the coupling $g$, the  Van Vleck perturbation theory expressions for the eigensystem of the quantum Rabi model, Eqs.~\eqref{eigenenergies_VVPTg}-\eqref{eigenstates_VVPTg}, reproduce the RWA results 
\begin{equation}
\begin{aligned}
\label{JCMspectrum}
\omega_0=&-\omega_q/2\;,\\
\omega_{2n-1}=& \left(n-\frac{1}{2} \right)\omega_r - \frac{1}{2}\sqrt{\delta^2 + n 4 g_x^2}\;, \\
\omega_{2n}=& \left(n-\frac{1}{2} \right)\omega_r + \frac{1}{2}\sqrt{\delta^2 + n 4g_x^2}
\end{aligned}
\end{equation}
($n\geq 1$), with the detuning $\delta$ defined as $\delta:=\omega_q-\omega_r$ and $g_x= g\Delta/\omega_q$. The corresponding eigenstates, in the energy basis of the uncoupled system, are
\begin{equation}
\begin{aligned}
\label{JCMeigenstates}
|0\rangle &= |{\rm g},0\rangle\;,\\
|2n-1\rangle &= {\rm u}_{n}^-|{\rm e},n-1\rangle + {\rm v}_{n}^-|{\rm g},n\rangle\;,\\
|2n\rangle &= {\rm u}_{n}^+|{\rm e},n-1\rangle + {\rm v}_{n}^+|{\rm g},n\rangle\;,
\end{aligned}    
\end{equation}
with coefficients
\begin{equation}
\begin{aligned}\label{QLR_RWA}
{\rm u}_n^\pm &:=\;\frac{\delta \pm \sqrt{\delta ^2 + n4g_x^2}}{\sqrt{\left(\delta \pm \sqrt{\delta^2+n4g_x^2}\right)^2+n4g_x^2}}\;,\qquad{\rm and}\qquad
{\rm v}_n^\pm :=\;\frac{-\sqrt{n}2g_x}{\sqrt{\left(\delta \pm \sqrt{\delta^2+n4g_x^2}\right)^2+n4g_x^2}}\,.
\end{aligned}
\end{equation}

Within the RWA, the relevant matrix elements of the system's coupling operators ($\hat{Q}_L=\hat{a}+\hat{a}^\dag$ and $\hat{Q}_R=\sigma_z\epsilon/\omega_q-\sigma_x \Delta/\omega_q$, in the qubit energy basis) read
\be{Q_RWA}
Q_{L,01}&={\rm v}_1^-\;,\qquad Q_{R,01}={\rm u}_1^-\frac{\Delta}{\omega_q}\\
Q_{L,02}&={\rm v}_1^+\;,\qquad Q_{R,02}={\rm u}_1^+\frac{\Delta}{\omega_q}\\
\ee
At resonance, $Q_{L,01}=Q_{L,02}=-1/\sqrt{2}$ while $Q_{R,01}=-Q_{R,02}=-1/\sqrt{2}$. If $\omega_{10}\simeq\omega_{02}$, the change of sign of the qubit matrix element yields the suppressed fourth-order conductance at weak coupling, see Fig.~\ref{fig_kappa_vs_T}(a).

\subsection{Generalized rotating-wave approximation (GRWA)}
\label{appendix_GRWA}

Within the GRWA~\cite{Irish2007,Zhang2013}, the spectrum 
 $E_n=\hbar\omega_n$ of the biased Rabi model is approximated  by 
\begin{equation}
\begin{aligned}\label{eigensys_GRWA_app}
\omega_0 &=\;-\omega_{q,0}/2- g^2/\omega_r\;,\\
\omega_{2n-1}
&=\;-\frac{\omega_{q,n}}{2}+ \frac{\delta_n}{2}+n\omega_r - \frac{g^2}{\omega_r}
-\frac{1}{2}\sqrt{\delta_n^2+\Omega_{n}^2}\;,\\
\omega_{2n}
&=\;-\frac{\omega_{q,n}}{2}+ \frac{\delta_n}{2}+n\omega_r - \frac{g^2}{\omega_r}
+\frac{1}{2}\sqrt{\delta_n^2+\Omega_{n}^2}\;.
\end{aligned}
\end{equation}
Here, $\delta_n  :=(\omega_{q,n}+\omega_{q,n-1})/2  -\omega_r$, 
$\omega_{q,n}:=\;\sqrt{\tilde\Delta_{nn}^2 + \epsilon^2}$, and $\Omega_{n} :=\;\tilde\Delta_{nn-1}(c^+_n c^+_{n-1} +  c^-_n c^-_{n-1})$, where \hbox{$c^\pm_n :=\;\sqrt{(\omega_{q,n}\pm\epsilon)/2\omega_{q,n}}$}.
Here, we have introduced the dressed qubit gap $\tilde{\Delta}_{ij}= \Delta e^{-\tilde\alpha /2}\tilde\alpha^{(i-j)/2}\sqrt{j!/i!}\;\mathsf{L}^{i-j}_j(\tilde\alpha)$ ($i\geq j$), where $\tilde\alpha:=(2g/\omega_r)^2$ and where $\mathsf{L}^{k}_n$ are the generalized Laguerre polynomials defined by the recurrence relation
\be{}
\mathsf{L}^k_{j+1}(\tilde{\alpha})=\frac{(2j+1+k-\tilde{\alpha})\mathsf{L}^k_j(\tilde{\alpha})-(j+k)\mathsf{L}^k_{j-1}(\tilde{\alpha})}{j+1}\;,
\ee
with $\mathsf{L}^k_0(\tilde{\alpha})=1$ and $\mathsf{L}^k_1(\tilde{\alpha})=1+k-\tilde{\alpha}$. 
The corresponding energy eigenstates are
\be{eigenstates_GRWA_app}
\Ket{0}&=\ket{{\Psi_{+,0}}}\;,\\
\Ket{2n-1}&={\rm u}_n^{-}\ket{{\Psi_{-,n-1}}}+{\rm v}_n^{-}\ket{{\Psi_{+,n}}}\;,\\
\Ket{2n}&={\rm u}_n^{+}\ket{{\Psi_{-,n-1}}}+{\rm v}_n^{+}\ket{{\Psi_{+,n}}}\;,
\ee
with the weights given by
\begin{equation}
\begin{aligned}\label{GRWA_quantities}
{\rm u}_n^\pm &:=\;\frac{\delta_n \pm \sqrt{\delta_n ^2+\Omega_{n}^2}}{\sqrt{\left(\delta_n \pm \sqrt{\delta_n ^2+\Omega_{n}^2}\right)^2+\Omega_{n}^2}}\;,\qquad{\rm and}\qquad
{\rm v}_n^\pm :=\;\frac{-\Omega_{n}}{\sqrt{\left(\delta_n \pm \sqrt{\delta_n ^2+\Omega_{n}^2}\right)^2+\Omega_{n}^2}}\,.
\end{aligned}
\end{equation}
The states 
\be{Psi_pm}
\Ket{\Psi_{\pm,j}}
=\,c^\mp_j\Ket{-_z j_-}\pm c^\pm_j\Ket{+_z j_+}
\ee
are superpositions of the displaced states $\ket{\pm_z j_\pm}=\exp[ g\sigma_z(a-a^\dag)/\omega_r]\ket{\pm_z}|j\rangle\equiv \ket{\pm_z}D(\pm g/\omega_r)|j\rangle$, where $\{\ket{\pm_z}\}$ is the qubit localized basis, i.e. $\{\ket{+_z},\ket{-_z}\}\equiv\{\ket{\circlearrowright },\ket{\circlearrowleft}\}$, see the main text, and $D(x)=\exp[x(a-a^\dag)]$ is the displacement operator. 

The two-level system truncation of the Rabi model
 gives for the gap $\omega_{10} =\omega_1 - \omega_0$ of the effective TLS
\begin{equation}
\begin{aligned}\label{}
\omega_{10} &=\;\omega_{q,0} - \frac{\delta_1}{2}
-\frac{1}{2}\sqrt{\delta_1^2+\Omega_1^2}\;.
\end{aligned}
\end{equation}
The relevant matrix elements are $Q_{l,01}=\bra{0}\hat{Q}_l\ket{1}$, where $\hat{Q}_L=a+a^\dag$ and $\hat{Q}_R=\sigma_z$ (in the qubit localized basis). Using $D(x)aD^\dag(x)=a+x$ and $D(x)a^\dag D^\dag(x)=(D(x)a D^\dag(x))^\dag= a^\dag + x^*$,  
Eqs.~\eqref{eigenstates_GRWA_app}  and~\eqref{Psi_pm}
give
\be{matrix_elements_GRWA_app}
Q_{L,01} &=\, \frac{4 g}{\omega_r}{\rm u}_1^-  c^-_0 c^+_0 + {\rm v}_1^-(c^-_0 c^-_1 + c^+_0 c^+_1)
\;,\qquad
Q_{R,01} &=\,-2{\rm u}_1^- c^-_0 c^+_0
\;.
\ee

\subsubsection{Zero bias, $\epsilon=0$}

The first two energy levels and the corresponding eigenstates read
\begin{equation}
\begin{aligned}\label{}
\omega_0 &=\;-\frac{1}{2}\tilde\Delta_{00}-\frac{g^2}{\omega_r}\;,\,\quad\qquad\quad\qquad\qquad\qquad\qquad\qquad\qquad\qquad\qquad\qquad\qquad\; \Ket{0}=\ket{{\Psi_{+,0}}}\;,\\
\omega_1 &=\; \frac{1}{2}\left(\frac{\tilde\Delta_{00}-\tilde\Delta_{11}}{2} + \omega_r\right)-\frac{g^2}{\omega_r}-\frac{1}{2}\sqrt{\left(\frac{\tilde\Delta_{00}+\tilde\Delta_{11}}{2} - \omega_r\right)^2+(\tilde\Delta_{10})^2}\;,\qquad \Ket{1}={\rm u}_1^{-}\ket{{\Psi_{-,0}}}+{\rm v}_1^{-}\ket{{\Psi_{+,1}}}\;, 
\end{aligned}
\end{equation}
and, since $c^\pm_j(\epsilon=0)=1/\sqrt{2}$,
\be{}
\Ket{\Psi_{\pm,j}}
=\,\frac{1}{\sqrt{2}}\left(\Ket{-_z j_-}\pm \Ket{+_z j_+}\right)\,.
\ee
The gap $\omega_{10} =\omega_1 - \omega_0$ of the effective qubit is
\begin{equation}
\begin{aligned}\label{DeltaeffGRWAzerobias}
\omega_{10}
&=\;\tilde\Delta-\frac{1}{2}\left(\tilde\Delta-\omega_r -\frac{\tilde\alpha\tilde\Delta}{2}\right)
-\frac{1}{2}\sqrt{\left(\tilde\Delta-\omega_r -\frac{\tilde\alpha\tilde\Delta}{2}\right)^2+\tilde\alpha\tilde\Delta^2}
\;,
\end{aligned}
\end{equation}
where we have defined $\tilde\Delta:=\Delta\exp(-\tilde\alpha/2)=\tilde\Delta_{00}$ and used $\tilde\Delta_{11}=\Delta\exp(-\tilde\alpha/2)(1-\tilde\alpha)$ and $\tilde\Delta_{01}=\sqrt{\tilde\alpha}\tilde\Delta$. 
The matrix elements of $\hat{Q}_l$ in the basis $\{\Ket{0},\Ket{1}\}$ read
\be{QLR_GRWA}
Q_{L,01} &=\, \frac{2 g}{\omega_r}{\rm u}_1^-
+{\rm v}_1^-\;,\qquad\,
Q_{R,01} =\,-{\rm u}_1^-\;.
\ee

From Eq.~\eqref{GRWA_quantities}, one can see that, due to the downward-renormalization of the qubit frequencies $\omega_{{\rm q},n}$, the coefficients ${\rm u}_1^-$ and ${\rm v}_1^-$ peak at a bare value of $\Delta>\omega_r$ for which the condition $\delta_n=0$ is satisfied. Note that the GRWA is perturbative in $\tilde\Delta$. However, this line of reasoning accounts qualitatively for the maxima $\Delta>\omega_r$ at zero bias in the USC regime found in Fig.~\ref{fig_kappa_vs_D}(a).

\end{widetext}


%

\end{document}